\documentclass[12pt]{article}

\usepackage[margin=1in]{geometry}

\usepackage{enumitem} 

\usepackage{centernot}

\usepackage{placeins}
\usepackage{flafter}

\usepackage{times}

\usepackage{setspace} 
\usepackage{caption}
\usepackage{comment}

\usepackage{amsmath}
\usepackage{amssymb}
\usepackage{amsthm} 
\usepackage{mathtools} 
\usepackage{bbm} 
\usepackage{centernot} 

\usepackage{dcolumn}
\newcolumntype{d}[1]{@{}D{.}{.}{#1}@{}} 

\usepackage{algorithm}
\usepackage{algorithmic}

\usepackage[utf8]{inputenc} 
\usepackage[english]{babel}

\usepackage{graphicx} 

\usepackage{float}

\usepackage{array} 
\usepackage{arydshln}

\usepackage{multirow}

\newtheorem{theorem}{Theorem}
\newtheorem{assumption}{Assumption}

\newtheorem{corollary}{Corollary}

\usepackage{appendix}

\numberwithin{equation}{section}

\newcommand\blfootnote[1]{
  \begingroup
  \renewcommand\thefootnote{}\footnote{#1}
  \addtocounter{footnote}{-1}
  \endgroup
}

\usepackage[hang]{footmisc}
\usepackage{hyperref}
\hypersetup{colorlinks=true, allcolors=black}

\usepackage[authoryear,round,sort,comma,longnamesfirst]{natbib} 
\usepackage{myapalike}
\begin{document}

\begin{titlepage}
\begin{center}
\linespread{1.2}
\Large{\textbf{A fully nonlinear structural vector autoregressive model identified via independent innovation analysis}}\\

\vspace{0.5cm} 
\Large{Savi Virolainen}\\
\large{University of Helsinki}\\
\vspace{1.0cm}

\begin{abstract}
\noindent We develop a fully nonlinear structural vector autoregressive framework in which the contemporaneous structural mapping may be nonlinear and non-additive. Identification is achieved by exploiting variation in the conditional distributions of the mutually independent structural shocks induced by an observed exogenous variable. Specifically, a general contrastive learning framework that makes use of this variation together with the assumed exponential-family structure is employed to recover the shocks. Existing independent innovation analysis results identify such shocks only up to arbitrary componentwise invertible transformations, which is generally insufficient for structural econometric analysis. We strengthen this result by imposing a structured exponential-family specification for the conditional shock distributions. With the imposed sufficient statistics, the remaining ambiguity is reduced to a one-parameter transformed-scale map for each shock. We then show that, under a logistic specification used for the natural parameters, the identification is further strengthened up to permutation and componentwise sign changes. Once the shocks have been recovered, the fully nonlinear structural vector autoregression can be estimated using feed-forward neural networks, motivated by their universal approximation capabilities. The empirical application studies asymmetries in the responses of U.S. industrial production to the real oil price shock. We find modest asymmetries with respect to the sign of the shock and state of the economy. The accompanying R package iiasvar implements the introduced methods.

\end{abstract}

\vfill

\blfootnote{The author thanks the Research Council of Finland (Grant 347986) and the Yrjö Jahnsson Foundation (Grant 20247849) for the financial support.}
\blfootnote{Contact address: Savi Virolainen, Faculty of Social Sciences, University of Helsinki, P. O. Box 17, FI–00014 University of Helsinki, Finland; e-mail: savi.virolainen@helsinki.fi. ORCiD ID: 0000-0002-5075-6821.}

\end{center}
\end{titlepage}

\section{Introduction}
Structural vector autoregressive (SVAR) models play a central role in empirical macroeconomics by providing a framework for interpreting economic fluctuations through latent structural shocks and their dynamic propagation. In conventional SVAR analysis, identification is typically achieved by imposing restrictions on the contemporaneous relationship between reduced-form innovations and structural shocks. Much of the literature has focused on linear specifications, where the reduced-form innovations are assumed to be linear combinations of the structural shocks and identification may be achieved, for example, through exclusion restrictions, external instruments, heteroskedasticity, or non-Gaussianity \citep[see, e.g.,][]{Kilian+Lutkepohl:2017}. 

While linear SVAR models are useful in many applications, they are not able to capture nonlinear features of economic dynamics. A growing literature has therefore considered nonlinear extensions of SVAR models. In nonlinear systems, the effects of shocks may depend on the state of the economy, the signs or sizes of shocks, or other nonlinear features of the transmission mechanism. More generally, the relationship between structural shocks and observables may itself be nonlinear, allowing for nonlinear interactions between shocks and lagged variables as well as non-additive transmission of the effects of the shocks. 

At the same time, the recent theoretical work of \cite{Kolesar+Plagborg-Moller:2025} has demonstrated that statistical identification strategies developed under linearity may fail when the structural relationship between the reduced-form innovations and shocks is nonlinear, even if the shocks themselves satisfy otherwise standard assumptions. In particular, both identification by heteroskedasticity (\citealp{Rigobon:2003}, \citealp{Lanne+Lutkepohl:2010}, \citealp{Lutkepohl+Netsunajev:2017}, \citealp{Lewis:2021}, \citealp{Virolainen:2025}, and others) and identification by non-Gaussianity (\citealp{Lanne+Meitz+Saikkonen:2017}, \citealp{Lanne+Luoto:2021}, \citealp{Lanne+Liu+Luoto:2023}, and others) can fail in such a case. While \cite{Virolainen2:2024} has recently extended identification by non-Gaussianity to smooth-transition SVAR models, also his framework assumes that the mapping from structural shocks to reduced-form innovations is linear in each point of time, with nonlinearity arising only through time-variation in the parameter values. This motivates the development of identification frameworks specifically designed for fully nonlinear SVAR models.

This paper develops a fully nonlinear SVAR framework in which the shocks are identified statistically by imposing structure on the shock process itself and by utilizing independent innovation analysis based on general contrastive learning (IIA-GCL). Our framework builds on the independent innovation analysis approach of \citet{Morioka+Halva+Hyvarinen:2021}, who study identification of structural shocks in highly general nonlinear vector autoregressive models. Their IIA-GCL framework exploits conditional mutual independence and exponential-family structure together with variation in the conditional distributions of the structural shocks that is induced by an observed auxiliary variable. Under suitable conditions, this auxiliary variable modulation facilitates recovery of the structural shocks even when the contemporaneous structural mapping is fully nonlinear and shocks do not enter additively. However, the IIA-GCL framework of \citet{Morioka+Halva+Hyvarinen:2021} recovers the shocks only up to permutation and componentwise invertible transformations. From the perspective of nonlinear independent component analysis, this constitutes a strong identification result \citep[see, e.g.,][]{Hyvarinen+Pajunen:1999}. However, the remaining ambiguity is generally too large for structural econometric analysis, as economically meaningful shock interpretations are not preserved under arbitrary componentwise invertible transformations.

We strengthen the identification result of \citet{Morioka+Halva+Hyvarinen:2021} substantially by imposing a structured conditional shock distribution based on a particular exponential-family specification whose natural parameters vary systematically with the observed auxiliary variable. This additional structure reduces the componentwise nonlinear ambiguity to permutation, sign, and a one-parameter class of transformed-scale maps. We then show that the remaining transformed-scale ambiguity is generically eliminated under our logistic natural-parameter specification. The resulting framework thus identifies structural shocks up to permutation and componentwise sign changes in a fully nonlinear SVAR model, where the contemporaneous structural mapping itself need not be linear or additive.

Once the structural shocks have been recovered, estimation of the nonlinear SVAR model becomes a nonlinear regression problem. As proposed by \citet{Morioka+Halva+Hyvarinen:2021}, the model can be implemented in practice as a feed-forward neural network, motivated by their universal approximation capabilities \citep[see][]{Hornik+Stinchcombe+White:1989}. This flexible approach allows the dynamic relationships between the shocks and included variables to be learned from the data without having to specify any particular functional form for this relationship. Following the statistical identification literature \citep[see, e.g.,][]{Lanne+Meitz+Saikkonen:2017}, we propose labelling the identified shocks based on their estimated effects on the variables, here represented by the generalized impulse response functions \citep{Koop+Pesaran+Potter:1996}.

We illustrate the use of our methods in an empirical application studying asymmetries in the effects of the real oil price shock. Specifically, we consider a monthly bivariate system of U.S. industrial production growth and real oil price growth, covering the period from 1974:2 to 2026:1. Since the periods of elevated macroeconomic uncertainty can plausibly feature changes in the variances and the shapes of macroeconomic shock distributions, we assume that shock distributions are modulated by the (lagged) macro uncertainty index of \citet{Jurado+Ludvigson+Ng:2015}. We find modest asymmetries in the effects of the oil price shock with respect to its sign and the state of the economy. In particular, our findings suggest stronger effects for negative than positive oil price shocks, differing from the post-1973 (aggregate industrial production) result of \citet{Herrera+Lagalo+Wada:2011} for shocks of typical magnitude. Moreover, we find the effects of the oil price shock stronger when the recent industrial production growth is low than when it is high.

The remainder of the paper is organized as follows. Section~\ref{sec:model} introduces the nonlinear SVAR framework and the conditional shock distribution assumptions. Section~\ref{sec:identification} presents the identification result based on our modified IIA-GCL and summarizes the related Monte Carlo experiment. Section~\ref{sec:FFNN-SVAR} discusses the feed-forward neural-network implementation of the SVAR model and structural analysis. Section~\ref{sec:empirical} presents the empirical application, and Section~\ref{sec:conclusion} concludes. The appendices provide the proofs of the identification results, computational details, and further details on the empirical application. The accompanying R package iiasvar \citep{iiasvar} implements the introduced methods.

\section{The model}\label{sec:model}
\subsection{The nonlinear SVAR model}\label{sec:NVAR}
Consider the $d$-dimensional time series $\{y_t\}_{t\in\mathbb{N}}$ of interest, and the general nonlinear SVAR model of autoregressive order $p$:
\begin{equation}\label{eq:genvar}
y_t = f(\boldsymbol{y}_{t-1}, e_t),
\end{equation}
where $f:\mathbb{R}^{d(p+1)}\rightarrow \mathbb{R}^d$ represents the (unknown) functional form of the model, $\boldsymbol{y}_{t-1}=(y_{t-1},\ldots,y_{t-p})$, and $e_t=(e_{1t},\ldots,e_{dt})$ is a vector of unobserved structural shocks. In contrast to linear SVAR models, the function $f$ is allowed to be fully nonlinear in both past observations and the contemporaneous shocks. This general formulation accommodates a wide range of nonlinear dynamics, including state-dependent impulse responses and nonlinear propagation mechanisms.

Our main goal is to recover the structural shocks $e_t$ from the observed process $y_t$. As is well known, without further restrictions, such recovery is not possible due to the inherent non-uniqueness of nonlinear decompositions. The identification strategy developed in this paper builds on the independent innovation analysis (IIA) framework of \cite{Morioka+Halva+Hyvarinen:2021}, which exploits mutual independence together with variation in the shock distributions to resolve this ambiguity.

Following \cite{Morioka+Halva+Hyvarinen:2021}, it is convenient to consider the model in the following augmented form ("the mixing model"):
\begin{equation}\label{eq:ftilde}
\begin{bmatrix}
y_t \\
\boldsymbol{y}_{t-1}
\end{bmatrix}
= \tilde{f}\left( 
\begin{bmatrix}
e_t \\
\boldsymbol{y}_{t-1}
\end{bmatrix}
\right)
=
\begin{bmatrix}
f(\boldsymbol{y}_{t-1},e_t) \\
\boldsymbol{y}_{t-1}
\end{bmatrix},
\end{equation}
where $\tilde{f}:\mathbb{R}^{d(p+1)}\rightarrow \mathbb{R}^{d(p+1)}$ is the augmented model, which incorporates~\eqref{eq:genvar} in the first $d$ entries and identity mapping for $\boldsymbol{y}_{t-1}$ in the remaining $dp$ entries. It is assumed that the augmented model is invertible (i.e., bijective) and twice continuously differentiable. In particular, the inverse of the augmented model $\tilde{f}$ ("the demixing model") can be expressed as:
\begin{equation}\label{eq:gtilde}
\begin{bmatrix}
e_t \\
\boldsymbol{y}_{t-1}
\end{bmatrix}
=
\tilde{g}\left( 
\begin{bmatrix}
y_t \\
\boldsymbol{y}_{t-1}
\end{bmatrix}
\right)
=
\begin{bmatrix}
g(y_t,\boldsymbol{y}_{t-1}) \\
\boldsymbol{y}_{t-1}
\end{bmatrix},
\end{equation}
where $\tilde{g}:\mathbb{R}^{d(p+1)}\rightarrow \mathbb{R}^{d(p+1)}$ is the augmented demixing model of the (true) augmented model $\tilde{f}$, and $g(y_t,\boldsymbol{y}_{t-1})=(g_1(y_t,\boldsymbol{y}_{t-1}),\ldots,g_d(y_t,\boldsymbol{y}_{t-1}))\in\mathbb{R}^d$ represents a mapping from temporally consecutive observations to the shocks at each time period.

\subsection{Assumptions about the shocks}\label{sec:shockassumptions}

As described by \cite{Morioka+Halva+Hyvarinen:2021}, identification of the shocks requires assumptions on how their distributions vary over time. In the present framework, this variation is governed by an exogenous auxiliary variable $u_t$, which may be interpreted as describing the state of the shock process. The key idea is that, while mutual independence separates the shocks cross-sectionally, time variation in their conditional distributions provides the additional information needed for identification.

These requirements are formalized in the following assumption, which is close in spirit to Assumption~A1 of \cite{Morioka+Halva+Hyvarinen:2021}, but is stated here in a form adapted to the nonlinear SVAR setting. In particular, we make explicit that the auxiliary variable $u_t$ is observed and exogenous, is not included in the lag vector $\boldsymbol y_{t-1}$, and that the structural shocks are conditionally independent of $\boldsymbol y_{t-1}$ given $u_t$. 
\begin{assumption}\label{as:shocks}
The following conditions hold:
\begin{enumerate}[label=(\roman*)]
    \item For each $i=1,\ldots,d$, the distribution of the shock $e_{it}$ is modulated by the observed exogenous auxiliary variable $u_t$, which is not included in the lag vector $\boldsymbol{y}_{t-1}$.

    \item Conditional on $u_t$, the shock vector $e_t$ is independent of the lag vector $\boldsymbol{y}_{t-1}$.

    \item For each $t$, the shocks $e_{1t},\ldots,e_{dt}$ are mutually conditionally independent given $u_t$.

    \item Conditional on $u_t$, the shock vector $e_t$ has density
    \begin{equation}\label{eq:shockdist}
        p(e_t\mid u_t) = \prod_{i=1}^d \frac{Q_i(e_{it})}{Z_i(u_t)} \exp\left\{\sum_{j=1}^k q_{ij}(e_{it})\lambda_{ij}(u_t)\right\},
    \end{equation}
    where $Q_i$ are the base measures, $Z_i$ are the normalizing constants, $k\geq 2$ is the number of sufficient statistics, $q_{ij}(e_{it})$ are the sufficient statistics, and $\lambda_{ij}(u_t)$ are the corresponding natural parameters.
\end{enumerate}
\end{assumption}

Assumption~\ref{as:shocks} gives the general exponential-family setting underlying IIA. As in \cite{Morioka+Halva+Hyvarinen:2021}, this assumption allows the conditional shock distribution to be learned in a flexible manner. However, the resulting identification is then only up to permutation and component-wise invertible transformations. In this paper, we sharpen the identification result further by imposing additional structure with restrictions on the sufficient statistics.

More specifically, we assume a particular exponential-family form with two sufficient statistics. The family is chosen to satisfy several criteria simultaneously. First, the shocks should have zero conditional mean, so that they retain the usual interpretation as structural shocks. Second, the family should fit into the exponential-family framework of Assumption~\ref{as:shocks}. Third, it should involve at least two sufficient statistics, which is essential for obtaining enough variation to facilitate identification. Fourth, the sufficient statistics should be smooth enough for the subsequent theoretical analysis. Finally, the family should allow the auxiliary variable $u_t$ to induce economically meaningful changes in the conditional shock distributions.
These requirements lead to the following assumption.
\begin{assumption}\label{as:suffstats}
Assumption~\ref{as:shocks} is satisfied with $k=2$, $Q_i(e_{it})=1$, $q_{i1}(e_{it})=e_{it}^2$, $q_{i2}(e_{it})=\sqrt{e_{it}^2+\varepsilon}-\sqrt{\varepsilon}$, $\varepsilon>0$, and $\lambda_{i1}(u_t)<0$ for all $i=1,\ldots,d$ and $u_t$.
\end{assumption}

Assumption~\ref{as:suffstats} strengthens the generic IIA framework of \cite{Morioka+Halva+Hyvarinen:2021} by fixing the sufficient statistics, thereby restricting the admissible family of conditional shock distributions. The first sufficient statistic is quadratic as in the Gaussian distribution, while the second introduces an additional nonlinear term that allows the conditional density to deviate from Gaussianity. The smoothing constant $\varepsilon>0$ ensures that $q_{i2}(e_{it})$ is differentiable at zero. While the resulting family is somewhat nonstandard, it is deliberately constructed to satisfy the set of above-described requirements that are difficult to meet simultaneously within standard parametric families. The practical appeal of the imposed distribution is that, through the natural parameters, it can accommodate changes in both scale and shape based on the level of the auxiliary variable $u_t$, and thereby allows the probability of large shocks to differ across economic states. Our specifications for the natural parameters are introduced next, and for intuition, we also illustrate how the resulting density can change with $u_t$.

\subsection{Specification of the natural parameters}\label{sec:natpar}

The density~\eqref{eq:shockdist} together with Assumption~\ref{as:suffstats} specifies the functional form of the conditional shock distribution, but leaves the natural parameters $\lambda_{ij}(u_t)$ unspecified. These parameters can be modeled as functions of the auxiliary variable $u_t$ as described below.

\begin{figure}[!t]
\centering
\includegraphics[width=\textwidth - 2cm]{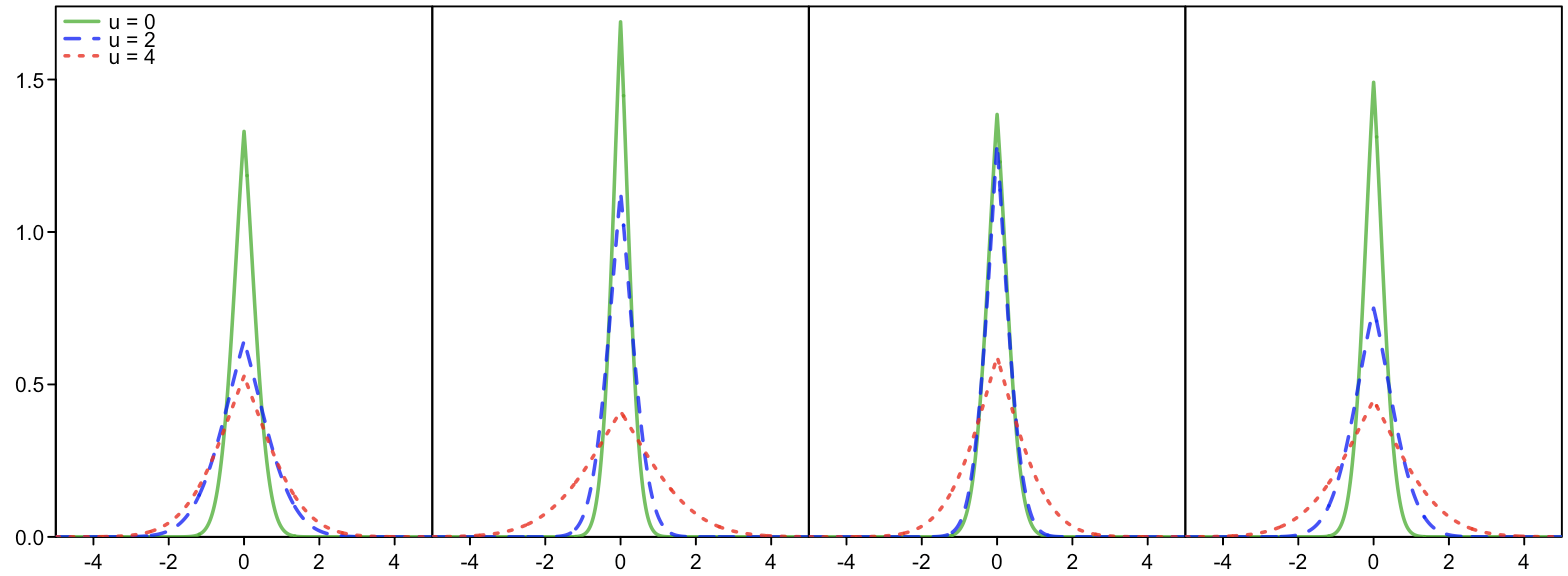}
\caption{Conditional densities implied by the logistic specification. The panels correspond, from left to right, to the parameter value configurations
$(a_{i1},b_{i1},\mathfrak{c}_{i1},\kappa_{i1},a_{i2},b_{i2},\mathfrak{c}_{i2},\kappa_{i2})=(-1.10,1.20,1.4,1.8,0.25,-0.12,2.4,1.4)$,
$(-1.40,1.85,2.5,2.0,0.20,-0.18,3.0,1.5)$,
$(-1.05,1.45,3.0,1.7,0.30,-0.22,1.7,1.7)$, and
$(-1.20,1.55,2.0,7.5,0.30,-0.20,3.6,5.0)$.
Within each panel, the density is plotted for $u_t\in\{0,2,4\}$ with a green solid line, a blue dashed line, and a red dotted line, respectively.}
\label{fig:density_logistic}
\end{figure}

Under Assumption~\ref{as:suffstats}, the conditional density of $e_{it}$ can be written as
\begin{equation}\label{eq:shock_dist_det}
p(e_{it}\mid u_t)\propto \exp\!\left\{ -\tau_i(u_t)^{-1} e_{it}^2 -\tau_i(u_t)^{-1/2}\rho_i(u_t)^{-1} \left(\sqrt{e_{it}^2+ \varepsilon}-\sqrt{\varepsilon}\right) \right\},
\end{equation}
with the natural parameter specifications
\begin{align}
\lambda_{i1}(u_t) &= -\tau_i(u_t)^{-1} \ \ \text{and}\label{eq:lambda1} \\
\lambda_{i2}(u_t) &= -\tau_i(u_t)^{-1/2}\rho_i(u_t)^{-1},\label{eq:lambda2}
\end{align}
where $\tau_i(u_t)>0$ governs the overall scale of the distribution and $\rho_i(u_t)>0$ governs its shape. The quadratic term determines the asymptotic tail behavior, which remains Gaussian, while the second term affects the shape of the distribution closer to the center. In particular, it allows the probability of large (but not asymptotically extreme) realizations to vary with $u_t$, thereby modifying the ``shoulders'' of the distribution, which becomes closer to Gaussian  the larger the value of $\rho_i(u_t)$ is.

We consider the following two-regime logistic  specifications for $\tau_i(u_t)$ and $\rho_i(u_t)$:
\begin{align}
\Lambda_{ij}(u_t) &= \left[1 + \exp\lbrace -\kappa_{ij}(u_t - \mathfrak{c}_{ij})\rbrace\right]^{-1}, \ \ j=1,2, \\
\tau_i(u_t) &= \exp\left\{(1 - \Lambda_{i1}(u_t))a_{i1} + \Lambda_{i1}(u_t)b_{i1}\right\},\label{eq:tau} \\
\rho_i(u_t) &= \exp\left\{(1 - \Lambda_{i2}(u_t))a_{i2} + \Lambda_{i2}(u_t)b_{i2}\right\},\label{eq:rho}
\end{align}
where $a_{ij},b_{ij},\mathfrak{c}_{ij}\in\mathbb{R}$ and $\kappa_{ij}>0$, $j=1,2$, are parameters and the exponential transformation forces positivity. This specification maps the auxiliary variable into a bounded range for the scale and shape parameters, capturing transitions between different states of the shock distribution.

Figure~\ref{fig:density_logistic} illustrates the logistic specification by depicting the implied conditional density $p(e_{it}\mid u_t)$ for four different parameter value configurations and for $u_t\in\{0,2,4\}$. The panels are chosen to represent distinct qualitative scenarios. In particular, from left to right, the figure includes a balanced benchmark configuration, a case with a stronger and later scale transition, a configuration in which the shape transition begins before the scale transition, and a case with much sharper logistic transitions. Across all four panels, the specification accommodates the interpretation that larger values of $u_t$ may be associated with a larger dispersion of the conditional shock distribution, although the timing and magnitude of the change differ across parameter configurations.

\section{Identification via independent innovation analysis}\label{sec:identification}

This section introduces our approach to identifying structural shocks based on the independent innovation analysis (IIA) of \cite{Morioka+Halva+Hyvarinen:2021}. The key idea of IIA is to exploit variation in the distributions of the shocks induced by an auxiliary variable $u_t$, and to recover the shocks by learning a representation that captures this variation. Depending on the nature of the auxiliary variable $u_t$, \cite{Morioka+Halva+Hyvarinen:2021} propose three different learning frameworks: general contrastive learning (IIA-GCL) for observed and possibly continuous $u_t$, time-contrastive learning (IIA-TCL) for observed discrete $u_t$, and a hidden Markov model-based approach (IIA-HMM) for unobserved $u_t$. We focus on the IIA-GCL framework, which allows the auxiliary variable to be specified directly and interpreted as an observed exogenous driver of distributional variation in the shocks.

Our contribution builds on the IIA-GCL framework of \cite{Morioka+Halva+Hyvarinen:2021}, but imposes the exponential-family structure introduced in Section~\ref{sec:shockassumptions} on the conditional distribution of the shocks within the learning problem. This restriction plays a central role in our identification strategy. While the original IIA-GCL framework identifies the shocks up to permutation and componentwise invertible transformations, the imposed exponential-family structure allows us to reduce this ambiguity substantially, and the identification results in Section~\ref{sec:ident_res} show that the remaining transformed-scale ambiguity is generically eliminated under the logistic natural-parameter specification introduced in Section~\ref{sec:natpar}.

\subsection{General contrastive learning framework (IIA-GCL)}\label{sec:IIA-GCL}

For an observed and possibly continuous exogenous variable $u_t$, \cite{Morioka+Halva+Hyvarinen:2021} propose applying general contrastive learning (GCL) to independent innovation analysis. In the IIA-GCL framework, a feature extractor and a logistic classifier are trained jointly to distinguish the real data $(y_t,\boldsymbol{y}_{t-1},u_t)$ from a version where randomization is performed on $u_t$. The objective is to learn a representation of $(y_t,\boldsymbol{y}_{t-1})$ that captures the dependence of the shock distribution on $u_t$.

More specifically, define:
\begin{equation}\label{eq:y_tilde}
\tilde{\boldsymbol{y}}_t = (y_t,\boldsymbol{y}_{t-1},u_t)
\quad \text{and} \quad
\tilde{\boldsymbol{y}}_t^\ast = (y_t,\boldsymbol{y}_{t-1},u_t^\ast),
\end{equation}
where $u_t^\ast$ is drawn from the marginal distribution of $u_t$ independently of $(y_t,\boldsymbol{y}_{t-1})$. In practice, $u_t^\ast$ is obtained by randomly permuting the sample of $u_t$, $t=1,\ldots,T$. This preserves the marginal distribution of $u_t$ while breaking its dependence on $(y_t,\boldsymbol{y}_{t-1})$. Consequently, $\tilde{\boldsymbol{y}}_t$ follows the joint density $p(y_t,\boldsymbol{y}_{t-1},u_t)$, whereas $\tilde{\boldsymbol{y}}_t^\ast$ follows $p(y_t,\boldsymbol{y}_{t-1})p(u_t^*)$.

A binary classifier is then trained to discriminate between $\tilde{\boldsymbol{y}}_t$ and $\tilde{\boldsymbol{y}}_t^\ast$. Let $D_t=1$ indicate that an observation is real and $D_t=0$ that it is randomized. The classifier is based on a scalar-valued nonlinear regression function $r(\tilde{\boldsymbol{y}}_t)$, whose logistic transformation gives the posterior probability that the observation is real:
\begin{equation}
p(D_t=1 \mid y_t,\boldsymbol{y}_{t-1},u_t) 
 = (1 + \exp\{-r(y_t,\boldsymbol{y}_{t-1},u_t)\})^{-1}.
\end{equation}
At the population level, the classifier maximizes the objective
\begin{equation}\label{eq:popobj}
\mathcal{L}(r) = \mathbb{E}_{\tilde{\boldsymbol{y}}_t}\big[\log \Lambda^{\text{std}}(r(\tilde{\boldsymbol{y}}_t))\big] + \mathbb{E}_{\tilde{\boldsymbol{y}}_t^\ast}\big[\log (1-\Lambda^{\text{std}}(r(\tilde{\boldsymbol{y}}_t^\ast)))\big],
\end{equation}
where $\Lambda^{\text{std}}(x)=(1+e^{-x})^{-1}$ is the standard logistic function and $\mathbb{E}_x[\cdot]$ denotes expectation with respect to the distribution of $x$. In practice, this corresponds to minimizing the empirical binary cross-entropy loss:
\begin{equation}\label{eq:lossbin}
\text{LOSS}_{\text{bin}} = -\sum_{t=1}^T \left[\log \Lambda^{\text{std}}(r(y_t,\boldsymbol{y}_{t-1},u_t)) + \log (1 - \Lambda^{\text{std}}(r(y_t,\boldsymbol{y}_{t-1},u_t^\ast)))\right].
\end{equation}

A key property of this classification problem is that, when $\mathcal{L}(r)$ is maximized over all measurable functions, the maximizer is given by the log-density ratio \citep[see, e.g.,][Appendix~A, and the references therein]{Morioka+Halva+Hyvarinen:2021}:
\begin{equation}\label{eq:ldr}
r^\circ(\tilde{\boldsymbol{y}}_t) = \log\frac{p(y_t,\boldsymbol{y}_{t-1},u_t)}{p(y_t,\boldsymbol{y}_{t-1})\,p(u_t)}.
\end{equation}
Thus, the population-optimal classifier measures the extent to which the joint distribution of $(y_t,\boldsymbol{y}_{t-1},u_t)$ departs from the product distribution that would prevail under independence between $u_t$ and $(y_t,\boldsymbol{y}_{t-1})$. The role of the learned representation is therefore to capture the features of the data through which this dependence operates.

To exploit the assumed structure of the shocks, we restrict the regression function $r$ to a class of functions that is consistent with the exponential family specification introduced in Section~\ref{sec:shockassumptions}. Specifically, we consider functions of the form:
\begin{equation}\label{eq:r}
r(\tilde{\boldsymbol{y}}_t) = \sum_{i=1}^d\sum_{j=1}^2q_{ij}(h_i(y_t,\boldsymbol{y}_{t-1}))\lambda_{ij}(u_t) - \sum_{i=1}^d \log Z_i(u_t) + \phi(\boldsymbol{y}_{t-1},u_t) + \eta(\boldsymbol{h}(y_t,\boldsymbol{y}_{t-1}),\boldsymbol{y}_{t-1}),
\end{equation}
where $q_{ij}(h_i(y_t,\boldsymbol{y}_{t-1}))$ are the sufficient statistics given by Assumption~\ref{as:suffstats} and $\lambda_{ij}(u_t)$ are the corresponding natural parameter functions; $Z_i(u_t)=\int \exp\{\sum_{j=1}^2 q_{ij}(x)\lambda_{ij}(u_t)\}\operatorname{d}x$ are the normalizing constants; $\boldsymbol{h}(y_t,\boldsymbol{y}_{t-1})=(h_1(y_t,\boldsymbol{y}_{t-1}),\ldots,h_d(y_t,\boldsymbol{y}_{t-1}))\in\mathbb{R}^d$ is a candidate demixing map; $\phi(\boldsymbol{y}_{t-1},u_t)$ is a function of $(\boldsymbol{y}_{t-1},u_t)$; and $\eta(\boldsymbol{h}(y_t,\boldsymbol{y}_{t-1}),\boldsymbol{y}_{t-1})$ is a function of $(\boldsymbol{h}(y_t,\boldsymbol{y}_{t-1}),\boldsymbol{y}_{t-1})$ that does not depend on $u_t$.

The first term in~\eqref{eq:r} captures the interaction between the learned features and the auxiliary variable through the natural parameters of the conditional shock distributions, and it is this term that carries the identifying information. The remaining terms absorb contributions to the log-density ratio that are not directly informative about the shocks. In particular, the normalizing constants ensure a proper conditional density, $\phi(\boldsymbol{y}_{t-1},u_t)$ captures dependence involving only $(\boldsymbol{y}_{t-1},u_t)$, and $\eta(\boldsymbol{h}(y_t,\boldsymbol{y}_{t-1}),\boldsymbol{y}_{t-1})$ collects terms that may depend on the learned representation and lagged variables but not on $u_t$.

In the theoretical analysis, the functions $\boldsymbol{h}$, $\phi$, and $\eta$ are treated as flexible population functions of their arguments, subject to the regularity conditions imposed in the identification results below. In practical implementations, they can be parametrized using feed-forward neural networks, motivated by their universal approximation properties \citep{Hornik+Stinchcombe+White:1989}. The important point is that the restricted regression function~\eqref{eq:r} is constructed to match the separable part of the population log-density ratio~\eqref{eq:ldr} induced by the conditional shock distribution. This restriction is what makes the learned representation informative about the structural shocks. The next subsection formalizes this intuition and states the corresponding population identification results.

\subsection{The identification results}\label{sec:ident_res}

We now establish that, under suitable conditions, the IIA-GCL framework described above identifies the structural shocks. The argument proceeds in three steps. Theorem~\ref{thm:IIA-GCL} first shows that the restricted IIA-GCL representation rules out the arbitrary componentwise nonlinear indeterminacies present in the original IIA-GCL result of \citet{Morioka+Halva+Hyvarinen:2021}. With the fixed sufficient statistics in Assumption~\ref{as:suffstats}, the remaining ambiguity is reduced to permutation, componentwise sign, and a one-parameter family of componentwise transformed-scale maps. Corollary~\ref{cor:nonclosure_ident} then gives a general condition on the admissible natural-parameter class under which this remaining transformed-scale ambiguity is resolved and the shocks are identified up to permutation and componentwise sign. Finally, Corollary~\ref{cor:logistic_ident} verifies this condition generically for the logistic natural-parameter specification introduced in Section~\ref{sec:natpar}. Together, these results imply that, under the logistic specification, the shocks are generically identified up to permutation and componentwise sign.
\begin{theorem}\label{thm:IIA-GCL}
Let
\begin{equation}\label{eq:qfixed}
\boldsymbol{q}(x) = \big(q_1(x),q_2(x)\big) = \big(x^2,\, \sqrt{x^2+\varepsilon}-\sqrt{\varepsilon}\big),
\end{equation}
where $\varepsilon>0$ is fixed and known. Assume the following:
\begin{enumerate}[label=(\alph*)]
    \item Observations $(y_t,\boldsymbol{y}_{t-1},u_t)$ are generated by the nonlinear SVAR~\eqref{eq:genvar}. The augmented model $\tilde{f}$ defined in~\eqref{eq:ftilde} is bijective, twice continuously differentiable, and its inverse $\tilde{g}$ defined in~\eqref{eq:gtilde} is twice continuously differentiable.\label{cond:smooth}

    \item The shocks $e_t$ satisfy Assumptions~\ref{as:shocks} and~\ref{as:suffstats}, with the fixed sufficient statistics $q_1$ and $q_2$ in~\eqref{eq:qfixed}.\label{cond:sufstat} 
    
    \item Let $\nu$ denote the marginal law of $u_t$ and let $\mathcal U_0$ denote its support. There exists a nonempty open connected set $\mathcal Y_0\subset\mathbb R^{dp}$ and conditional densities specified for every $u\in\mathcal U_0$ such that the conditional distribution of $(e_t,\boldsymbol y_{t-1})$ given $u_t=u$ assigns probability one to $\mathbb R^d\times\mathcal Y_0$ and has Lebesgue density
    \begin{equation}\label{eq:pointwise_conditional_factorization}
    p(e_t,\boldsymbol y_{t-1}\mid u)
    =p(e_t\mid u)p(\boldsymbol y_{t-1}\mid u)
    \end{equation}
    on $\mathbb R^d\times\mathcal Y_0$. Here $p(e_t\mid u)$ is the exponential-family density in Assumptions~\ref{as:shocks} and~\ref{as:suffstats}, and both factors in~\eqref{eq:pointwise_conditional_factorization} are finite and strictly positive on their respective domains. Define
    \begin{equation}\label{eq:common_supports}
    \mathcal Z_0=\mathbb R^d\times\mathcal Y_0
    \quad\text{and}\quad
    \mathcal X_0=\tilde f(\mathcal Z_0).
    \end{equation}
    For $\boldsymbol x=(y_t,\boldsymbol y_{t-1})\in\mathcal X_0$, let $p_{\mathcal X}(\boldsymbol x\mid u)$ denote the conditional Lebesgue density induced by~\eqref{eq:pointwise_conditional_factorization} and the augmented model $\tilde f$, and let $p_{\mathcal X}(\boldsymbol x)$ denote its marginal density. Both densities are finite and strictly positive for every $(\boldsymbol x,u)\in\mathcal X_0\times\mathcal U_0$. The population log-density ratio~\eqref{eq:ldr} is then represented pointwise on this domain by
    \begin{equation}\label{eq:pointwise_ldr_theorem}
    r^\circ(\boldsymbol x,u)
    =\log p_{\mathcal X}(\boldsymbol x\mid u)-\log p_{\mathcal X}(\boldsymbol x).
    \end{equation}
    This function is jointly continuous on $\mathcal X_0\times\mathcal U_0$. Finally, $\mathcal U_0$ contains $2d+1$ distinct values $u^{(0)},u^{(1)},\ldots,u^{(2d)}$ for which the true variability matrix
    \begin{equation}\label{eq:L}
    L^0 = \left[\boldsymbol{\lambda}^0(u^{(1)})-\boldsymbol{\lambda}^0(u^{(0)}):\cdots : \boldsymbol{\lambda}^0(u^{(2d)})-\boldsymbol{\lambda}^0(u^{(0)})\right]
    \end{equation}
    is invertible, where
    \begin{equation*}
    \boldsymbol{\lambda}^0(u)
    =
    \big(\lambda_{11}^0(u),\lambda_{12}^0(u), \ldots,
    \lambda_{d1}^0(u),\lambda_{d2}^0(u)\big)
    \end{equation*}
    is the true vector of natural-parameter functions.\label{cond:variability}

    \item The population objective $\mathcal{L}(r)$ in~\eqref{eq:popobj} has at least one maximizer over the class of functions
    of the form~\eqref{eq:r}, with the fixed sufficient statistics $q_1$ and $q_2$ in~\eqref{eq:qfixed} and with learned natural-parameter functions. Every such maximizer satisfies the regularity requirements stated below. For an arbitrary maximizer, denote its feature extractor by
    \begin{equation*}
    \boldsymbol{h}^\ast(y_t,\boldsymbol{y}_{t-1}) = \big(h_1^\ast(y_t,\boldsymbol{y}_{t-1}),\ldots, h_d^\ast(y_t,\boldsymbol{y}_{t-1})\big),
    \end{equation*}
    and its learned natural-parameter vector by
    \begin{equation*}
    \boldsymbol{\lambda}^\ast(u)
    =
    \big(\lambda_{11}^\ast(u),\lambda_{12}^\ast(u), \ldots,
    \lambda_{d1}^\ast(u),\lambda_{d2}^\ast(u)\big).
    \end{equation*}
    Let $r^\ast$ denote the scalar function generated pointwise by these learned constituents through~\eqref{eq:r}. The function $r^\ast$ is jointly continuous on $\mathcal X_0\times\mathcal U_0$. The augmented mapping
    \begin{equation*}
    (y_t,\boldsymbol{y}_{t-1})
    \mapsto
    \big(\boldsymbol{h}^\ast(y_t,\boldsymbol{y}_{t-1}),\boldsymbol{y}_{t-1}\big)
    \end{equation*}
    is a $C^2$ diffeomorphism from $\mathcal{X}_0$ onto its image.\label{cond:regulatory}
\end{enumerate}
Then there exist a permutation $\pi$ of $\{1,\ldots,d\}$, constants $\iota_1,\ldots,\iota_d\in\{-1,+1\}$, and constants
$c_1,\ldots,c_d>0$ such that, for all $(y_t,\boldsymbol{y}_{t-1})\in\mathcal{X}_0$,
\begin{align}
h_i^\ast(y_t,\boldsymbol{y}_{t-1}) &= \iota_i\,T_{c_i}\!\left(g_{\pi(i)}(y_t,\boldsymbol{y}_{t-1})\right),\ \ i=1,\ldots,d, \ \ \text{where}\\
T_c(x) &= \operatorname{sgn}(x)\sqrt{\left(\sqrt{\varepsilon} + c\left(\sqrt{x^2+\varepsilon}-\sqrt{\varepsilon}\right)\right)^2 - \varepsilon}, \ \ \operatorname{sgn}(0)=0.\label{eq:Tc}
\end{align}
Consequently, without further restrictions on the learned natural-parameter family, the structural shocks are identified up to permutation, componentwise sign, and the componentwise transformations $T_{c_i}$.
\end{theorem}
Theorem~\ref{thm:IIA-GCL} is proven in Appendix~\ref{sec:proof_thm1}. It is a population identification result for the restricted IIA-GCL classification problem. Condition~\ref{cond:smooth} ensures that the mapping between observables and shocks is invertible and sufficiently smooth, whereas Condition~\ref{cond:sufstat} imposes the distributional structure that restricts the admissible componentwise transformations. Condition~\ref{cond:variability} supplies the density and common-support regularity needed to compare the population log-density-ratio representation~\eqref{eq:ldr} pointwise over the auxiliary-variable support, and requires sufficiently rich variation in the natural parameters. Finally, Condition~\ref{cond:regulatory} ensures that each learned population maximizer is sufficiently regular for the almost-sure equality of the optimal regression functions to extend to the full common support.

Relative to the original IIA-GCL result of \citet[][Theorem~1]{Morioka+Halva+Hyvarinen:2021}, where the shocks are identified only up to arbitrary componentwise invertible transformations, Theorem~\ref{thm:IIA-GCL} substantially strengthens the identification by fixing the sufficient statistics~\eqref{eq:qfixed}. However, it does not give identification up to permutation and sign alone but leaves the componentwise transformed-scale maps $T_{c_i}$ in~\eqref{eq:Tc}. This remaining ambiguity arises because the chosen sufficient statistics have an algebraic closure property. Applying $T_c$ to a shock changes the sufficient-statistic vector in a way that can, in principle, be offset by a corresponding transformation of the natural-parameter path. Whether this compensation is admissible depends on the class of natural-parameter functions imposed on the true and learned conditional shock densities. The next corollary gives a general non-closure condition under which such compensation is impossible, so that the transformed-scale ambiguity collapses to componentwise sign.
\begin{corollary}\label{cor:nonclosure_ident}
Suppose the conditions of Theorem~\ref{thm:IIA-GCL} hold. Let $\mathfrak L$ be a single admissible class of two-dimensional componentwise natural-parameter paths, imposed on both the true and learned conditional shock densities and common to all component labels. Thus, for each $i=1,\ldots,d$, the paths
\begin{equation}
\boldsymbol\lambda_i^0(u)=\big(\lambda_{i1}^0(u),\lambda_{i2}^0(u)\big)
\quad\text{and}\quad
\boldsymbol\lambda_i^\ast(u)=\big(\lambda_{i1}^\ast(u),\lambda_{i2}^\ast(u)\big)
\end{equation}
belong to $\mathfrak L$.

For $c>0$, define
\begin{equation}\label{eq:Bc}
B(c)=
\begin{bmatrix}
c^2 & 2\sqrt{\varepsilon}\,c(1-c)\\
0 & c
\end{bmatrix}.
\end{equation}
Assume that $\mathfrak L$ is not closed under a nontrivial transformed-scale representation of any true component. Specifically, for each $k=1,\ldots,d$, there do not exist a constant $c>0$ with $c\neq1$, an alternative admissible path $\bar{\boldsymbol\lambda}(u)=\big(\bar\lambda_1(u),\bar\lambda_2(u)\big)\in\mathfrak L$, and a constant vector $\boldsymbol a=(a_1,a_2)\in\mathbb R^2$ such that
\begin{equation}\label{eq:natpar_nonclosure}
\boldsymbol\lambda_k^0(u)=B(c)'\bar{\boldsymbol\lambda}(u)+\boldsymbol a
\end{equation}
for every $u\in\mathcal U_0$.

Then any population maximizer satisfies
\begin{equation}
h_i^\ast(y_t,\boldsymbol y_{t-1}) = \iota_i g_{\pi(i)}(y_t,\boldsymbol y_{t-1}), \quad i=1,\ldots,d,
\end{equation}
on the common support $\mathcal X_0$, where $\pi$ is a permutation of $\{1,\ldots,d\}$ and $\iota_i\in\{-1,+1\}$. Hence, under the natural-parameter non-closure condition~\eqref{eq:natpar_nonclosure}, the shocks are identified up to permutation and componentwise sign changes.
\end{corollary}
Corollary~\ref{cor:nonclosure_ident} is proven in Appendix~\ref{sec:proof_cor1}. The non-closure condition rules out the possibility that a nontrivial componentwise transformation $T_{c_i}$ can be absorbed into another admissible natural-parameter path in $\mathfrak L$. Under this condition, the constants $c_i$ in Theorem~\ref{thm:IIA-GCL} must equal one, so that the remaining transformed-scale ambiguity collapses to componentwise sign. Thus, the admissible natural-parameter class plays two roles. It determines how the conditional shock distribution is allowed to vary with the auxiliary variable, and it determines whether the transformed-scale ambiguity left by Theorem~\ref{thm:IIA-GCL} is admissible.

The non-closure condition is stated abstractly because Theorem~\ref{thm:IIA-GCL} itself does not require a parametric model for the natural-parameter paths. In empirical work, however, it is often useful to estimate the natural parameters within a chosen admissible class. The following corollary specializes the non-closure argument to the logistic specification introduced in Section~\ref{sec:natpar} and shows that, under this specification, sign identification holds generically.
\begin{corollary}\label{cor:logistic_ident}
Suppose the conditions of Theorem~\ref{thm:IIA-GCL} hold. Suppose, in addition, that the true and learned componentwise natural-parameter paths are restricted to the logistic class $\mathfrak L_{\mathrm{log}}$ defined by Equations~\eqref{eq:lambda1}--\eqref{eq:rho} in Section~\ref{sec:natpar}. Assume that the support $\mathcal U_0$ of $u_t$ contains a nonempty open interval.

Then, for Lebesgue-almost every value of the true logistic parameter vector in its full parameter space, any population maximizer satisfies
\begin{equation}
h_i^\ast(y_t,\boldsymbol y_{t-1}) = \iota_i g_{\pi(i)}(y_t,\boldsymbol y_{t-1}), \quad i=1,\ldots,d,
\end{equation}
on the common support $\mathcal X_0$, where $\pi$ is a permutation of $\{1,\ldots,d\}$ and $\iota_i\in\{-1,+1\}$. Hence, under the logistic natural-parameter specification, the shocks are generically identified up to permutation and componentwise sign changes.
\end{corollary}
Corollary~\ref{cor:logistic_ident} is proven in Appendix~\ref{sec:proof_cor2}. The proof shows that the set of true logistic parameter values for which the non-closure condition of Corollary~\ref{cor:nonclosure_ident} can fail is Lebesgue-null. Consequently, under the maintained assumptions, the modified IIA-GCL stage generically recovers the structural shocks up to permutation and componentwise sign.\footnote{This is a population identification statement. In finite samples, the transformed-scale ambiguity removed by Corollary~\ref{cor:logistic_ident} may still be weakly separated from ordinary rescaling, particularly when the smoothing constant $\varepsilon$ is very small. It is therefore often useful to normalize the estimated shocks before estimating the nonlinear SVAR mapping~\eqref{eq:genvar} (see Section~\ref{sec:FFNN-SVAR}).}

\subsection{A Monte Carlo experiment}\label{sec:mc_evidence}

We examine the finite-sample performance of our modified IIA-GCL method with a small-scale Monte Carlo (MC) experiment. The experiment focuses on a two-variable nonlinear SVAR in which the true structural map is a feed-forward neural network with two hidden layers, four neurons in each hidden layer, and sigmoid activation (see Section~\ref{sec:FFNN-SVAR-subsec}). The benchmark design is correctly specified in the sense that the structural shocks follow the exponential-family conditional distribution in Assumption~\ref{as:suffstats}, with logistic natural-parameter functions of the form~\eqref{eq:tau}--\eqref{eq:rho}. To examine robustness to distributional misspecification, we also consider two designs in which the same nonlinear structural map is used but the conditional shock distributions are replaced by time-varying Student-$t$ and skewed-$t$ distributions \citep[see][for the latter]{Hansen:1994}.

The estimator is evaluated by comparing the recovered shocks with the true shocks after aligning the remaining sign and permutation indeterminacies. Specifically, for each replication we compute the componentwise absolute correlations between the true and recovered shocks under the permutation and sign choices that maximize the mean absolute correlations. We then report the mean correlations (and their standard deviations) across Monte Carlo replications. Details on the IIA-GCL training procedure are provided in Appendix~\ref{sec:iia_gcl_training_details}, whereas further details on the Monte Carlo experiment are reported in Appendix~\ref{sec:mc_details}.

\begin{table}[!t]
\centering
\begin{tabular}{lcccc}
\hline\\[-1.3ex]
DGP & $T=250$ & $T=500$ & $T=1000$ & $T=2000$\\
\hline\\[-1.3ex]
Exponential-family shocks & 0.87 (0.04) & 0.88 (0.04) & 0.89 (0.01) & 0.90 (0.02)\\
Student-$t$ shocks & 0.87 (0.05) & 0.88 (0.03) & 0.88 (0.02) & 0.89 (0.01)\\
Skewed-$t$ shocks & 0.88 (0.06) & 0.88 (0.05) & 0.89 (0.03) & 0.90 (0.02)\\
\hline
\end{tabular}
\caption{Monte Carlo shock-recovery correlations under minimum-BCE selection based on $100$ Monte Carlo replications. The entries report the Monte Carlo mean of the average aligned componentwise absolute correlation between the true and recovered shocks, with the Monte Carlo standard deviation in parentheses.}
\label{tab:mc_main_results}
\end{table}

The results in Table~\ref{tab:mc_main_results} show a reasonable recovery already in small samples, but the recovery improves only slowly with the sample size. The MC standard deviations mainly decrease with sample size, but not uniformly in our results based on $100$ MC replications. The results are very similar between the benchmark specification (Exponential-family) and the two misspecified cases (Student-$t$ and Skewed-$t$), suggesting robustness to misspecification of the shock distribution. The fitted demixer $\boldsymbol{h}$ uses a parsimonious one-hidden-layer architecture with $12$ neurons, which we find to provide more stable optimization across random initializations and better shock recovery at the sample sizes considered than comparable deeper specifications. Recovery may therefore eventually level off as network approximation error becomes relatively more important. With larger samples, increasing the capacity of the demixer (and the nuisance networks $\phi$ and $\eta$) could potentially improve recovery.

Because the neural-network optimization problem is nonconvex, different optimization starts can converge to different local solutions. Different starts may recover nearly the same shocks while obtaining slightly different terminal binary cross-entropies (BCE) through different nuisance-network fits (see~\eqref{eq:lossbin}). Conversely, two starts with similar BCEs may recover materially different shocks. For this reason, we group the local solutions into basins according to the similarity of the recovered shocks, after which the shock basin needs to be selected. In the benchmark results reported in Table~\ref{tab:mc_main_results}, we select the shock basin containing the solution with the smallest BCE. Since the selected shock basin can affect the recovery, and in empirical applications it may be useful to include diagnostic criteria in the basin selection rule, we report the MC results in Appendix~\ref{sec:basin_selection} for various basin selection methods. In this MC experiment, the considered alternative basin selection methods produce results similar to those reported in Table~\ref{tab:mc_main_results}, i.e., a reasonable recovery in small samples that improves only slowly with the sample size.

\section{A neural network SVAR and structural analysis}\label{sec:FFNN-SVAR}

\subsection{A feed-forward neural network SVAR}\label{sec:FFNN-SVAR-subsec}

After identifying the structural shocks, the next step is to approximate and estimate the unknown function $f$ in the SVAR model~\eqref{eq:genvar}, which maps lagged observations and contemporaneous shocks to the observed variables. Following \citet{Morioka+Halva+Hyvarinen:2021}, we approximate this function using a feed-forward neural network (FFNN), which provides a flexible parametric representation capable of capturing complex nonlinear relationships. In particular, sufficiently large feed-forward networks can approximate a wide class of measurable functions arbitrarily well \citep{Hornik+Stinchcombe+White:1989}, making them an appealing choice when the functional form of $f$ is left unrestricted.
 
In practice, it is convenient to normalize the recovered shocks prior to estimating the structural mapping. Under the logistic natural-parameter specification and the generic conditions of Corollary~\ref{cor:logistic_ident}, the remaining population ambiguity is limited to permutation and componentwise sign. Nevertheless, finite-sample estimates may still differ in scale across components, and the absolute scaling of the recovered shocks is inconsequential for the subsequent neural-network approximation because it can be absorbed by the nonlinear function $f_\theta$. For these reasons, we rescale each recovered shock component by its sample standard deviation and denote the resulting normalized shocks by $\tilde e_t$. In addition, a sign convention is fixed for each component, which is without loss of generality.

We consider the nonlinear SVAR model
\begin{equation}\label{eq:nn_svar}
y_t = f_{\theta}(\boldsymbol{y}_{t-1}, \tilde{e}_t),
\end{equation}
where $f_{\theta}:\mathbb{R}^{d(p+1)}\to\mathbb{R}^d$ is an FFNN with parameter vector $\theta$.
An FFNN maps an input vector to an output vector through a sequence of layers. Each layer consists of a collection of nodes (or neurons), where every neuron in a given layer is connected to all neurons in the preceding layer via a system of weights and biases. Specifically, each layer computes an affine transformation of its inputs, followed by an elementwise nonlinear activation function.

To illustrate the structure of an FFNN, let $z_t = (\boldsymbol{y}_{t-1}, \tilde e_t) \in \mathbb{R}^{d(p+1)}$ denote the input vector, and consider a network with two hidden layers containing $n_1$ and $n_2$ neurons, respectively. Then the network output can be written as
\begin{equation}\label{eq:nn_example}
f_{\theta}(z_t) = W_3 \,\sigma_2\!\left(W_2 \,\sigma_1\!\left(W_1 z_t + b_1\right) + b_2\right) + b_3,
\end{equation}
where $W_1 \in \mathbb{R}^{n_1 \times d(p+1)}$ and $b_1 \in \mathbb{R}^{n_1}$ are the weight matrix and bias vector of the first hidden layer; $W_2 \in \mathbb{R}^{n_2 \times n_1}$ and $b_2 \in \mathbb{R}^{n_2}$ correspond to the second hidden layer; and $W_3 \in \mathbb{R}^{d \times n_2}$ and $b_3 \in \mathbb{R}^{d}$ correspond to the output layer. The functions $\sigma_1:\mathbb{R}^{n_1}\to\mathbb{R}^{n_1}$ and $\sigma_2:\mathbb{R}^{n_2}\to\mathbb{R}^{n_2}$ are activation functions applied elementwise.
The activation functions introduce nonlinearity into the model, allowing the network to represent complex nonlinear relationships between inputs and outputs. A common choice is the leaky rectified linear unit (leaky ReLU), defined elementwise as $\sigma(x) = \max\{\alpha x, x\}$ for some small $\alpha > 0$, which mitigates certain optimization issues associated with standard ReLU (defined as $\max\{0, x\}$) while retaining computational simplicity \citep{Maas+Hannun+Ng:2013}.

More generally, a feed-forward neural network with $\mathcal{H}$ hidden layers can be defined recursively as
\begin{equation}
h_0(z_t) = z_t, \ \ h_\ell(z_t) = \sigma_\ell\!\left(W_\ell h_{\ell-1}(z_t) + b_\ell\right), \ \ \ell=1,\ldots,\mathcal{H},
\end{equation}
with output
\begin{equation}
f_{\theta}(z_t) = W_{\mathcal{H}+1} h_H(z_t) + b_{\mathcal{H}+1},
\end{equation}
where each hidden layer $\ell$ contains $n_\ell$ neurons, so that $W_\ell \in \mathbb{R}^{n_\ell \times n_{\ell-1}}$, $b_\ell \in \mathbb{R}^{n_\ell}$, $n_0 = d(p+1)$, and $n_{\mathcal{H}+1} = d$. The parameter vector $\theta$ collects all weights and biases across layers.
The flexibility of the network is governed by its architecture, in particular the number of layers $H$ and the number of neurons $n_\ell$ in each layer. Increasing these quantities improves the network's flexibility, but also increases the number of parameters and the risk of overfitting. 

The use of a feed-forward architecture is natural in the present framework, where the structural model is specified in terms of a finite set of lagged variables and contemporaneous shocks. While recurrent neural networks, such as long short-term memory networks, are often used in time series applications \citep[see, e.g.,][and references therein]{Altmeyer+Agusti+Costa:2022}, they are designed to learn latent state representations over long sequences of observations. In contrast, the present model explicitly specifies the relevant state vector as $\boldsymbol{y}_{t-1}$, so an FFNN provides a direct and transparent representation of the structural SVAR.

The network parameters $\theta$ are estimated by minimizing a prediction loss, typically using numerical gradient-based methods. In particular, gradients of the loss function with respect to the network parameters can be computed efficiently using the backpropagation algorithm, which applies the chain rule through the layered structure of the network. The resulting gradients are then used within iterative optimization schemes such as stochastic gradient descent or its variants. See, for example, \citealp{Hastie+Tibshirani+Friedman:2009}; \citealp{Goodfellow+Bengio+Courville:2016}, for a more detailed discussion on neural networks and their training in general. Appendix~\ref{sec:svar_training_details} provides details on the FFNN-SVAR training employed in our empirical application and implemented in the accompanying R package iiasvar \citep{iiasvar}.

\subsection{Generalized impulse response function}\label{sec:GIRF}

To analyze the dynamic effects of structural shocks in our nonlinear SVAR model, it needs to be taken into account that the response of the variables may depend on both the history of the process as well as on the sign and size of the shock. To accommodate these features, we consider the generalized impulse response function (GIRF) \citep{Koop+Pesaran+Potter:1996}, which is a nonlinear counterpart of the conventional impulse response function. The GIRF is defined as:
\begin{equation}\label{eq:girf}
\text{GIRF}(h,\delta_i,\mathcal{F}_{t-1}) = \text{E}[y_{t+h}|\delta_i,\mathcal{F}_{t-1}] - \text{E}[y_{t+h}|\mathcal{F}_{t-1}],
\end{equation}
where $h$ is the horizon. The first term on the right side of (\ref{eq:girf}) is the expected realization of the process at time $t+h$ conditionally on the $i$th structural shock of sign and size $\delta_i \in\mathbb{R}$ at time $t$, and the previous observations. The latter term is the expected realization of the process conditionally on the previous observations only. The GIRF thus expresses the expected difference in the future outcomes when the $i$th structural shock of sign and size $\delta_i$ arrives at time $t$ as opposed to all shocks being random. 

The dynamics of our model depend only on the current shocks and the previous $p$ observations, so the conditioning set $\mathcal{F}_{t-1}$ can be replaced by the lag vector $\boldsymbol{y}_{t-1}=(y_{t-1},...,y_{t-p})$. Because the contemporaneous dynamics may depend on the entire structural shock vector, the response generally depends not only on the specified shock $e_{it}=\delta_i$ but also on the remaining contemporaneous shocks. Since (\ref{eq:girf}) conditions only on the $i$th shock, the conditional expectation averages over the distribution of the remaining contemporaneous shocks, yielding the average effect of the $i$th structural shock of sign and size $\delta_i$ rather than the response associated with any particular realization of the other contemporaneous shocks.\footnote{Unlike in conventional nonlinear SVAR models, in our SVAR model~\ref{eq:genvar}, the effects of a shock may generally depend on the full contemporaneous shock vector and not just on the sign and size of the shock of interest. Nevertheless, our GIRFs average over the distribution of the rest of the shocks to obtain impulse response with interpretations similar to those of conventional nonlinear SVAR models.}

In practice, we evaluate GIRFs for histories drawn from the observed sample \citep[cf.][]{Lanne+Virolainen:2025}. For a given history $\boldsymbol{y}_{t-1}$, the response to a structural shock is computed by simulating the model forward under the estimated dynamics. The impact shock is set to a given value $\delta_i$, and in particular we set $\delta_i$ equal to the recovered structural shock associated with the history, so that the responses reflect empirically relevant combinations of states and shock signs and sizes. Future shocks are generated from the fitted conditional distribution, which depends on the auxiliary variable $u_t$. For the initial period, we use the observed value of $u_t$ corresponding to the given history. Along the simulated paths, the observed future values $u_{t+h}$ are used whenever they are available in the sample. If the simulation horizon extends beyond the available observations, the missing values of $u_{t+h}$ are generated from a fitted univariate autoregressive model for $u_t$, conditional on the available past observations of $u_t$.

This procedure yields, for each history, a path-specific response that depends on both the initial state and the magnitude of the shock. Consequently, state-dependent effects can be analyzed by comparing GIRFs across different subsets of histories. Algorithm~\ref{algo:girf} \citep[adapted from the algorithm in Appendix~B of][]{Lanne+Virolainen:2025} formalizes the computation of the GIRF for a given history. The GIRFs are computed under the fitted model, combining the estimated dynamics $f_{\theta}$ and the fitted conditional distribution of shocks, and are therefore subject to the associated estimation uncertainty.\footnote{The conditional shock distribution is fitted in the IIA-GCL training for shocks that have not been normalized for sign and scale. To simulate observations from the (componentwise) normalized shock distributions, we first simulate a shock from the nonnormalized shock distribution, and then normalize it \textit{ex post}.}

\begin{algorithm}[!ht]
\caption{Generalized impulse response function}
\begin{algorithmic}[1]\label{algo:girf}  
  \STATE Set the horizon $H$ and the number of replications $R$. For the $j$th Monte Carlo replication, let $y_{t+h}^{(j)}(\delta_i,\boldsymbol{y}_{t-1})$ denote a realization of the process at time $t+h$ conditional on the history $\boldsymbol{y}_{t-1}$ and the $i$th structural shock of sign and size $\delta_i \in\mathbb{R}$ arriving at time $t$, and let $y_{t+h}^{(j)}(\boldsymbol{y}_{t-1})$ denote an alternative realization conditional on the history $\boldsymbol{y}_{t-1}$ only.
  \FOR{each $j$ in $1,2,\ldots,R$}
    \FOR{each $h$ in $0,1,\ldots,H$}
      \STATE Take the observed value of $u_{t+h}$ from the data if available; otherwise, simulate $u_{t+h}$ from a fitted univariate autoregressive model for $u_t$ conditional on the previous observations.
\label{step:girf_shock_starts}
      \STATE Draw $e_{t+h}$ from its fitted conditional distribution given $u_{t+h}$.
      \IF{$h=0$}
        \STATE Impose the sign and size $\delta_i$ to the $i$th element of $e_{t+h}$ to obtain the modified structural shock vector $\tilde{e}_{t+h}$.
      \ENDIF \label{step:girf_shock_ends}
      \STATE Calculate $y_{t+h}^{(j)}(\delta_i,\boldsymbol{y}_{t-1})$ from the fitted model~\eqref{eq:nn_svar} using the previous observations $(y_{t}^{(j)}(\delta_i,\boldsymbol{y}_{t+h-1}),\ldots,y_{t}^{(j)}(\delta_i,\boldsymbol{y}_{t-1}),\boldsymbol{y}_{t-1})$ and the structural shock vector $e_{t+h}$ (or $\tilde{e}_{t+h}$ if $h=0$) obtained from Steps~\ref{step:girf_shock_starts}--\ref{step:girf_shock_ends}.
      \STATE Calculate $y_{t+h}^{(j)}(\boldsymbol{y}_{t-1})$ from the fitted model~\eqref{eq:nn_svar} by using the previous observations $(y_{t+h-1}^{(j)}(\boldsymbol{y}_{t-1}),\ldots,y_{t}^{(j)}(\boldsymbol{y}_{t-1}),\boldsymbol{y}_{t-1})$ and the structural shock vector $e_{t+h}$ obtained from Steps~\ref{step:girf_shock_starts}--\ref{step:girf_shock_ends}.
    \ENDFOR
    \STATE Calculate $y_{t+h}^{(j)}(\delta_i,\boldsymbol{y}_{t-1}) - y_{t+h}^{(j)}(\boldsymbol{y}_{t-1})$.
  \ENDFOR  
\STATE Calculate the sample mean of $y_{t+h}^{(j)}(\delta_i,\boldsymbol{y}_{t-1}) - y_{t+h}^{(j)}(\boldsymbol{y}_{t-1})$ across the Monte Carlo replications $j=1,\ldots,R$ for all $h=0,1,\ldots,H$ to obtain the GIRF related to the history $\boldsymbol{y}_{t-1}$ for the $i$th structural shock of sign and size $\delta_i$.
\end{algorithmic}
\end{algorithm}

\subsection{Shock labelling}\label{sec:shocklabelling}

After normalizing the sign and scale of the structural shocks, they remain identified up to permutation and are subject to the usual labelling problem. In other words, in line with the statistical identification literature, labelling the identified shocks as economic shocks requires external information. To address this, we follow a procedure similar in spirit to \cite{Lanne+Meitz+Saikkonen:2017} and propose labelling the shocks based on their estimated impulse response functions.

Specifically, we compute GIRFs for histories drawn from the observed sample and use the recovered shocks associated with these histories. To make the GIRFs comparable across different signs and sizes of the shocks, we normalize each response by dividing by the corresponding recovered shock. This transformation rescales the responses to correspond to a positive unit shock. The normalized responses are then aggregated across histories by taking the median response in each time period, yielding a representative response profile for each shock. Finally, the shocks of interest are labelled based on the patterns of these median responses across variables and horizons. The interpretation of each shock depends on how it affects the variables in the system over time and therefore on the specific empirical application.

\section{Empirical application}\label{sec:empirical}

We illustrate the use of our methods to study the effects of real oil price shocks on U.S. industrial production. In particular, our empirical application is motivated by \cite{Herrera+Lagalo+Wada:2011}, who study whether the response of U.S. industrial production to real oil price shocks is asymmetric. Their results show that the evidence depends on the estimation period and the level of aggregation. In post-1973 data, evidence against symmetry is weak for aggregate industrial production, but it is stronger for some sector-specific series.

We consider a bivariate monthly U.S. dataset covering the period from February 1974 through January 2026 ($624$ observations). Industrial production is measured by the industrial production index (IPI), which is logarithmized and detrended by taking first differences. Following \cite{Herrera+Lagalo+Wada:2011}, the nominal oil price is measured by the U.S. crude-oil composite refiner acquisition cost, which is deflated by the consumer price index to obtain the real oil price (OIL), which is then likewise logarithmized and detrended by taking first differences.

\begin{figure}[!ht]
    \centerline{\includegraphics[width=\textwidth - 2cm]{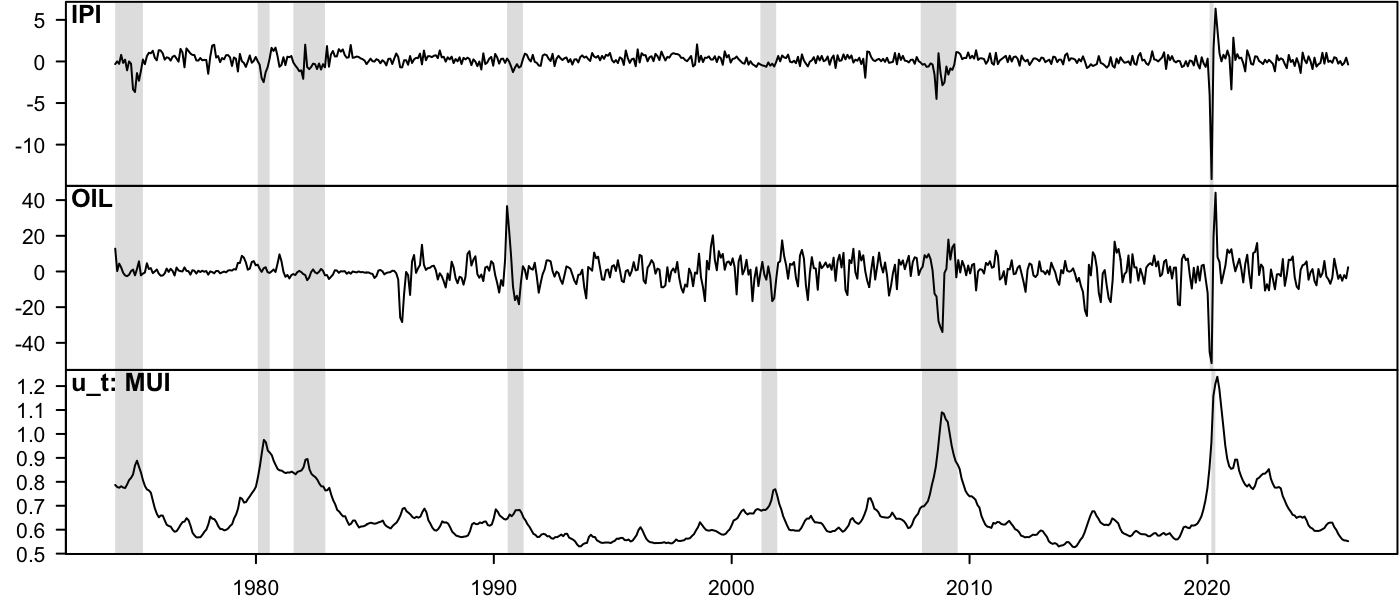}}
    \caption{Monthly U.S. time series from 1974:2 through 2026:1. The top panel shows the first difference of the logarithm of the seasonally adjusted industrial production index (multiplied by $100$). The middle panel shows the first difference of the logarithm of the real oil price, constructed by deflating the U.S. crude-oil composite refiner acquisition cost by the consumer price index (multiplied by $100$). The bottom panel shows the one-month-ahead macro uncertainty index of \citet{Jurado+Ludvigson+Ng:2015}, lagged by one month and reported in its original units. The shaded areas indicate the periods of U.S. recessions defined by the NBER.}
\label{fig:seriesplot}
\end{figure}

As the auxiliary variable $u_t$, we use the first lag of the one-month-ahead macro uncertainty index (MUI) of \citet{Jurado+Ludvigson+Ng:2015}. The MUI summarizes uncertainty about the unpredictable component of a broad set of macroeconomic indicators, and periods of elevated macroeconomic uncertainty can plausibly feature changes in the variances and the shapes of macroeconomic shock distributions. It therefore provides an economically meaningful source of variation in the conditional shock distributions. Using the first lag, in turn, makes the auxiliary variable predetermined relative to the time-$t$ shocks, facilitating plausibility of Assumption~\ref{as:shocks}. The series of the considered variables are depicted in Figure~\ref{fig:seriesplot}.\footnote{The IPI series is obtained from \url{https://fred.stlouisfed.org/series/INDPRO}, the consumer price index from \url{https://fred.stlouisfed.org/series/CPIAUCSL}, the nominal oil price series from \url{https://www.eia.gov/dnav/pet/pet_pri_rac2_dcu_nus_m.htm}, and the macro uncertainty series from \url{https://www.sydneyludvigson.com/macro-and-financial-uncertainty-indexes}. The availability of the oil price data determines the starting period and the availability of the macro uncertainty data the ending period of our sample.}

\subsection{Training and model selection}\label{sec:emp_training}

Training our IIA-SVAR model is conducted in two stages. In Stage~1, the structural shocks are estimated with the IIA-GCL method described in Section~\ref{sec:IIA-GCL}. Specifically, the demixing network $\boldsymbol{h}$ and the logistic classifier~\eqref{eq:r} are trained jointly by minimizing the binary cross-entropy loss~\eqref{eq:lossbin}. In Stage~2, the recovered shocks are first normalized with respect to permutation, sign, and scale. Then, the nonlinear SVAR model~\eqref{eq:nn_svar} is estimated by training the neural network $f_{\theta}$ to predict the observations $y_t$ from the lag vector $\boldsymbol{y}_{t-1}$ and normalized recovered shocks $\tilde{e}_t$.

Model specification consists of selecting the autoregressive order $p$ as well as the architectures of the involved feed-forward neural networks. In particular, for the demixing network $\boldsymbol{h}$, the nuisance neural networks $\phi$ and $\eta$ (see~\eqref{eq:r}), as well as the SVAR network $f_{\theta}$, one must choose the number of hidden layers $\mathcal{H}$, the number of neurons in each hidden layer $n_\ell$, and the activation function $\sigma_{\ell}$, $\ell=1,...,\mathcal{H}$. The Akaike information criterion selects $p=3$ for a conventional linear VAR, indicating that three lags could be sufficient to capture the majority of the autocorrelation structure of the data, so we also set $p=3$ for our nonlinear SVAR model. 

Table~\ref{tab:nn_architectures} summarizes the selected network architectures. In particular, the demixing and SVAR networks are assigned greater capacity than the nuisance networks because they perform the primary shock-recovery and SVAR-modeling tasks. Relatively shallow architectures are employed to balance flexibility and parsimony. The nuisance and SVAR networks use leaky-ReLU activation, whereas the demixer uses sigmoid activation because, in our finite-sample experiments (not shown), it produced more stable shock estimates across optimizer initializations without materially weakening shock recovery.\footnote{In particular, the demixer and nuisance networks follow the architecture from our Monte Carlo experiment (see Section~\ref{sec:mc_evidence}).}

\begin{table}[!ht]
\centering
\begin{tabular}{lccccc}
\hline \\[-1.3ex]
 & Demixer $\boldsymbol{h}$ & Nuisance $\phi$ & Nuisance $\eta$ & SVAR $f_\theta$ \\
\hline \\[-1.3ex]
Hidden layers $\mathcal{H}$        & 1 & 1 & 1 & 2 \\
Neurons in layer $n_\ell$          & 12 & 8 & 8 & 12 \\
Activation function $\sigma_\ell$ & Sigmoid & Leaky ReLU & Leaky ReLU & Leaky ReLU \\
\hline
\end{tabular}
\caption{Selected neural network architectures. The sigmoid activation is defined as $\sigma_\ell(x)=(1+e^{-x})^{-1}$ and Leaky-ReLU as $\sigma_\ell(x) = \max\{\alpha x, x\}$ for some small $\alpha > 0$, where we use $\alpha=0.01$.}\label{tab:nn_architectures}
\end{table}

\begin{figure}[!ht]
    \centerline{\includegraphics[width=\textwidth - 1cm]{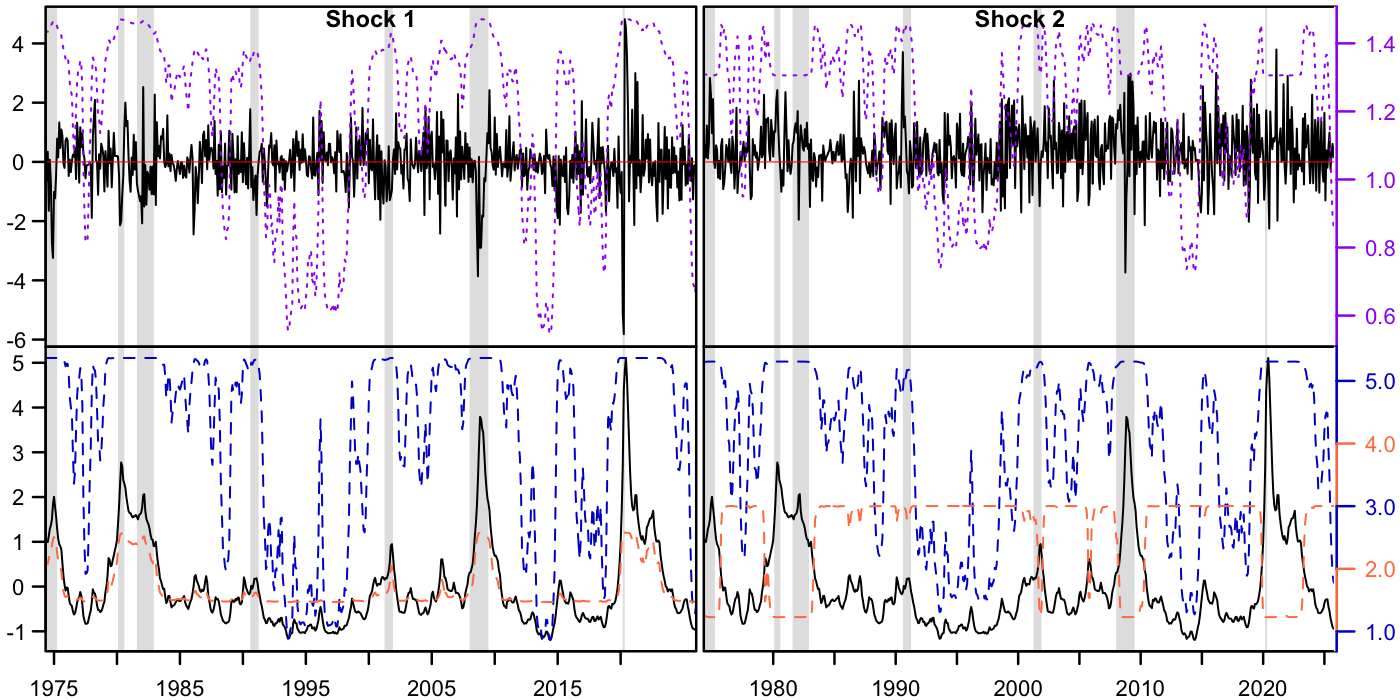}}
    \caption{Each column corresponds to one recovered structural shock component obtained from the Stage-1 IIA-GCL estimator. The top panels display the recovered shocks $\hat{e}_{it}$ (black solid line) together with the standard deviations of their fitted distributions (purple dotted line). The bottom panels display the (standardized) auxiliary variable $u_t$ (solid black line) together with the estimated modulation functions governing the conditional distribution of the shock component, namely the scale parameter function $\hat{\tau}_i(u_t)$ (blue dashed line) and the shape parameter function $\hat{\rho}_i(u_t)$ (orange dashed line). The modulation functions are plotted on a common rescaled axis for visualization, with their original scale indicated by the colored right-hand axis. The shaded areas indicate the U.S. recessions defined by the NBER.}
\label{fig:shockplot}
\end{figure}

Training the IIA-GCL model for our dataset, with further details on the training procedure provided in Appendix~\ref{sec:iia_gcl_training_details}, recovers the normalized shocks presented in the top panels of Figure~\ref{fig:shockplot}.  The bottom panels of Figure~\ref{fig:shockplot} depict the fitted scale (blue dashed line) and shape (orange dashed line) parameter values for each shock along with the auxiliary variable $u_t$, showing how their values vary with $u_t$. For both shocks, higher macroeconomic uncertainty is associated with larger values of the scale parameter. Interestingly, for Shock~1, elevated uncertainty is associated with larger values of the shape parameter, whereas for Shock~2 it is the other way around. An increase in the shape parameter implies that the shock distribution becomes more Gaussian (see Section~\ref{sec:natpar}), while both the scale and shape parameters jointly affect its dispersion. To provide a more direct measure of time-varying dispersion, the estimated standard deviations of the fitted shock distributions are shown in the top panels of Figure~\ref{fig:shockplot} (purple dotted line). These estimates indicate that the dispersion of both shock distributions tends to increase with macroeconomic uncertainty. The diagnostics presented in Appendix~\ref{sec:empapp_shockdiag} (Figure~\ref{fig:shock_acf}) show that there is not much auto- or crosscorrelation in the recovered shocks.\footnote{
We also check that Condition~\ref{cond:variability} of Theorem~\ref{thm:IIA-GCL} (the variability condition) is satisfied in the estimate by randomly sampling $100000$ length $2d+1$ subsets of observed $u_t$ and constructing empirical analogues of the matrix $L$ in~\eqref{eq:L}. In the subset that has the largest smallest singular value of $L$, the smallest singular value is $0.04$ and the ratio between the largest and smallest singular value is $29$, suggesting that the variability condition is well satisfied in the estimate.}

Using the recovered (normalized) shocks, we then train the nonlinear SVAR model~\eqref{eq:nn_svar} implemented as the FFNN described in Table~\ref{tab:nn_architectures}. Details on the training procedure are provided in Appendix~\ref{sec:svar_training_details}. The related prediction error diagnostics presented in Appendix~\ref{sec:svar_diagnostics} suggest a reasonable fit. In particular, there is no obvious systematic variation nor substantial auto- or crosscorrelation in the prediction errors, and apart from two particularly large outliers for oil prices (related to the Gulf War oil shock in August 1990 and the COVID-19 shock in March 2020), the prediction errors seem reasonable.

\subsection{Labelling the shocks}\label{sec:emp_shocklabelling}

\begin{figure}[!t]
    \centerline{\includegraphics[width=\textwidth - 1cm]{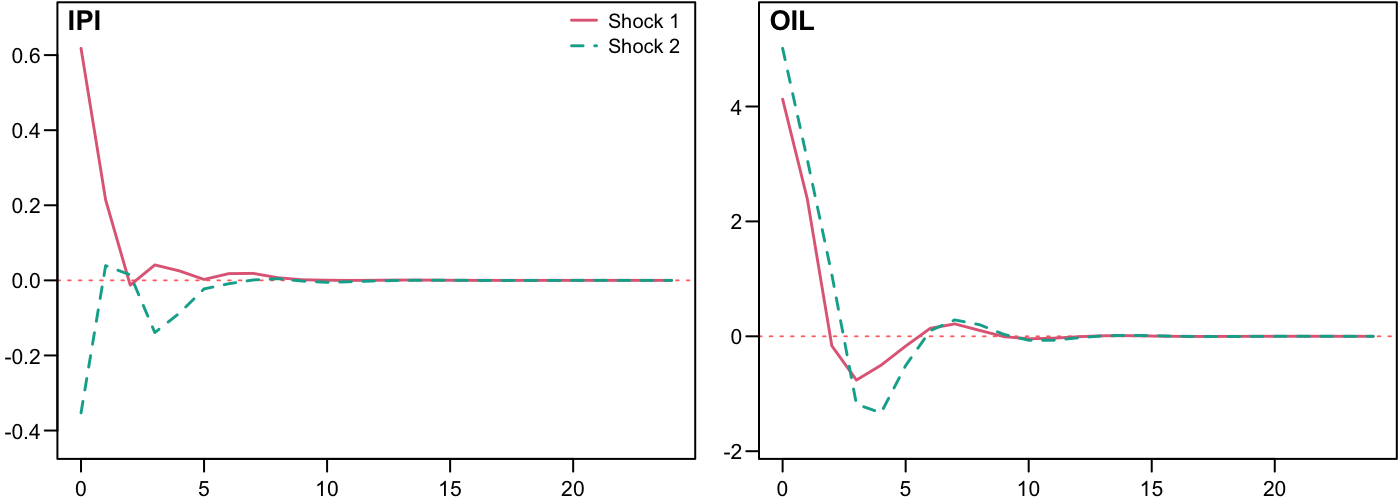}}
    \caption{The median GIRFs across all the length $p$ histories in the data for the recovered structural shocks, where the responses are normalized to correspond to a positive unit shock. Each panel shows the response of one variable to each shock, with the left panel for IPI and the right panel for OIL, covering the horizons of $h=0,1,\ldots,24$ months. The responses are in the approximate percentage growth rate scale on which the variables are presented in Figure~\ref{fig:seriesplot}.}
\label{fig:labellingplot}
\end{figure}

Following the procedure described in Section~\ref{sec:shocklabelling}, we label the structural shocks based on their median GIRFs across all length $p$ histories in the data. These median GIRFs computed for our IIA-SVAR model fitted in Section~\ref{sec:emp_training} are presented in Figure~\ref{fig:labellingplot}. Shock~1 moves output and oil prices in the same direction, while Shock~2 moves them in opposite directions. Moreover, Shock~1 has a larger impact effect on production than Shock~2, while Shock~2 has a larger impact effect on oil prices. Hence, the median effects of Shock~2 seem to resemble those of a (real) oil price shock, so we deem it as the oil price shock. 

\subsection{Impulse response analysis}

Median GIRFs provide compact summary statistics on the effects of the shocks useful for labelling them, but in general the effects of the shocks may depend on the related history and sign and size of the shock. Here we are interested in studying asymmetries in the effects of the oil price shock. 
The GIRFs are computed using Algorithm~\ref{algo:girf} with a procedure comparable to \cite{Lanne+Virolainen:2025} and \cite{Virolainen2:2024}. Specifically, for each length $p$ history in the data, we take the corresponding oil price shock recovered from the data and compute the corresponding GIRF. Each GIRF is assigned to a sign group according to the sign of the recovered shock. The GIRFs are then normalized to correspond to a positive unit impact response of OIL (i.e., roughly a one-percentage-point increase in the real oil price). Consequently, the comparisons below summarize model-implied heterogeneity across the observed history-shock pairs rather than ceteris-paribus effects of changing only the shock sign or history. The reported $90\%$ GIRF intervals are empirical quantile ranges across these pairs and do not quantify parameter uncertainty.

\begin{figure}[!t]
    \centerline{\includegraphics[width=\textwidth - 1cm]{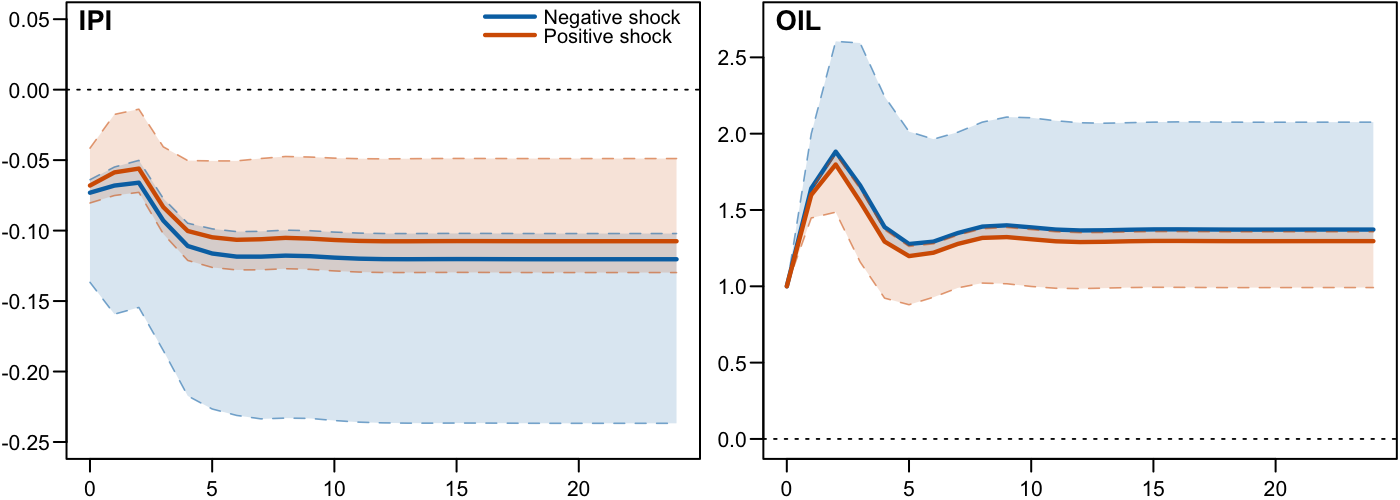}}
    \caption{Generalized impulse response functions to the identified oil price shock $h=0,1,\ldots,24$ months ahead, computed with Algorithm~\ref{algo:girf}. The left panel shows the responses of IPI and the right panel the responses of OIL, both accumulated to ($100\times$)log-levels. The blue solid line shows the responses for negative shocks and the red solid line shows the responses for positive shocks. There are $212$ negative and $409$ positive shocks. The shaded areas (also indicated by dashed lines) are the symmetric $90\%$ GIRF intervals that contain $90\%$ of the GIRFs across the histories and shocks of the given sign. All the GIRFs have been scaled so that the instantaneous increase in OIL is unity.}
\label{fig:posnegplot}
\end{figure}

Firstly, motivated by \cite{Herrera+Lagalo+Wada:2011}, we study how the effects of the oil price shock vary between positive and negative shocks. Figure~\ref{fig:posnegplot} illustrates the distribution of the GIRFs for positive and negative shocks, where the median GIRF is depicted with a solid line (blue for negative and red for positive shocks), with the shaded areas quantifying the dispersion of the GIRFs across the related histories (and shocks). The median responses of both IPI and OIL are slightly stronger for negative than positive shocks. However, the GIRF intervals show that the dispersion of the responses to negative shocks is clearly skewed towards stronger responses, while for positive shocks it is skewed towards weaker responses. The effects of negative shocks thereby seem overall stronger than the effects of positive shocks. Our findings thus differ from the post-1973 (aggregate IPI) result of \citet{Herrera+Lagalo+Wada:2011} for shocks of typical magnitude. In particular, they find no evidence against symmetry in that case, although they do find evidence of asymmetry for large shocks.

\begin{figure}[!t]
    \centerline{\includegraphics[width=\textwidth - 1cm]{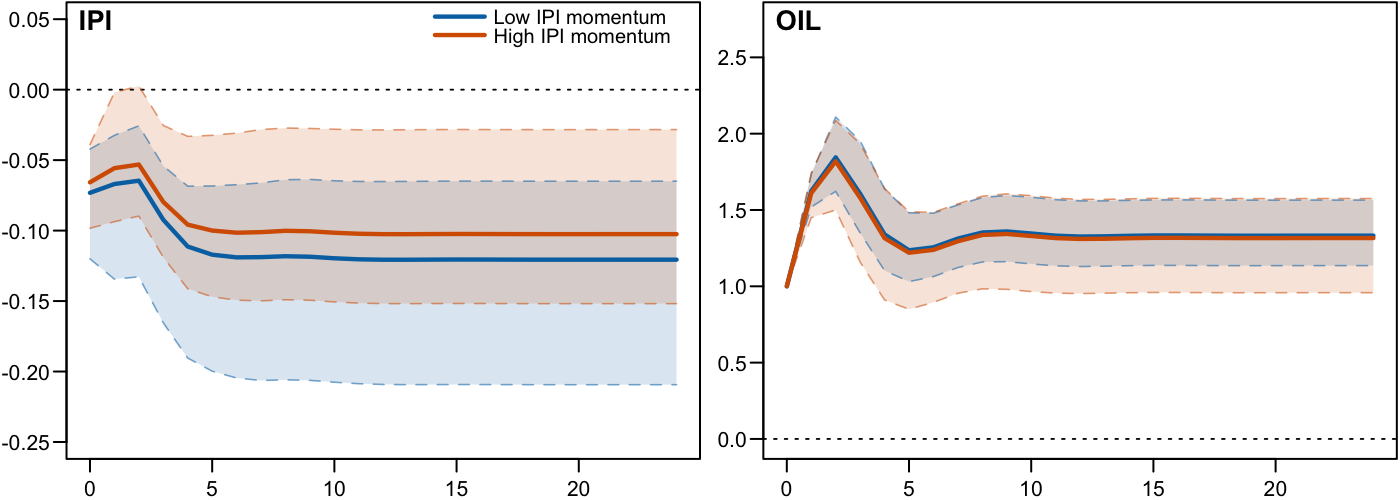}}
    \caption{Generalized impulse response functions to the identified oil price shock $h=0,1,\ldots,24$ months ahead, computed with Algorithm~\ref{algo:girf}. The left panel shows the responses of IPI and the right panel the responses of OIL, both accumulated to ($100\times$)log-levels. The blue solid line shows the responses for histories with low IPI momentum and the red solid line shows the responses for histories with high IPI momentum. There are $117$ histories with low IPI momentum and $122$ histories with high IPI momentum. The shaded areas (also indicated by dashed lines) are the symmetric $90\%$ GIRF intervals that contain $90\%$ of the GIRFs across the histories (and shocks) of the given momentum. All the GIRFs have been scaled so that the instantaneous increase in OIL is unity.}
\label{fig:momentumplot}
\end{figure}

Secondly, our IIA-SVAR model has the interesting property that it allows effects of the shocks to vary depending on the joint features of the length $p$ history of the included variables and the contemporaneous shocks, where this dependency is learned from the data during the Stage-2 neural network training. We find that the effects of the oil price shock seem to particularly depend on the momentum of the industrial production growth in the previous $p$ periods. Specifically, we define the momentum as
$\text{IPI}^{\text{mom}}_t = \sum_{i=1}^p \text{IPI}_{t-i}$,
where $\text{IPI}_{t-i}$ is the approximate percentage growth rate of the industrial production index at lag $i$. Then, we separately consider histories with low momentum, $\text{IPI}^{\text{mom}}_t<-0.5$, and histories with high momentum, $\text{IPI}^{\text{mom}}_t>1.5$.

Figure~\ref{fig:momentumplot} shows that even though the responses of oil prices are similar for low and high momentum histories, the responses of production are somewhat different. In particular, they are generally stronger when the recent IPI momentum is low than when it is high. Overall, we therefore find modest asymmetries in the effects of the oil price shock with respect to its sign and the state of the economy.

\section{Conclusion}\label{sec:conclusion}
We develop a fully nonlinear structural vector autoregressive framework identified through variation in the conditional distributions of the mutually independent structural shocks, induced by an observed exogenous variable. In particular, a general contrastive learning framework that makes use of this variation together with the assumed exponential-family structure is employed to recover the shocks. The framework allows the contemporaneous structural mapping to be fully nonlinear and non-additive, thereby substantially extending the class of admissible structural dynamics relative to conventional nonlinear SVAR models.

Our identification strategy builds on the independent innovation analysis framework of \citet{Morioka+Halva+Hyvarinen:2021} based on general contrastive learning. Their identification results recover the structural shocks only up to permutation and componentwise invertible transformations, which is generally insufficient for structural econometric analysis. By imposing a structured exponential-family specification for the conditional shock distributions, we reduce this ambiguity substantially. The general identification result leaves only a transformed-scale ambiguity, and we show that this remaining ambiguity is generically eliminated under our logistic natural-parameter specification, yielding identification up to permutation and componentwise sign changes.

Once the structural shocks have been recovered, the fully nonlinear SVAR model can be estimated flexibly using feed-forward neural networks, motivated by their universal approximation capabilities \citep[see][]{Hornik+Stinchcombe+White:1989}. The resulting framework permits structural analysis through generalized impulse response functions \citep{Koop+Pesaran+Potter:1996} that accommodate nonlinear and state-dependent dynamics. The accompanying R package iiasvar \citep{iiasvar} implements the introduced methods.

We illustrate the use of methods in an empirical application considering a monthly bivariate system of U.S. industrial production growth and real oil price growth, covering the period from 1974:2 to 2026:1. To facilitate identification, the shock distributions are assumed to be modulated by the (lagged) macro uncertainty index of \citet{Jurado+Ludvigson+Ng:2015}. We find modest asymmetries in the effects of the oil price shock with respect to its sign and the state of the economy. In particular, our findings suggest stronger effects for negative than positive oil price shocks, differing from the post-1973 (aggregate industrial production) result of \citet{Herrera+Lagalo+Wada:2011} for shocks of typical magnitude. Moreover, we find the effects of the oil price shock stronger when the recent industrial production growth is low than when it is high.


\section*{Declaration of the use of AI and AI-assisted technologies}
During the preparation of this work, the author used ChatGPT and Codex to improve the language, write the practical R implementation of the method, and brainstorm ideas, including the preparation of mathematical proofs. After using these tools, the author reviewed and edited the content as needed and takes full responsibility for the content of the publication.

\bibliography{/Users/savi/Documents/texrefs/masterrefs.bib}

\pagebreak
\begin{appendices}
\renewcommand{\thefigure}{\thesection.\arabic{figure}}
\renewcommand{\thetable}{\thesection.\arabic{table}}
\setcounter{figure}{0}    
\setcounter{table}{0}

\newtheorem{appendixlemma}{Lemma}[section]
\newtheorem{appendixproposition}{Proposition}[section]
\newtheorem{appendixtheorem}{Theorem}[section]
\newtheorem{appendixassumption}{Assumption}[section]
\newtheorem{appendixcondition}{Condition}[section]
\newtheorem{appendixcorollary}{Corollary}[section]

\setcounter{page}{1}

\section{Proofs}\label{sec:proofs}

\subsection{Proof of Theorem~\ref{thm:IIA-GCL}}\label{sec:proof_thm1}

The proof proceeds in five steps. In \textbf{Step~1}, we start from the binary classification problem, derive the population-optimal discriminator as a log-density ratio under the true model, and express it in terms of the latent shocks and the true natural-parameter functions $\boldsymbol{\lambda}^0(u_t)$. In \textbf{Step~2}, we equate this true log-density-ratio representation with an arbitrary population-maximizing element of the restricted model class. The learned representation is allowed to have its own learned natural-parameter functions $\boldsymbol{\lambda}^\ast(u_t)$. We then rewrite the resulting identity in latent coordinates by composing the learned feature extractor with the true mixing map. This gives the basic identity linking the true shocks and the learned latent variables.

In \textbf{Step~3}, we exploit variation in $u_t$ by evaluating this identity at $u^{(\ell)}$, $\ell=0,\ldots,2d$. Subtracting the equation at $u^{(0)}$ eliminates the nuisance terms that do not vary with $u_t$ and yields a finite-difference system involving both the true variability matrix $L^0$ and a learned variability matrix $L^\ast$. Following the same structural logic as \cite{Morioka+Halva+Hyvarinen:2021}, we first show that $L^\ast$ is nonsingular and then use derivative arguments to conclude that, locally, each learned latent variable is a function of exactly one true shock component, up to a permutation.

In \textbf{Step~4}, we use the specific fixed sufficient statistics $q_1(x)=x^2$ and $q_2(x)=\sqrt{x^2+\varepsilon}-\sqrt{\varepsilon}$. These statistics rule out arbitrary componentwise invertible transformations locally by implying that each local scalar ambiguity must act affinely on the second sufficient statistic $q_2$, equivalently satisfying a corresponding squared relation. Finally, in \textbf{Step~5}, we globalize the local componentwise representations and the local affine coefficients. The global scalar domains then contain zero, which implies that the affine intercepts vanish, that the slopes are positive, and hence that the remaining scalar ambiguities have the form $S_i=\iota_iT_{c_i}$. This yields the global representation $h_i^\ast=\iota_i\,T_{c_i}\circ g_{\pi(i)}$ on $\mathcal{X}_0$, proving the theorem.

\noindent\textbf{Step~1. Derivation of the population regression function under the true model.}
Following \cite{Morioka+Halva+Hyvarinen:2021}, we begin from the population binary classification problem underlying our modified IIA-GCL method introduced in Section~\ref{sec:IIA-GCL}. Consider the classification problem that distinguishes the random vectors $\tilde{\boldsymbol{y}}_t=(y_t,\boldsymbol{y}_{t-1},u_t)$ from the corresponding $u$-shuffled observations $\tilde{\boldsymbol{y}}_t^\ast=(y_t,\boldsymbol{y}_{t-1},u_t^\ast)$ defined in~\eqref{eq:y_tilde}, where $u_t^\ast$ is independent of $(y_t,\boldsymbol{y}_{t-1})$. Formally, this classification problem corresponds to maximizing the population logistic objective $\mathcal{L}(r)$ in~\eqref{eq:popobj} over measurable functions $r$.

Let $\nu$ be the marginal law of $u_t$ introduced in Condition~\ref{cond:variability}. With respect to the common dominating measure given by Lebesgue measure on $\mathcal X_0$ times $\nu$ on $\mathcal U_0$, the real class has density $p_{\mathcal X}(\boldsymbol x\mid u)$ and the randomized class has density $p_{\mathcal X}(\boldsymbol x)$, where $\boldsymbol x=(y_t,\boldsymbol y_{t-1})$. By standard results for logistic classification and density-ratio estimation (see, e.g., \citealp{Hastie+Tibshirani+Friedman:2009}; \citealp{Gutmann+Hyvarinen:2012}, as also invoked by \citealp{Morioka+Halva+Hyvarinen:2021}, Appendix~A), a population-optimal discriminator is represented by the pointwise log-density ratio in~\eqref{eq:pointwise_ldr_theorem}; throughout the proof, $r^\circ$ denotes this representative. Suppressing the subscript $\mathcal X$ and returning to the time-series notation gives
\begin{equation}\label{eq:logdensratio}
r^\circ(\tilde{\boldsymbol{y}}_t) = \log p(y_t,\boldsymbol{y}_{t-1}\mid u_t) - \log p(y_t,\boldsymbol{y}_{t-1}).
\end{equation}

We now rewrite the first term in~\eqref{eq:logdensratio} by using the augmented inverse map. Since the augmented model $\tilde{f}$ is bijective by assumption, its inverse $\tilde{g}$ satisfies
\begin{equation}
\tilde{g}(y_t,\boldsymbol{y}_{t-1}) = \bigl(g(y_t,\boldsymbol{y}_{t-1}),\boldsymbol{y}_{t-1}\bigr) = (e_t,\boldsymbol{y}_{t-1}).
\end{equation}
Hence, by the change-of-variables formula,
\begin{equation}
\log p(y_t,\boldsymbol{y}_{t-1}\mid u_t) = \log p(e_t,\boldsymbol{y}_{t-1}\mid u_t) + \log\big|\det J\tilde{g}(y_t,\boldsymbol{y}_{t-1})\big|,
\end{equation}
where $J\tilde{g}(\cdot)$ denotes the Jacobian matrix of $\tilde{g}$. By the pointwise conditional factorization~\eqref{eq:pointwise_conditional_factorization} and the product form~\eqref{eq:shockdist},
\begin{equation}
p(e_t,\boldsymbol{y}_{t-1}\mid u_t) = p(e_t\mid u_t)p(\boldsymbol{y}_{t-1}\mid u_t), \ \ \text{where} \ \ 
p(e_t\mid u_t) = \prod_{i=1}^d p_i(e_{it}\mid u_t).
\end{equation}
Therefore,
\begin{equation}
\log p(y_t,\boldsymbol{y}_{t-1}\mid u_t) = \sum_{i=1}^d \log p_i(e_{it}\mid u_t) + \log p(\boldsymbol{y}_{t-1}\mid u_t) + \log\big|\det J\tilde{g}(y_t,\boldsymbol{y}_{t-1})\big|.
\end{equation}

By Assumption~\ref{as:suffstats}, the base measures satisfy $Q_i(e_{it})=1$ and the sufficient statistics are fixed as in~\eqref{eq:qfixed}. Hence, for each $i=1,\ldots,d$,
\begin{equation}
\log p_i(e_{it}\mid u_t) = \sum_{j=1}^2 q_j(e_{it})\lambda_{ij}^0(u_t) - \log Z_i^0(u_t).
\end{equation}
Thus,
\begin{equation}
\begin{split}
\log p(y_t,\boldsymbol{y}_{t-1}\mid u_t) =& \sum_{i=1}^d\sum_{j=1}^2 q_j(e_{it})\lambda_{ij}^0(u_t) - \sum_{i=1}^d \log Z_i^0(u_t) + \log p(\boldsymbol{y}_{t-1}\mid u_t)\\
& + \log\big|\det J\tilde{g}(y_t,\boldsymbol{y}_{t-1})\big|.
\end{split}
\end{equation}
Similarly, again by the change-of-variables formula,
\begin{equation}
\log p(y_t,\boldsymbol{y}_{t-1}) = \log p(e_t,\boldsymbol{y}_{t-1}) + \log\big|\det J\tilde{g}(y_t,\boldsymbol{y}_{t-1})\big|.
\end{equation}
Substituting these two expressions into the log-density ratio~\eqref{eq:logdensratio}, the Jacobian terms cancel, and we obtain
\begin{equation}\label{eq:rcirc}
r^\circ(\tilde{\boldsymbol{y}}_t) = \sum_{i=1}^d\sum_{j=1}^2 q_j(e_{it})\lambda_{ij}^0(u_t) - \sum_{i=1}^d \log Z_i^0(u_t) +
\phi^0(\boldsymbol{y}_{t-1},u_t) + \eta^0(e_t,\boldsymbol{y}_{t-1}),
\end{equation}
where $\phi^0(\boldsymbol{y}_{t-1},u_t) = \log p(\boldsymbol{y}_{t-1}\mid u_t)$ and $\eta^0(e_t,\boldsymbol{y}_{t-1}) = -\log p(e_t,\boldsymbol{y}_{t-1})$.

For later use, define the stacked $(2d\times 1)$ sufficient-statistic vector
\begin{equation}\label{eq:Qstack}
\boldsymbol{Q}(z) = \big(q_1(z_1),q_2(z_1),\ldots,q_1(z_d),q_2(z_d)\big), \quad z=(z_1,\ldots,z_d).
\end{equation}
Then~\eqref{eq:rcirc} can equivalently be written as
\begin{equation}\label{eq:rcirc_vector}
r^\circ(\tilde{\boldsymbol{y}}_t) = \boldsymbol{\lambda}^0(u_t)'\boldsymbol{Q}(e_t) - \sum_{i=1}^d \log Z_i^0(u_t) + \phi^0(\boldsymbol{y}_{t-1},u_t) + \eta^0(e_t,\boldsymbol{y}_{t-1}).
\end{equation}

\noindent\textbf{Step~2. The latent-coordinate identity.}
Let $r^\ast$ be the regression function represented by any maximizer of the population objective~\eqref{eq:popobj} over the restricted class~\eqref{eq:r}. By Step~1, the unrestricted population maximizer $r^\circ$ is given by~\eqref{eq:rcirc_vector}. Moreover, since $e_t=g(y_t,\boldsymbol{y}_{t-1})$, the representation~\eqref{eq:rcirc_vector} belongs to the restricted class~\eqref{eq:r} by taking $\boldsymbol{h}=g$, $\boldsymbol{\lambda}=\boldsymbol{\lambda}^0$, $Z_i=Z_i^0$, $\phi=\phi^0$, and $\eta=\eta^0$. Condition~\ref{cond:variability} ensures that $(y_t,\boldsymbol y_{t-1},u_t)$ belongs to $\mathcal X_0\times\mathcal U_0$ almost surely under both classifier classes. Thus, the displayed representation is valid on the full probability-one comparison domain, and the restricted and unrestricted maximum values coincide. The population logistic objective is pointwise strictly concave in the regression function wherever the two class densities are positive, so any restricted-class maximizer satisfies $r^\ast=r^\circ$ almost surely with respect to the sum of the two class distributions.

Conditions~\ref{cond:variability} and~\ref{cond:regulatory} strengthen this almost-sure equality to pointwise equality on $\mathcal X_0\times\mathcal U_0$. Indeed, the strict positivity of $p_{\mathcal X}(\boldsymbol x\mid u)$ and $p_{\mathcal X}(\boldsymbol x)$, together with the definition of $\mathcal U_0$ as the support of $\nu$, implies that both class distributions have full support on this product. If the two jointly continuous functions differed at any point, they would therefore differ on a relatively open neighborhood having positive probability under the sum of the two class distributions, contradicting almost-sure equality. Hence,
\begin{equation}\label{eq:rstar_pointwise}
r^\ast(y_t,\boldsymbol y_{t-1},u)
=
r^\circ(y_t,\boldsymbol y_{t-1},u)
\end{equation}
for every $(y_t,\boldsymbol y_{t-1},u)\in\mathcal X_0\times\mathcal U_0$.

By construction, the restricted-class maximizer can be represented as
\begin{equation}\label{eq:rstar_learned}
\begin{split}
r^\ast(\tilde{\boldsymbol{y}}_t) =& \sum_{i=1}^d\sum_{j=1}^2 q_j\!\left(h_i^\ast(y_t,\boldsymbol{y}_{t-1})\right) \lambda_{ij}^\ast(u_t) - \sum_{i=1}^d \log Z_i^\ast(u_t) + \phi^\ast(\boldsymbol{y}_{t-1},u_t)\\
& + \eta^\ast\!\left(\boldsymbol{h}^\ast(y_t,\boldsymbol{y}_{t-1}),\boldsymbol{y}_{t-1}\right).
\end{split}
\end{equation}
To simplify notation, absorb the normalizing constants into the $u_t$-dependent nuisance functions by defining
\begin{equation}\label{eq:phibar_def}
\begin{split}
\bar\phi^0(\boldsymbol{y}_{t-1},u_t) &= \phi^0(\boldsymbol{y}_{t-1},u_t) - \sum_{i=1}^d \log Z_i^0(u_t), \ \ \text{and} \\ 
\bar\phi^\ast(\boldsymbol{y}_{t-1},u_t) &= \phi^\ast(\boldsymbol{y}_{t-1},u_t) - \sum_{i=1}^d \log Z_i^\ast(u_t).
\end{split}
\end{equation}
Equating the true representation~\eqref{eq:rcirc_vector} and the learned representation~\eqref{eq:rstar_learned}, we obtain
\begin{equation}\label{eq:rstar_tworeps1}
\begin{split}
&\boldsymbol{\lambda}^0(u_t)'\boldsymbol{Q}(e_t) + \bar\phi^0(\boldsymbol{y}_{t-1},u_t) + \eta^0(e_t,\boldsymbol{y}_{t-1})\\
&\qquad = \boldsymbol{\lambda}^\ast(u_t)'\boldsymbol{Q}\!\left(\boldsymbol{h}^\ast(y_t,\boldsymbol{y}_{t-1})\right) + \bar\phi^\ast(\boldsymbol{y}_{t-1},u_t) + \eta^\ast\!\left(\boldsymbol{h}^\ast(y_t,\boldsymbol{y}_{t-1}),\boldsymbol{y}_{t-1}\right).
\end{split}
\end{equation}

We now rewrite~\eqref{eq:rstar_tworeps1} in latent coordinates. Define
\begin{equation}\label{eq:ve}
\boldsymbol{v}(e_t,\boldsymbol{y}_{t-1}) = \big(v_1(e_t,\boldsymbol{y}_{t-1}),\ldots,v_d(e_t,\boldsymbol{y}_{t-1})\big) \equiv \boldsymbol{h}^\ast\!\left(f(\boldsymbol{y}_{t-1},e_t),\boldsymbol{y}_{t-1}\right).
\end{equation}
Thus, $\boldsymbol{v}(e_t,\boldsymbol{y}_{t-1})$ is the latent representation produced by the learned extractor when the observable $y_t$ is generated from the true model with shock vector $e_t$. Since $y_t=f(\boldsymbol{y}_{t-1},e_t)$, the true inverse satisfies
\begin{equation}
g\!\left(f(\boldsymbol{y}_{t-1},e_t),\boldsymbol{y}_{t-1}\right) = e_t.
\end{equation}
Using this identity together with~\eqref{eq:rstar_tworeps1} and~\eqref{eq:ve}, we obtain the latent-coordinate identity
\begin{equation}\label{eq:rstar_tworeps2}
\begin{split}
& \boldsymbol{\lambda}^0(u_t)'\boldsymbol{Q}(e_t) + \bar\phi^0(\boldsymbol{y}_{t-1},u_t) + \eta^0(e_t,\boldsymbol{y}_{t-1})\\
&\qquad = \boldsymbol{\lambda}^\ast(u_t)'\boldsymbol{Q}\!\left(\boldsymbol{v}(e_t,\boldsymbol{y}_{t-1})\right) + \bar\phi^\ast(\boldsymbol{y}_{t-1},u_t) + \eta^\ast\!\left(\boldsymbol{v}(e_t,\boldsymbol{y}_{t-1}),\boldsymbol{y}_{t-1}\right).
\end{split}
\end{equation}
The identity~\eqref{eq:rstar_tworeps2} holds for every $(e_t,\boldsymbol{y}_{t-1})\in\mathcal Z_0$ and every $u_t\in\mathcal U_0$. This identity is the starting point for the finite-difference argument in Step~3 and, later, for the all-$u$ comparison in Corollary~\ref{cor:nonclosure_ident}. The key point is that the two exponential-family terms in~\eqref{eq:rstar_tworeps2} are multiplied by (potentially) different natural-parameter functions, $\boldsymbol{\lambda}^0(u_t)$ and $\boldsymbol{\lambda}^\ast(u_t)$. Therefore, the finite-difference system will involve both the true variability matrix $L^0$ and the learned variability matrix $L^\ast$.

\noindent\textbf{Step~3. Finite differencing and componentwise separation.}
By~\eqref{eq:common_supports} and the inverse relation between $\tilde f$ and $\tilde g$, the latent-coordinate domain is
\begin{equation}\label{eq:Z0def}
\mathcal Z_0=\tilde g(\mathcal X_0)=\mathbb R^d\times\mathcal Y_0.
\end{equation}
This product form also follows directly from the assumptions on the shocks: their conditional density is strictly positive on $\mathbb R^d$, and conditional independence of $e_t$ and $\boldsymbol y_{t-1}$ given $u_t$ combines this full shock support with the common lag domain $\mathcal Y_0$. In particular, $\mathcal Z_0$ is open and connected, every shock-coordinate fiber is all of $\mathbb R$, and the latent-coordinate identity~\eqref{eq:rstar_tworeps2} holds pointwise on $\mathcal Z_0\times\mathcal U_0$.

For each $\ell=1,\ldots,2d$, evaluate~\eqref{eq:rstar_tworeps2} at $u_t=u^{(\ell)}$ and at $u_t=u^{(0)}$, and subtract the latter equation from the former. The terms $\eta^0(e_t,\boldsymbol{y}_{t-1})$ and
$\eta^\ast(\boldsymbol{v}(e_t,\boldsymbol{y}_{t-1}),\boldsymbol{y}_{t-1})$, which do not depend on $u_t$, cancel. Thus, for $\ell=1,\ldots,2d$, 
\begin{equation}\label{eq:ident_differenced_new}
\left[\boldsymbol{\lambda}^0(u^{(\ell)}) - \boldsymbol{\lambda}^0(u^{(0)})\right]'\boldsymbol{Q}(e_t) - \left[\boldsymbol{\lambda}^\ast(u^{(\ell)}) - \boldsymbol{\lambda}^\ast(u^{(0)})\right]'\boldsymbol{Q}\!\left(\boldsymbol{v}(e_t,\boldsymbol{y}_{t-1})\right) = \psi_\ell(\boldsymbol{y}_{t-1}),
\end{equation}
where
\begin{equation}\label{eq:psiell_new}
\psi_\ell(\boldsymbol{y}_{t-1}) = \bar\phi^\ast(\boldsymbol{y}_{t-1},u^{(\ell)}) - \bar\phi^\ast(\boldsymbol{y}_{t-1},u^{(0)}) -
\left( \bar\phi^0(\boldsymbol{y}_{t-1},u^{(\ell)}) - \bar\phi^0(\boldsymbol{y}_{t-1},u^{(0)}) \right).
\end{equation}
Define the learned variability matrix
\begin{equation}\label{eq:Lstar}
L^\ast = \left[\boldsymbol{\lambda}^\ast(u^{(1)})-\boldsymbol{\lambda}^\ast(u^{(0)}) : \cdots : \boldsymbol{\lambda}^\ast(u^{(2d)})-\boldsymbol{\lambda}^\ast(u^{(0)}) \right],
\end{equation}
and denote $\boldsymbol{\psi}(\boldsymbol{y}_{t-1}) = \big(\psi_1(\boldsymbol{y}_{t-1}),\ldots,\psi_{2d}(\boldsymbol{y}_{t-1})\big)$. Stacking~\eqref{eq:ident_differenced_new} over $\ell=1,\ldots,2d$ gives
\begin{equation}\label{eq:finite_difference_system}
(L^0)'\boldsymbol{Q}(e_t) - (L^\ast)' \boldsymbol{Q}\!\left(\boldsymbol{v}(e_t,\boldsymbol{y}_{t-1})\right) = \boldsymbol{\psi}(\boldsymbol{y}_{t-1}).
\end{equation}
Equivalently,
\begin{equation}\label{eq:finite_difference_system_alt}
(L^\ast)'\boldsymbol{Q}\!\left(\boldsymbol{v}(e_t,\boldsymbol{y}_{t-1})\right) + \boldsymbol{\psi}(\boldsymbol{y}_{t-1}) = (L^0)'\boldsymbol{Q}(e_t).
\end{equation}

We next show that $L^\ast$ is nonsingular. Differentiating~\eqref{eq:finite_difference_system_alt} with respect to $e_{it}$ gives
\begin{equation}\label{eq:first_derivative_system}
(L^\ast)'\frac{\partial}{\partial e_{it}}\boldsymbol{Q}\!\left(\boldsymbol{v}(e_t,\boldsymbol{y}_{t-1})\right)= (L^0)'
\frac{\partial}{\partial e_{it}} \boldsymbol{Q}(e_t), \quad i=1,\ldots,d.
\end{equation}
For the fixed sufficient statistics in~\eqref{eq:qfixed}, $\partial q_1(x) /\partial x=2x$ and $\partial q_2(x)/\partial x =\frac{x}{\sqrt{x^2+\varepsilon}}$. Since $\mathcal{Z}_0$ is open, after fixing any point in $\mathcal{Z}_0$, we may vary the $i$th shock coordinate locally while holding the remaining shock coordinates and $\boldsymbol{y}_{t-1}$ fixed. Choose two admissible values $a_i$ and $b_i$ for this coordinate such that $a_i b_i\neq 0$ and $|a_i|\neq |b_i|$. For a scalar value $a$ of the $i$th shock coordinate, the derivative vector $\frac{\partial}{\partial e_{it}}\boldsymbol{Q}(e_t)$ is a $(2d\times 1)$ column vector whose entries are zero except in the two entries corresponding to the $i$th block of $\boldsymbol{Q}(e_t)$, where the nonzero entries are $(\partial q_1(a) /\partial a, \partial q_2(a) /\partial a) = (2a, a/\sqrt{a^2+\varepsilon})$. Thus, for component $i$, the two derivative columns obtained at $e_{it}=a_i$ and $e_{it}=b_i$ have the following nonzero two-dimensional block:
\begin{equation}\label{eq:Ri0}
R_i^0 = \begin{bmatrix}
2a_i & 2b_i\\
\dfrac{a_i}{\sqrt{a_i^2+\varepsilon}} &
\dfrac{b_i}{\sqrt{b_i^2+\varepsilon}}
\end{bmatrix}.
\end{equation}
The determinant of this block is
\begin{equation}
\det R_i^0 = 2a_i b_i \left(\frac{1}{\sqrt{b_i^2+\varepsilon}} - \frac{1}{\sqrt{a_i^2+\varepsilon}}\right) \neq 0,
\end{equation}
by the choice of $a_i$ and $b_i$. For each $i=1,\ldots,d$, let $\boldsymbol{d}_{i,a}^0$ and $\boldsymbol{d}_{i,b}^0$ denote the two $(2d\times 1)$ derivative columns
\begin{align*}
\boldsymbol{d}_{i,a}^0
&=
\left.\frac{\partial}{\partial e_{it}}\boldsymbol{Q}(e_t)\right|_{e_{it}=a_i},
&
\boldsymbol{d}_{i,b}^0
&=
\left.\frac{\partial}{\partial e_{it}}\boldsymbol{Q}(e_t)\right|_{e_{it}=b_i},
\end{align*}
with the remaining coordinates held fixed at the selected local point. Collect these columns in
\begin{equation*}
R^0
=
\left[\boldsymbol{d}_{1,a}^0:\boldsymbol{d}_{1,b}^0:
\cdots:\boldsymbol{d}_{d,a}^0:\boldsymbol{d}_{d,b}^0\right].
\end{equation*}
Since $\partial\boldsymbol{Q}(e_t)/\partial e_{it}$ has nonzero entries only in the $i$th two-dimensional block of $\boldsymbol{Q}(e_t)$, the matrix $R^0$ has the block-diagonal form
\begin{equation}\label{eq:R0_def}
R^0 = \begin{bmatrix}
R_1^0 & 0 & \cdots & 0\\
0 & R_2^0 & \cdots & 0\\
\vdots & \vdots & \ddots & \vdots\\
0 & 0 & \cdots & R_d^0
\end{bmatrix}.
\end{equation}
Since each diagonal block $R_i^0$ is nonsingular, $R^0$ is nonsingular.

Similarly, for each $i=1,\ldots,d$, let $\boldsymbol{d}_{i,a}^\ast$ and $\boldsymbol{d}_{i,b}^\ast$ denote the learned-side derivative columns
\begin{align*}
\boldsymbol{d}_{i,a}^\ast
&=
\left.\frac{\partial}{\partial e_{it}}
\boldsymbol{Q}\!\left(\boldsymbol{v}(e_t,\boldsymbol{y}_{t-1})\right)
\right|_{e_{it}=a_i},\\
\boldsymbol{d}_{i,b}^\ast
&=
\left.\frac{\partial}{\partial e_{it}}
\boldsymbol{Q}\!\left(\boldsymbol{v}(e_t,\boldsymbol{y}_{t-1})\right)
\right|_{e_{it}=b_i},
\end{align*}
with the remaining coordinates fixed at the same selected local points as in the construction of $R^0$. Define
\begin{equation*}
R^\ast
=
\left[\boldsymbol{d}_{1,a}^\ast:\boldsymbol{d}_{1,b}^\ast:
\cdots:\boldsymbol{d}_{d,a}^\ast:\boldsymbol{d}_{d,b}^\ast\right].
\end{equation*}
Unlike $R^0$, the matrix $R^\ast$ need not be block diagonal, since $\boldsymbol{v}(e_t,\boldsymbol{y}_{t-1})$ may initially depend on all shock coordinates.
Stacking~\eqref{eq:first_derivative_system} at these points gives
\begin{equation}\label{eq:Lstar_invertibility_system}
(L^\ast)' R^\ast = (L^0)' R^0.
\end{equation}
The right-hand side is nonsingular because $L^0$ is nonsingular by Condition~\ref{cond:variability} and $R^0$ is nonsingular by construction. Hence the left-hand side is nonsingular, which implies that $L^\ast$ is nonsingular.

We now use cross-derivatives to show componentwise separation. Write
\begin{equation}\label{eq:x}
x=(x_1,\ldots,x_{d+dp}) = (e_{1t},\ldots,e_{dt},\boldsymbol{y}_{t-1})
\end{equation}
for the latent-coordinate argument of $\boldsymbol{v}$. In this notation, write
\begin{equation}
\boldsymbol{v}(x) = \big(v_1(x),\ldots,v_d(x)\big),
\end{equation}
so that $v_a(x)$ denotes the $a$th learned coordinate as a function of the latent-coordinate argument $x$. For any $c\in\{1,\ldots,d\}$ and any $m\in\{1,\ldots,d+dp\}$ with $m\neq c$, first differentiate~\eqref{eq:finite_difference_system_alt} with respect to the shock coordinate $x_c$. This removes $\boldsymbol{\psi}(\boldsymbol{y}_{t-1})$ without requiring it to be differentiable in the lag coordinates. Next differentiate the resulting identity with respect to $x_m$. The corresponding mixed derivative of the right-hand side is zero because $\boldsymbol{Q}(e_t)$ is componentwise in the shock coordinates. Hence,
\begin{equation}
(L^\ast)'\frac{\partial^2}{\partial x_c\partial x_m}\boldsymbol{Q}\!\left(\boldsymbol{v}(x)\right)=0.
\end{equation}
Because $L^\ast$ is nonsingular,
\begin{equation}\label{eq:cross_second_zero}
\frac{\partial^2}{\partial x_c\partial x_m}\boldsymbol{Q}\!\left(\boldsymbol{v}(x)\right) = 0.
\end{equation}
For each $a=1,\ldots,d$, define 
\begin{equation}
v_a^c(x)=\frac{\partial v_a(x)}{\partial x_c}, \ \ v_a^m(x)=\frac{\partial v_a(x)}{\partial x_m}, \ \ \text{and} \ \ v_a^{cm}(x)=\frac{\partial^2 v_a(x)}{\partial x_c\partial x_m}.
\end{equation} 
By the chain rule, the two entries of~\eqref{eq:cross_second_zero} corresponding to the $a$th learned component are
\begin{equation}\label{eq:two_cross_equations}
\begin{bmatrix}
\left.\partial q_1(z)/\partial z\right|_{z=v_a(x)} & \left.\partial^2 q_1(z)/\partial z^2\right|_{z=v_a(x)}\\
\left.\partial q_2(z)/\partial z\right|_{z=v_a(x)} & \left.\partial^2 q_2(z)/\partial z^2\right|_{z=v_a(x)}
\end{bmatrix}
\big(v_a^{cm}(x),v_a^c(x)v_a^m(x)\big) = 0.
\end{equation}
For the sufficient statistics in~\eqref{eq:qfixed}, we have
\begin{equation}
\frac{\partial q_1(z)}{\partial z}=2z, \ \ \frac{\partial^2 q_1(z)}{\partial z \partial z}=2, \ \
\frac{\partial q_2(z)}{\partial z}=\frac{z}{\sqrt{z^2+\varepsilon}}, \ \ \text{and} \ \ \frac{\partial^2 q_2(z)}{\partial z \partial z}=\frac{\varepsilon}{(z^2+\varepsilon)^{3/2}}.
\end{equation}
Therefore the determinant of the $(2\times2)$ matrix in~\eqref{eq:two_cross_equations} is
\begin{equation}
\left.\left(\frac{\partial q_1(z)}{\partial z}\frac{\partial^2 q_2(z)}{\partial z \partial z} - \frac{\partial q_2(z)}{\partial z}\frac{\partial^2 q_1(z)}{\partial z \partial z} \right)\right|_{z=v_a(x)} = -\frac{2v_a(x)^3}{(v_a(x)^2+\varepsilon)^{3/2}},
\end{equation}
which is nonzero whenever $v_a(x)\neq0$. Hence, at such points, the coefficient matrix in~\eqref{eq:two_cross_equations} is invertible, which gives both $v_a^c(x)v_a^m(x)=0$ and $v_a^{cm}(x)=0$. If $v_a(x)=0$, the second row of~\eqref{eq:two_cross_equations} gives the product-zero relation directly because $q_2'(0)=0$ and $q_2''(0)=1/\sqrt{\varepsilon}>0$. Consequently,
\begin{equation}\label{eq:product_zero}
v_a^c(x)v_a^m(x)=0
\end{equation}
for every $x\in\mathcal Z_0$, every learned component $a$, every shock-coordinate index $c$, and every $m\neq c$.

Equation~\eqref{eq:product_zero} has the following implication. Fix a point $x$ at which the Jacobian of the augmented map
\begin{equation}
(e_t,\boldsymbol{y}_{t-1}) \mapsto \big(\boldsymbol{v}(e_t,\boldsymbol{y}_{t-1}),\boldsymbol{y}_{t-1}\big)
\end{equation}
is nonsingular. If, for the $a$th coordinate function $v_a$, we have $v_a^c(x) = \partial v_a(x)/\partial x_c\neq0$ for some shock coordinate $c\in\{1,\ldots,d\}$, then~\eqref{eq:product_zero}, applied with this $c$ and each $m\neq c$, gives
\begin{equation}
v_a^m(x) = \frac{\partial v_a(x)}{\partial x_m}=0 \ \ \text{for all } m\neq c.
\end{equation}
Thus, at such a point, the $a$th learned coordinate can have a nonzero first derivative with respect to at most one shock coordinate; if it has such a nonzero shock derivative, then all its derivatives with respect to the lag coordinates are zero.

The augmented map
\begin{equation}
(e_t,\boldsymbol{y}_{t-1}) \mapsto \big(\boldsymbol{v}(e_t,\boldsymbol{y}_{t-1}),\boldsymbol{y}_{t-1}\big)
\end{equation}
is the composition of the true augmented mixing map and the learned augmented map. By Conditions~\ref{cond:smooth} and~\ref{cond:regulatory} of Theorem~\ref{thm:IIA-GCL}, it is a $C^2$ diffeomorphism on $\mathcal{Z}_0$. Its Jacobian has the block form
\begin{equation}
\begin{bmatrix}
J_e(x) & J_y(x)\\
0 & I
\end{bmatrix},
\end{equation}
where $J_e(x)=\left[\partial v_a(x)/\partial e_{ct}\right]_{a,c=1}^d$ and $J_y(x)$ is the corresponding matrix of derivatives with respect to the lag coordinates. Hence, $J_e(x)$ is nonsingular for every $x\in\mathcal{Z}_0$. By the preceding product-zero argument, each row of $J_e(x)$ has at most one nonzero entry. Since $J_e(x)$ is nonsingular, no row or column can be zero. Therefore each row and each column of $J_e(x)$ has exactly one nonzero entry.

Fix a point $x^\circ\in\mathcal{Z}_0$. For each learned coordinate $v_i$, let $\pi_{x^\circ}(i)$ denote the unique shock-coordinate index such that
\begin{equation}
\frac{\partial v_i(x^\circ)}{\partial e_{\pi_{x^\circ}(i)t}}\neq0.
\end{equation}
By continuity, after possibly shrinking the neighborhood of $x^\circ$, this derivative remains nonzero throughout the neighborhood. Applying~\eqref{eq:product_zero} in this neighborhood then implies that all other first derivatives of $v_i$ vanish there, including the derivatives with respect to the lag coordinates. Hence, locally around $x^\circ$, the learned coordinate $v_i$ is a function only of the shock coordinate $e_{\pi_{x^\circ}(i)t}$. Moreover, because the derivative with respect to this selected shock coordinate remains nonzero, the corresponding local scalar function is one-to-one on its local scalar domain, after shrinking the scalar domain if necessary. Since the columns of $J_e(x^\circ)$ also contain exactly one nonzero entry, the map $i\mapsto\pi_{x^\circ}(i)$ is a permutation of $\{1,\ldots,d\}$.

Consequently, for every $x^\circ\in\mathcal{Z}_0$, there exists a neighborhood $\mathcal{U}_{x^\circ}\subset\mathcal{Z}_0$, a permutation $\pi_{x^\circ}$ of $\{1,\ldots,d\}$, and scalar one-to-one functions $S_{1,x^\circ},\ldots,S_{d,x^\circ}$ such that, on $\mathcal{U}_{x^\circ}$,
\begin{equation}\label{eq:componentwise_S_local}
v_i(e_t,\boldsymbol{y}_{t-1}) = S_{i,x^\circ}(e_{\pi_{x^\circ}(i)t}), \quad i=1,\ldots,d.
\end{equation}

\noindent\textbf{Step~4. Algebraic restriction imposed by the fixed sufficient statistics.}
We now use the local componentwise representation obtained at the end of Step~3. Fix $x^\circ\in\mathcal{Z}_0$, and let $\mathcal{U}_{x^\circ}$ be a neighborhood on which the local componentwise representation~\eqref{eq:componentwise_S_local} holds. To simplify notation within this step, write $\pi=\pi_{x^\circ}$ and $S_i=S_{i,x^\circ}$.

Recall from~\eqref{eq:qfixed} that $\boldsymbol{q}(a) = (q_1(a),q_2(a))$ denotes the two-dimensional sufficient-statistic block for a scalar argument $a$. Thus, for a $d$-dimensional vector $z=(z_1,\ldots,z_d)$, the $(2d\times 1)$ stacked vector defined in~\eqref{eq:Qstack} can be written blockwise as
\begin{equation}
\boldsymbol{Q}(z)
=
\big(\boldsymbol{q}(z_1),\ldots,\boldsymbol{q}(z_d)\big).
\end{equation}
Since $L^\ast$ is nonsingular, the finite-difference system~\eqref{eq:finite_difference_system_alt} can be solved for $\boldsymbol{Q}(\boldsymbol{v}(e_t,\boldsymbol{y}_{t-1}))$. Define
\begin{equation}
A = \big((L^\ast)'\big)^{-1}(L^0)'.
\end{equation}
Then, on $\mathcal{U}_{x^\circ}$,
\begin{equation}\label{eq:Q_affine_local}
\boldsymbol{Q}\!\left(\boldsymbol{v}(e_t,\boldsymbol{y}_{t-1})\right) = A\boldsymbol{Q}(e_t) - \big((L^\ast)'\big)^{-1}\boldsymbol{\psi}(\boldsymbol{y}_{t-1}).
\end{equation}
By the local componentwise representation~\eqref{eq:componentwise_S_local}, the left-hand side of~\eqref{eq:Q_affine_local} depends on $e_t$ but not on $\boldsymbol{y}_{t-1}$. Since $A\boldsymbol{Q}(e_t)$ also does not depend on $\boldsymbol{y}_{t-1}$, the last term in~\eqref{eq:Q_affine_local} is locally constant on $\mathcal{U}_{x^\circ}$. Denote the resulting constant vector by $\boldsymbol{b}$. Then
\begin{equation}\label{eq:Q_affine_local_const}
\boldsymbol{Q}\!\left(\boldsymbol{v}(e_t,\boldsymbol{y}_{t-1})\right) = A\boldsymbol{Q}(e_t)+\boldsymbol{b}.
\end{equation}
Partition $A$ into $(2\times2)$ blocks $A_{ik}$, $i,k=1,\ldots,d$, conformably with the two-entry blocks of $\boldsymbol{Q}$, and partition $\boldsymbol{b}$ into two-dimensional blocks $\boldsymbol{b}_i=(b_{i1},b_{i2})$. Thus, $A_{ik}$ maps the $k$th two-entry block $\boldsymbol{q}(e_{kt})$ of $\boldsymbol{Q}(e_t)$ into the $i$th two-entry block of $A\boldsymbol{Q}(e_t)$. Taking rows $2i-1$ and $2i$ of the vector equation~\eqref{eq:Q_affine_local_const}, and using the local representation $v_i(e_t,\boldsymbol{y}_{t-1})=S_i(e_{\pi(i)t})$, gives
\begin{equation}\label{eq:block_affine_pre}
\boldsymbol{q}\!\left(S_i(e_{\pi(i)t})\right) = \sum_{k=1}^d A_{ik}\boldsymbol{q}(e_{kt})+\boldsymbol{b}_i.
\end{equation}

For $k\neq\pi(i)$, the left-hand side of~\eqref{eq:block_affine_pre} does not depend on $e_{kt}$. Differentiating~\eqref{eq:block_affine_pre} with respect to $e_{kt}$ therefore gives
\begin{equation}\label{eq:Aijqderiv}
A_{ik}\frac{\partial}{\partial e_{kt}}\boldsymbol{q}(e_{kt})=0.
\end{equation}
This equality holds for all local admissible values of $e_{kt}$. Since the local domain is open, choose two such values $a_k^{(1)}$ and $a_k^{(2)}$ satisfying $a_k^{(1)}a_k^{(2)}\neq0$ and $|a_k^{(1)}|\neq|a_k^{(2)}|$. Evaluating the previous vector equality at $e_{kt}=a_k^{(1)}$ and $e_{kt}=a_k^{(2)}$ gives two zero-vector equations. Placing the two corresponding derivative vectors side by side as columns and using~\eqref{eq:Aijqderiv} gives
\begin{equation}
A_{ik}\begin{bmatrix}
2a_k^{(1)} & 2a_k^{(2)}\\
\dfrac{a_k^{(1)}}{\sqrt{(a_k^{(1)})^2+\varepsilon}} &
\dfrac{a_k^{(2)}}{\sqrt{(a_k^{(2)})^2+\varepsilon}}
\end{bmatrix}=0.
\end{equation}
The determinant of the $(2\times2)$ matrix above is
\begin{equation}
2a_k^{(1)}a_k^{(2)} \left(\frac{1}{\sqrt{(a_k^{(2)})^2+\varepsilon}} - \frac{1}{\sqrt{(a_k^{(1)})^2+\varepsilon}}\right),
\end{equation}
which is nonzero by the choice of $a_k^{(1)}$ and $a_k^{(2)}$. Hence the matrix is nonsingular, and therefore $A_{ik}=0$ for all $k\neq\pi(i)$.

Consequently, for each $i=1,\ldots,d$, only the block $A_{i,\pi(i)}$ remains in~\eqref{eq:block_affine_pre}. Hence,
\begin{equation}\label{eq:block_affine}
\boldsymbol{q}\!\left(S_i(a)\right) = A_{i,\pi(i)}\boldsymbol{q}(a)+\boldsymbol{b}_i,
\end{equation}
for scalar arguments $a$ in the local domain of $S_i$. The matrix $A_{i,\pi(i)}$ is nonsingular because $A$ is nonsingular and, after permuting its block columns according to $\pi$, it is block diagonal with diagonal blocks $A_{1,\pi(1)},\ldots,A_{d,\pi(d)}$.

We next characterize the scalar transformations $S_i$ satisfying~\eqref{eq:block_affine}. Let $s_\varepsilon=\sqrt{\varepsilon}$. Since $q_2(a)=\sqrt{a^2+\varepsilon}-s_\varepsilon$, we have
\begin{equation}\label{eq:qcurve}
q_1(a)=a^2=q_2(a)^2+2s_\varepsilon q_2(a).
\end{equation}
Writing
\begin{equation}
A_{i,\pi(i)}=\begin{bmatrix}
\alpha_{i,11} & \alpha_{i,12}\\
\alpha_{i,21} & \alpha_{i,22}
\end{bmatrix},
\end{equation}
the second row of~\eqref{eq:block_affine} is
\begin{equation}\label{eq:q2_affine_general}
q_2(S_i(a)) = \alpha_{i,21}q_1(a)+\alpha_{i,22}q_2(a)+b_{i2}.
\end{equation}
Using~\eqref{eq:qcurve}, Equation~\eqref{eq:q2_affine_general} can be rewritten as
\begin{equation}\label{eq:q2_affine_general_q2}
q_2(S_i(a)) = \alpha_{i,21}q_2(a)^2 + (2s_\varepsilon\alpha_{i,21}+\alpha_{i,22})q_2(a) + b_{i2}.
\end{equation}
Substituting~\eqref{eq:q2_affine_general_q2} into~\eqref{eq:qcurve} evaluated at $S_i(a)$ shows that, if $\alpha_{i,21}\neq0$, then $q_1(S_i(a))$ would contain the fourth-order term $\alpha_{i,21}^2q_2(a)^4$. In contrast, the first row of~\eqref{eq:block_affine} gives
\begin{equation}
q_1(S_i(a)) = \alpha_{i,11}q_1(a)+\alpha_{i,12}q_2(a)+b_{i1},
\end{equation}
which, by~\eqref{eq:qcurve}, is a polynomial of degree at most two in $q_2(a)$. Hence $\alpha_{i,21}=0$.

Since $\alpha_{i,21}=0$, equation~\eqref{eq:q2_affine_general} reduces to
\begin{equation}\label{eq:q2_affine}
q_2(S_i(a)) = \alpha_{i,22}q_2(a)+b_{i2}.
\end{equation}
Since $S_i$ is continuous and one-to-one, $\alpha_{i,22}\neq0$. Equation~\eqref{eq:q2_affine} gives the local affine form of the remaining scalar ambiguity. Using the definition of $q_2$, i.e., $q_2(a)=\sqrt{a^2+\varepsilon}-s_\varepsilon$, we obtain
\begin{equation}\label{eq:Si_squared_general}
S_i(a)^2 =
\left(s_\varepsilon+b_{i2}
+\alpha_{i,22}\left(\sqrt{a^2+\varepsilon}-s_\varepsilon\right)\right)^2-\varepsilon.
\end{equation}
Thus, on each local scalar domain, the fixed sufficient statistics restrict the scalar ambiguity to the affine transformation~\eqref{eq:q2_affine} of the second sufficient statistic, equivalently to the squared relation~\eqref{eq:Si_squared_general}. The next step globalizes the local componentwise representation. Once the scalar functions and the affine coefficients have been shown to be global, the fact that the scalar shock domains contain zero will imply $b_{i2}=0$, $\alpha_{i,22}>0$, and hence the stated $T_{c_i}$ form.

\noindent\textbf{Step~5. Globalization of the local representation.}
It remains to globalize the local componentwise representations from Step~3 and the local affine restrictions from Step~4 on the connected support $\mathcal{Z}_0$. Recall from Step~3 that, around each point $x^\circ\in\mathcal{Z}_0$, there exists a local permutation $\pi_{x^\circ}$ such that each learned coordinate depends locally on exactly one shock coordinate. We first show that this permutation is constant on $\mathcal{Z}_0$.

For each learned coordinate $i$ and each shock coordinate $c=1,\ldots,d$, define
\begin{equation}
\mathcal{A}_{ic} =\left\{ x\in\mathcal{Z}_0:\frac{\partial v_i(x)}{\partial e_{ct}}\neq0 \right\}.
\end{equation}
The derivatives are continuous (by the $C^2$ regularity of the true augmented map in Condition~\ref{cond:smooth} and of the learned augmented map in Condition~\ref{cond:regulatory}), so each set $\mathcal{A}_{ic}$ is open. For each fixed $i$, the product-zero result~\eqref{eq:product_zero} from Step~3 implies that, for any $c\neq m$,
\begin{equation}
\frac{\partial v_i(x)}{\partial e_{ct}}\frac{\partial v_i(x)}{\partial e_{mt}} =0, \quad x\in\mathcal Z_0.
\end{equation}
Hence two distinct shock derivatives in row $i$ of $J_e(x)$ cannot be nonzero at the same point. Therefore the sets $\mathcal A_{i1},\ldots,\mathcal A_{id}$ are pairwise disjoint. On the other hand, $J_e(x)$ is nonsingular for every $x\in\mathcal Z_0$ (recall~\eqref{eq:Z0def}). Thus row $i$ of $J_e(x)$ cannot be the zero row, and at least one derivative $\partial v_i(x)/\partial e_{ct}$ must be nonzero at every $x\in\mathcal Z_0$. Consequently,
\begin{equation}
\bigcup_{c=1}^d \mathcal A_{ic}=\mathcal Z_0.
\end{equation}
Since $\mathcal{Z}_0$ is connected (as shown in the beginning of Step~3), exactly one of the sets $\mathcal A_{i1},\ldots,\mathcal A_{id}$ is nonempty for each fixed $i$.\footnote{If two of the sets were nonempty, then one of them and the union of all the others would form two disjoint nonempty open subsets whose union is $\mathcal Z_0$, contradicting connectedness. Since their union is $\mathcal Z_0$, at least one of them is nonempty. Hence exactly one of $\mathcal A_{i1},\ldots,\mathcal A_{id}$ is nonempty for each fixed $i$.}
Let $\kappa_i$ denote the unique index such that $\mathcal A_{i\kappa_i}$ is nonempty. On any local neighborhood $\mathcal U_{x^\circ}$ from Step~3, the representation $v_i(e_t,\boldsymbol y_{t-1}) = S_{i,x^\circ}(e_{\pi_{x^\circ}(i)t})$ implies that all shock derivatives in row $i$ except possibly the derivative with respect to $e_{\pi_{x^\circ}(i)t}$ are zero. Since $J_e(x)$ is nonsingular on $\mathcal Z_0$, row $i$ cannot be the zero row, so the nonzero derivative in row $i$ is precisely the derivative with respect to $e_{\pi_{x^\circ}(i)t}$. Hence, $\pi_{x^\circ}(i)=\kappa_i$ for every $x^\circ\in\mathcal Z_0$. The local permutation index is therefore independent of the choice of base point $x^\circ$, and we write this global index as $\pi(i)$.

Thus, throughout $\mathcal{Z}_0$,
\begin{equation}
\frac{\partial v_i(x)}{\partial e_{kt}}=0 \ \ \ \text{for } k\neq\pi(i), \ \ \text{and} \ \ \
\frac{\partial v_i(x)}{\partial\boldsymbol{y}_{t-1}}=0.
\end{equation}
The product form $\mathcal Z_0=\mathbb R^d\times\mathcal Y_0$ now gives the global componentwise representation directly. For any $a\in\mathbb R$, the fiber obtained by fixing $e_{\pi(i)t}=a$ is connected because it is a product of $\mathbb R^{d-1}$ and the connected set $\mathcal Y_0$. All derivatives of $v_i$ tangent to this fiber vanish by~\eqref{eq:product_zero}. Hence $v_i$ is constant on each such fiber. Consequently, there exist global scalar functions $S_i:\mathbb R\to\mathbb R$, $i=1,\ldots,d$, such that
\begin{equation}\label{eq:componentwise_S_global}
v_i(e_t,\boldsymbol{y}_{t-1}) = S_i(e_{\pi(i)t}), \quad i=1,\ldots,d,
\end{equation}
on $\mathcal Z_0$. Moreover, $S_i'$ equals the unique nonzero shock derivative in row $i$ of $J_e$ and is therefore nowhere zero. Hence each $S_i$ is strictly monotone and, in particular, one-to-one.

We next globalize the affine relation without using local-overlap arguments. Solving~\eqref{eq:finite_difference_system_alt} gives, throughout $\mathcal Z_0$,
\begin{equation}\label{eq:Q_affine_global}
\boldsymbol Q\!\left(\boldsymbol v(e_t,\boldsymbol y_{t-1})\right)
=A\boldsymbol Q(e_t)-\big((L^\ast)'\big)^{-1}\boldsymbol\psi(\boldsymbol y_{t-1}),
\end{equation}
where $A=\big((L^\ast)'\big)^{-1}(L^0)'$ as in Step~4. The first term on each side depends only on $e_t$. Because $\mathcal Z_0$ is a product domain, fixing $e_t$ and varying $\boldsymbol y_{t-1}$ over $\mathcal Y_0$ shows that the final term in~\eqref{eq:Q_affine_global} is a single constant vector $\boldsymbol b$ on $\mathcal Y_0$. The block argument in Step~4, now using the global permutation $\pi$, therefore gives
\begin{equation}\label{eq:block_affine_global}
\boldsymbol q(S_i(a))=A_{i,\pi(i)}\boldsymbol q(a)+\boldsymbol b_i,
\quad a\in\mathbb R.
\end{equation}
The coefficients in the local relation~\eqref{eq:q2_affine} are thus the same global coefficients from $A$ and $\boldsymbol b$. Consequently, for each $i=1,\ldots,d$,
\begin{equation}\label{eq:q2_affine_global}
q_2(S_i(a))=\alpha_{i,22}q_2(a)+b_{i2}
\end{equation}
for all $a\in\mathbb R$.

Since $q_2(a)=q_2(-a)$, the relation~\eqref{eq:q2_affine_global} implies $q_2(S_i(a))=q_2(S_i(-a))$ for all sufficiently small positive $a$. Because $q_2(x)$ is a one-to-one function of $x^2$, this implies
\begin{equation}
|S_i(a)|=|S_i(-a)|.
\end{equation}
Suppose, for contradiction, that $S_i(0)\neq0$. By continuity, $S_i(a)$ and $S_i(-a)$ have the same sign as $S_i(0)$ for all sufficiently small positive $a$. It follows that $|S_i(a)|=|S_i(-a)|$ implies $S_i(a)=S_i(-a)$, contradicting one-to-one-ness of $S_i$ because $a\neq -a$. Thus, $S_i(0)=0$. Evaluating~\eqref{eq:q2_affine_global} at $a=0$ gives $0=q_2(S_i(0))=\alpha_{i,22}q_2(0)+b_{i2}=b_{i2}$.

With $b_{i2}=0$, Equation~\eqref{eq:q2_affine_global} becomes
\begin{equation}
q_2(S_i(a))=\alpha_{i,22}q_2(a).
\end{equation}
For $a\neq0$, one-to-one-ness and $S_i(0)=0$ imply $S_i(a)\neq0$, so $q_2(S_i(a))>0$. Since also $q_2(a)>0$ for $a\neq0$, it follows that $\alpha_{i,22}>0$. With the notation $c_i=\alpha_{i,22}$ and using the definition of $q_2$, the relation $q_2(S_i(a))=c_iq_2(a)$ is equivalent to
\begin{equation}
S_i(a)^2 = \left(s_\varepsilon + c_i\left(\sqrt{a^2+\varepsilon}-s_\varepsilon\right)\right)^2 - \varepsilon = T_{c_i}(a)^2,
\end{equation}
where the last equality uses the definition of $T_c$ in~\eqref{eq:Tc}. Hence $S_i(a)=\pm T_{c_i}(a)$ for all $a\in\mathbb R$.

Since $S_i$ is continuous and one-to-one on $\mathbb R$, it is strictly monotone. If it is increasing, then $S_i(a)$ has the same sign as $a$ and $S_i(a)=T_{c_i}(a)$. If it is decreasing, then $S_i(a)$ has the opposite sign and $S_i(a)=-T_{c_i}(a)$. Therefore, there exists a constant $\iota_i\in\{-1,+1\}$ such that
\begin{equation}\label{eq:global_Tc}
S_i(a)=\iota_i\,T_{c_i}(a)
\end{equation}
for all $a\in\mathbb R$.

Recall that~\eqref{eq:componentwise_S_global} expresses each learned latent coordinate as
$v_i(e_t,\boldsymbol y_{t-1})=S_i(e_{\pi(i)t})$. Evaluating~\eqref{eq:global_Tc} at the scalar argument $a=e_{\pi(i)t}$ therefore gives $S_i(e_{\pi(i)t}) = \iota_i\,T_{c_i}(e_{\pi(i)t})$. Combining these two identities, we obtain
\begin{equation}
v_i(e_t,\boldsymbol{y}_{t-1}) = \iota_i\,T_{c_i}(e_{\pi(i)t}), \quad i=1,\ldots,d,
\end{equation}
for all $(e_t,\boldsymbol{y}_{t-1})\in\mathcal{Z}_0$. Finally, by definition, $\boldsymbol{v}(e_t,\boldsymbol{y}_{t-1}) = \boldsymbol{h}^\ast(f(\boldsymbol{y}_{t-1},e_t),\boldsymbol{y}_{t-1})$ and $e_t=g(y_t,\boldsymbol{y}_{t-1})$. Translating the preceding identity back to observable coordinates yields
\begin{equation}
h_i^\ast(y_t,\boldsymbol{y}_{t-1}) = \iota_i\,T_{c_i}\!\left( g_{\pi(i)}(y_t,\boldsymbol{y}_{t-1}) \right), \quad i=1,\ldots,d,
\end{equation}
for all $(y_t,\boldsymbol{y}_{t-1})\in\mathcal{X}_0$. This proves the theorem.\qed


\subsection{Proof of Corollary~\ref{cor:nonclosure_ident}}\label{sec:proof_cor1}

By Theorem~\ref{thm:IIA-GCL}, any population maximizer satisfies, on the common support $\mathcal X_0$,
\begin{equation}\label{eq:cor1_start_Tc}
h_i^\ast(y_t,\boldsymbol y_{t-1}) = \iota_iT_{c_i}\!\left(g_{\pi(i)}(y_t,\boldsymbol y_{t-1})\right), \quad i=1,\ldots,d,
\end{equation}
for some permutation $\pi$, signs $\iota_i\in\{-1,+1\}$, and constants $c_i>0$, where $T_c$ is defined in~\eqref{eq:Tc}. Since $T_1(x)=x$, it remains to show that the natural-parameter non-closure condition imposed on $\mathfrak L$ implies $c_i=1$ for every component $i=1,\ldots,d$.

Recall that the componentwise sufficient-statistic vector is
\begin{equation}\label{eq:q_cor_def}
\boldsymbol q(x)=\big(q_1(x),q_2(x)\big), \quad \text{where} \ \ q_1(x)=x^2 \ \ \text{and} \ \  q_2(x)=\sqrt{x^2+\varepsilon}-\sqrt{\varepsilon}.
\end{equation}
Both sufficient statistics depend on $x$ only through $x^2$. Hence, changing the sign of $x$ leaves them unchanged:
\begin{equation}
q_j(-x)=q_j(x), \quad j=1,2.
\end{equation}
By the definition of $T_c$,
\begin{align}
q_1(T_c(x)) &= c^2q_1(x)+2\sqrt{\varepsilon}\,c(1-c)q_2(x),\\
q_2(T_c(x)) &= c q_2(x).
\end{align}
Equivalently,
\begin{equation}\label{eq:q_Tc_B_cor}
\boldsymbol q(T_c(x)) = B(c)\boldsymbol q(x),
\end{equation}
where $B(c)$ is defined in~\eqref{eq:Bc}.

We now apply this identity in latent coordinates. Let $\boldsymbol v(e_t,\boldsymbol y_{t-1}) = \boldsymbol h^\ast(f(\boldsymbol y_{t-1},e_t),\boldsymbol y_{t-1})$ as in the proof of Theorem~\ref{thm:IIA-GCL}. Equation~\eqref{eq:cor1_start_Tc} implies
\begin{equation}\label{eq:v_Tc_cor}
v_i(e_t,\boldsymbol y_{t-1}) = \iota_iT_{c_i}(e_{\pi(i)t}), \quad i=1,\ldots,d.
\end{equation}
Using~\eqref{eq:q_cor_def}--\eqref{eq:q_Tc_B_cor}, the sign $\iota_i$ drops out and we obtain
\begin{equation}\label{eq:q_v_B_cor}
\boldsymbol q(v_i(e_t,\boldsymbol y_{t-1})) = B(c_i)\boldsymbol q(e_{\pi(i)t}), \quad i=1,\ldots,d.
\end{equation}

Write the stacked natural-parameter vectors component by component as
\begin{equation}\label{eq:lambda_blocks_cor}
\boldsymbol\lambda^0(u) = \big(\boldsymbol\lambda_1^0(u),\ldots,\boldsymbol\lambda_d^0(u)\big) \ \ \text{and} \ \
\boldsymbol\lambda^\ast(u) = \big(\boldsymbol\lambda_1^\ast(u),\ldots,\boldsymbol\lambda_d^\ast(u)\big),
\end{equation}
where
\begin{equation}
\boldsymbol\lambda_i^0(u) = \big(\lambda_{i1}^0(u),\lambda_{i2}^0(u)\big)\ \ \text{and} \ \ \boldsymbol\lambda_i^\ast(u) = \big(\lambda_{i1}^\ast(u),\lambda_{i2}^\ast(u)\big).
\end{equation}
Similarly, the stacked sufficient-statistic vector~\eqref{eq:Qstack} can be written as
\begin{equation}
\boldsymbol Q(e_t) = \big(\boldsymbol q(e_{1t}),\ldots,\boldsymbol q(e_{dt})\big).
\end{equation}

Now take the latent-coordinate identity~\eqref{eq:rstar_tworeps2} from the proof of Theorem~\ref{thm:IIA-GCL}. Evaluate it at some $u\in\mathcal U_0$ and subtract the same identity evaluated at the baseline value $u^{(0)}$. The terms that are invariant with respect to $u$ cancel, giving
\begin{align}
\begin{aligned}
&\left[\boldsymbol\lambda^0(u)-\boldsymbol\lambda^0(u^{(0)})\right]'\boldsymbol Q(e_t) + \Delta\bar\phi^0(\boldsymbol y_{t-1};u,u^{(0)})\\
&= \left[\boldsymbol\lambda^\ast(u)-\boldsymbol\lambda^\ast(u^{(0)})\right]'\boldsymbol Q(\boldsymbol v(e_t,\boldsymbol y_{t-1})) +  \Delta\bar\phi^\ast(\boldsymbol y_{t-1};u,u^{(0)}),\label{eq:cor_subtracted_ratio}
\end{aligned}
\end{align}
where
\begin{align}
\Delta\bar\phi^0(\boldsymbol y_{t-1};u,u^{(0)}) &= \bar\phi^0(\boldsymbol y_{t-1},u)-\bar\phi^0(\boldsymbol y_{t-1},u^{(0)}),\\
\Delta\bar\phi^\ast(\boldsymbol y_{t-1};u,u^{(0)}) &= \bar\phi^\ast(\boldsymbol y_{t-1},u)-\bar\phi^\ast(\boldsymbol y_{t-1},u^{(0)}).
\end{align}
The terms $\Delta\bar\phi^0$ and $\Delta\bar\phi^\ast$ may depend on $u$ and on the lag vector $\boldsymbol y_{t-1}$, but they do not depend on the current shock vector $e_t$. Thus all dependence on the current shocks in~\eqref{eq:cor_subtracted_ratio} is contained in the two sufficient-statistic terms involving $\boldsymbol Q$.

Using the block notation in~\eqref{eq:lambda_blocks_cor} and substituting~\eqref{eq:q_v_B_cor} into the learned sufficient-statistic term in~\eqref{eq:cor_subtracted_ratio}, we obtain
\begin{align}
\begin{aligned}
&\sum_{j=1}^d \left[\boldsymbol\lambda_j^0(u)-\boldsymbol\lambda_j^0(u^{(0)})\right]'\boldsymbol q(e_{jt}) +\Delta\bar\phi^0(\boldsymbol y_{t-1};u,u^{(0)}) \\
&=\sum_{i=1}^d\left[\boldsymbol\lambda_i^\ast(u)-\boldsymbol\lambda_i^\ast(u^{(0)})\right]'B(c_i)\boldsymbol q(e_{\pi(i)t}) + \Delta\bar\phi^\ast(\boldsymbol y_{t-1};u,u^{(0)}).\label{eq:cor_substituted_ratio}
\end{aligned}
\end{align}
We now isolate the terms associated with a single shock component. Since~\eqref{eq:cor_substituted_ratio} holds for all $(e_t,\boldsymbol y_{t-1})\in\mathcal Z_0$, it also holds after restricting attention to points where $\boldsymbol y_{t-1}$ and all current shock components except $e_{\pi(i)t}$ are held fixed. Along such a one-dimensional slice of the support, the terms $\Delta\bar\phi^0$ and $\Delta\bar\phi^\ast$ are constant because they do not depend on $e_t$, and the sufficient-statistic terms involving the other shock components are also constant because those shock components have been fixed. Hence the only terms that can vary with $e_{\pi(i)t}$ are the two terms multiplying $\boldsymbol q(e_{\pi(i)t})$.

The identity must hold for all values of $e_{\pi(i)t}$ in an open interval. Since the functions $x\mapsto 1$, $q_1(x)=x^2$, and $q_2(x)=\sqrt{x^2+\varepsilon}-\sqrt{\varepsilon}$ are linearly independent on any open interval, no nonzero linear combination of $q_1$ and $q_2$ can be constant there. Therefore the two coefficient vectors multiplying $\boldsymbol q(e_{\pi(i)t})$ must be equal; otherwise the difference between the two sides would contain a nonconstant function of $e_{\pi(i)t}$ that could not be offset by the fixed terms. Thus, for every $u\in\mathcal U_0$,
\begin{equation}\label{eq:lambda_B_relation_cor}
\boldsymbol\lambda_{\pi(i)}^0(u)-\boldsymbol\lambda_{\pi(i)}^0(u^{(0)}) = B(c_i)'\big(\boldsymbol\lambda_i^\ast(u)-\boldsymbol\lambda_i^\ast(u^{(0)})\big), \quad i=1,\ldots,d.
\end{equation}
Define the constant vector
\begin{equation}
\boldsymbol a_i
=
\boldsymbol\lambda_{\pi(i)}^0(u^{(0)})
-B(c_i)'\boldsymbol\lambda_i^\ast(u^{(0)}).
\end{equation}
Equation~\eqref{eq:lambda_B_relation_cor} is equivalent to
\begin{equation}
\boldsymbol\lambda_{\pi(i)}^0(u)
=B(c_i)'\boldsymbol\lambda_i^\ast(u)+\boldsymbol a_i,
\quad u\in\mathcal U_0.
\end{equation}
If $c_i\neq1$, this is a nontrivial transformed-scale representation of the true component $\pi(i)$ with alternative path $\boldsymbol\lambda_i^\ast\in\mathfrak L$. The use of a single class common to all component labels ensures that the learned path with label $i$ is admissible as an alternative path for the true component with label $\pi(i)$. This contradicts the non-closure condition~\eqref{eq:natpar_nonclosure}. Hence, $c_i=1$ for every $i=1,\ldots,d$.

Finally, $T_1(x)=x$. Therefore~\eqref{eq:cor1_start_Tc} reduces to
\begin{equation}
h_i^\ast(y_t,\boldsymbol y_{t-1}) = \iota_i g_{\pi(i)}(y_t,\boldsymbol y_{t-1}), \quad i=1,\ldots,d,
\end{equation}
proving the corollary.\qed


\subsection{Proof of Corollary~\ref{cor:logistic_ident}}\label{sec:proof_cor2}

The proof proceeds by reducing the logistic result to the general non-closure result in Corollary~\ref{cor:nonclosure_ident}. In \textbf{Step~1}, we show that it is enough to verify the non-closure condition~\eqref{eq:natpar_nonclosure} for the logistic class $\mathfrak L_{\mathrm{log}}$. In \textbf{Step~2}, we consider what it would mean for this condition to fail. In particular, a nontrivial transformed-scale representation would have to map one admissible pair of logistic natural-parameter paths into another admissible pair over an open interval of values of $u$. This gives an exact analytic self-aliasing identity for the finite-dimensional logistic specification. In \textbf{Step~3}, we use the singularity structure of the logistic natural-parameter paths to show that these exact self-aliasing configurations can occur only on an explicit Lebesgue-null exceptional set, so that Corollary~\ref{cor:nonclosure_ident} applies and yields identification up to permutation and componentwise sign changes.

\textbf{Step~1}. Set the admissible natural-parameter class in Corollary~\ref{cor:nonclosure_ident} equal to the logistic class $\mathfrak L_{\mathrm{log}}$, which is common to all component labels as required by that corollary. By assumption, both the true and learned componentwise natural-parameter paths belong to $\mathfrak L_{\mathrm{log}}$. Hence, all conditions of Corollary~\ref{cor:nonclosure_ident} are satisfied once the non-closure condition~\eqref{eq:natpar_nonclosure} is verified for $\mathfrak L_{\mathrm{log}}$. It is therefore enough to show that the parameter values for which a nontrivial transformed-scale representation of the form~\eqref{eq:natpar_nonclosure} exists within $\mathfrak L_{\mathrm{log}}$ are contained in a Lebesgue-null exceptional set. Once this is established, Corollary~\ref{cor:nonclosure_ident} applies with $\mathfrak L=\mathfrak L_{\mathrm{log}}$ and gives the desired identification up to permutation and componentwise sign changes.

\textbf{Step~2}. Let $\mathcal I\subseteq\mathcal U_0$ be a nonempty open interval, whose existence is assumed in Corollary~\ref{cor:logistic_ident}, and fix a component $i\in\{1,\ldots,d\}$. Since the non-closure condition in Corollary~\ref{cor:nonclosure_ident} is componentwise, it suffices at this stage to characterize what failure of non-closure would mean for a single component. For a path in $\mathfrak L_{\mathrm{log}}$, write (see Section~\ref{sec:natpar})
\begin{equation}\label{eq:cor2logisticfun}
\Lambda_{ij}(u) = \left[1+\exp\{-\kappa_{ij}(u-\mathfrak c_{ij})\}\right]^{-1}, \quad j=1,2,
\end{equation}
and let $\Delta_{ij}=b_{ij}-a_{ij}$, $j=1,2$. Then
\begin{align}
\tau_i(u) &= \exp\{a_{i1}+\Delta_{i1}\Lambda_{i1}(u)\},\\
\rho_i(u) &= \exp\{a_{i2}+\Delta_{i2}\Lambda_{i2}(u)\}.
\end{align}
The associated natural-parameter paths can be written explicitly as
\begin{align}
\lambda_{i1}(u)
&=-\exp\{-a_{i1}-\Delta_{i1}\Lambda_{i1}(u)\},\label{eq:cor2_explicit_paths}\\
\lambda_{i2}(u)
&=-\exp\left\{-\frac{a_{i1}}{2}-a_{i2}
-\frac{\Delta_{i1}}{2}\Lambda_{i1}(u)
-\Delta_{i2}\Lambda_{i2}(u)\right\}.\nonumber
\end{align}
For finite parameter values with $\kappa_{ij}>0$, these functions are real analytic on $\mathbb R$. The explicit exponential forms also define their complex continuations without requiring a choice of a complex square-root branch.

Suppose that the non-closure condition fails for component $i$. Then there exist a constant $\check c_i>0$ with $\check c_i\neq1$, an alternative admissible logistic path $\bar{\boldsymbol\lambda}_i(u)=\big(\bar\lambda_{i1}(u),\bar\lambda_{i2}(u)\big)\in\mathfrak L_{\mathrm{log}}$, and a constant vector satisfying the affine representation in Corollary~\ref{cor:nonclosure_ident} on $\mathcal U_0$. Fix a baseline value $u_0\in\mathcal I$ and subtract the affine representation at $u_0$ from its value at $u$. This gives
\begin{align}
\lambda_{i1}^0(u)-\lambda_{i1}^0(u_0) &= \check c_i^2 \left(\bar\lambda_{i1}(u)-\bar\lambda_{i1}(u_0)\right),\label{eq:log_self_alias1}\\
\lambda_{i2}^0(u)-\lambda_{i2}^0(u_0) &= 2\sqrt{\varepsilon}\,\check c_i(1-\check c_i) \left(\bar\lambda_{i1}(u)-\bar\lambda_{i1}(u_0)\right) + \check c_i \left(\bar\lambda_{i2}(u)-\bar\lambda_{i2}(u_0)\right)\label{eq:log_self_alias2}
\end{align}
for every $u\in\mathcal U_0$, and hence in particular for every $u\in\mathcal I$. All paths involved are real analytic, so these equalities are exact functional identities on the open interval $\mathcal I$.

Equation~\eqref{eq:log_self_alias1} implies that the true and alternative scale natural-parameter paths are affinely related. Specifically, there exists a constant $K_{i1}$ such that
\begin{equation}\label{eq:lambda1_affine_alias}
\lambda_{i1}^0(u) = \check c_i^2\bar\lambda_{i1}(u)+K_{i1}
\end{equation}
on the open interval. Substituting~\eqref{eq:lambda1_affine_alias} into~\eqref{eq:log_self_alias2} shows that the alternative second natural-parameter path must satisfy, for some constant $K_{i2}$,
\begin{equation}\label{eq:lambda2_additive_alias}
\bar\lambda_{i2}(u) = \check c_i^{-1}\lambda_{i2}^0(u) - 2\sqrt{\varepsilon}\,(1-\check c_i)\check c_i^{-2}\lambda_{i1}^0(u) + K_{i2}.
\end{equation}
Thus, if non-closure fails with $\check c_i\neq1$, the nontrivial transformed-scale representation forces an exact self-aliasing identity within the logistic class. In particular, an alternative admissible logistic second natural-parameter path must be representable as an affine combination of the true first and second natural-parameter paths. Because $\check c_i\neq1$, the coefficient multiplying $\lambda_{i1}^0(u)$ in~\eqref{eq:lambda2_additive_alias} is nonzero. The next step rules out such exact self-aliasing configurations generically.

\textbf{Step~3}. We now rule out the self-aliasing identities~\eqref{eq:lambda1_affine_alias} and~\eqref{eq:lambda2_additive_alias} outside an explicit Lebesgue-null exceptional set. Let $\Theta_{\mathrm{log}}=\mathbb R^6\times(0,\infty)^2$ denote the componentwise logistic parameter space. For each component $i=1,\ldots,d$, write
\begin{equation}
\theta_{i,\tau\rho}^0=
(a_{i1}^0,b_{i1}^0,\mathfrak c_{i1}^0,\kappa_{i1}^0,
a_{i2}^0,b_{i2}^0,\mathfrak c_{i2}^0,\kappa_{i2}^0)
\in\Theta_{\mathrm{log}},
\end{equation}
and define $\Delta_{ij}^0=b_{ij}^0-a_{ij}^0$, $j=1,2$. Define the conservative componentwise exceptional set
\begin{equation}\label{eq:cor2_exceptional_component}
\mathcal E_i=
\left\{
\theta_{i,\tau\rho}^0\in\Theta_{\mathrm{log}}:
\Delta_{i1}^0=0
\quad\text{or}\quad
\Delta_{i2}^0=0
\quad\text{or}\quad
\mathfrak c_{i1}^0=\mathfrak c_{i2}^0
\right\}.
\end{equation}
The corresponding exceptional set in the full parameter space is
\begin{equation}\label{eq:cor2_exceptional_full}
\mathcal E=
\bigcup_{i=1}^d
\left\{
(\theta_{1,\tau\rho}^0,\ldots,\theta_{d,\tau\rho}^0)
\in\Theta_{\mathrm{log}}^d:
\theta_{i,\tau\rho}^0\in\mathcal E_i
\right\}.
\end{equation}
This is a conservative exceptional set: membership in $\mathcal E$ does not by itself imply non-identification. Fix a component $i$ with $\theta_{i,\tau\rho}^0\notin\mathcal E_i$. Then
\begin{equation}
\Delta_{i1}^0\neq0, \quad \Delta_{i2}^0\neq0, \quad \mathfrak c_{i1}^0\neq\mathfrak c_{i2}^0.
\end{equation}

Consider the complex continuation of the logistic transition function
\begin{equation}
\Lambda_{ij}(u) = \left[1+\exp\{-\kappa_{ij}(u-\mathfrak c_{ij})\}\right]^{-1}.
\end{equation}
For finite parameter values with $\kappa_{ij}>0$, this continuation is meromorphic in the complex variable $u$ and has simple poles at
\begin{equation}\label{eq:logistic_poles}
\mathcal P_{ij} = \left\{\mathfrak c_{ij} + \frac{(2m+1)\pi \mathrm{i}}{\kappa_{ij}}: m\in\mathbb Z \right\},
\end{equation}
where $\mathrm{i}=\sqrt{-1}$. The sign of the imaginary part is immaterial because $m$ ranges over all integers. If $\gamma\neq0$, then $\exp\{\gamma\Lambda_{ij}(u)\}$ has an isolated essential singularity at every point of $\mathcal P_{ij}$. It follows from~\eqref{eq:cor2_explicit_paths} that, when $\Delta_{i1}\neq0$, the singularity set of $\lambda_{i1}$ is exactly $\mathcal P_{i1}$. For $\lambda_{i2}$, a point in $\mathcal P_{i1}\setminus\mathcal P_{i2}$ is an essential singularity when $\Delta_{i1}\neq0$, and a point in $\mathcal P_{i2}\setminus\mathcal P_{i1}$ is an essential singularity when $\Delta_{i2}\neq0$. This qualification is necessary because cancellations can occur when the two pole sets overlap. In the present argument, the true scale and shape pole sets are disjoint because their real locations differ.

Suppose, toward a contradiction, that the non-closure condition fails for component $i$. By Step~2, the identity
\begin{equation}\label{eq:cor2_step3_scale_alias}
\lambda_{i1}^0(u) = \check c_i^2\bar\lambda_{i1}(u)+K_{i1}
\end{equation}
holds on $\mathcal I$. Because $\Delta_{i1}^0\neq0$, the true scale path is nonconstant, and therefore the alternative scale path must also satisfy $\bar\Delta_{i1}\neq0$. Define
\begin{equation}
\Omega_{i1}=\mathbb C\setminus\left(\mathcal P_{i1}^0\cup\bar{\mathcal P}_{i1}\right).
\end{equation}
The removed set is discrete and has no finite accumulation point, so $\Omega_{i1}$ is connected. Both sides of~\eqref{eq:cor2_step3_scale_alias} are holomorphic on $\Omega_{i1}$ and agree on $\mathcal I\subset\Omega_{i1}$. The identity theorem therefore extends their equality throughout $\Omega_{i1}$.

If a point $u_\star$ belonged to $\mathcal P_{i1}^0\setminus\bar{\mathcal P}_{i1}$, the right-hand side of~\eqref{eq:cor2_step3_scale_alias} would be holomorphic on a disk around $u_\star$, while equality on the punctured disk would give a removable extension of $\lambda_{i1}^0$ through its essential singularity. This is impossible. Interchanging the true and alternative paths rules out a point in $\bar{\mathcal P}_{i1}\setminus\mathcal P_{i1}^0$. Hence the two scale pole sets coincide:
\begin{equation}
\bar{\mathcal P}_{i1}=\mathcal P_{i1}^0.
\end{equation}
Equality of these pole sets identifies their real locations and their smallest positive absolute imaginary displacements from the real axis. Since both steepness parameters are positive, consequently,
\begin{equation}
\bar{\mathfrak c}_{i1}=\mathfrak c_{i1}^0, \quad \bar\kappa_{i1}=\kappa_{i1}^0.
\end{equation}

It follows that both sides of~\eqref{eq:cor2_step3_scale_alias} are functions of the same logistic variable $z\equiv\Lambda_{i1}^0(u)$. Hence, on a nonempty open interval of values of $z$,
\begin{equation}
-\exp\{-a_{i1}^0-\Delta_{i1}^0 z\} = -\check c_i^2 \exp\{-\bar a_{i1}-\bar\Delta_{i1} z\} + K_{i1}.
\end{equation}
Since $\Delta_{i1}^0\neq0$, the functions $1$, $\exp\{-\Delta_{i1}^0 z\}$, and $\exp\{-\bar\Delta_{i1} z\}$ cannot satisfy this identity unless the constant term vanishes and the exponential rates coincide. Therefore,
\begin{equation}
K_{i1}=0 \ \ \text{and} \ \  \bar\Delta_{i1}=\Delta_{i1}^0,
\end{equation}
and the remaining multiplicative constant is absorbed by the alternative level parameter. Equivalently,
\begin{equation}\label{eq:cor2_step3_lambda1_scaling}
\bar\lambda_{i1}(u)=\check c_i^{-2}\lambda_{i1}^0(u).
\end{equation}

We now analyze the second self-aliasing identity~\eqref{eq:lambda2_additive_alias}. Define
\begin{equation}
\Omega_{i2}=\mathbb C\setminus\left(
\mathcal P_{i1}^0\cup\mathcal P_{i2}^0
\cup\bar{\mathcal P}_{i1}\cup\bar{\mathcal P}_{i2}
\right).
\end{equation}
This domain is connected for the same reason as $\Omega_{i1}$. Both sides of~\eqref{eq:lambda2_additive_alias} are holomorphic on $\Omega_{i2}$ and agree on $\mathcal I$, so the identity theorem extends their equality throughout $\Omega_{i2}$.

Since $\check c_i\neq1$, the coefficient multiplying $\lambda_{i1}^0(u)$ in~\eqref{eq:lambda2_additive_alias} is nonzero. First, consider the singularities at the true shape pole set $\mathcal P_{i2}^0$. The true scale and shape pole sets are disjoint because $\mathfrak c_{i1}^0\neq\mathfrak c_{i2}^0$. At any point $u_\star\in\mathcal P_{i2}^0$, the terms $\lambda_{i1}^0(u)$ and $K_{i2}$ are holomorphic, whereas $\lambda_{i2}^0(u)$ has an essential singularity because $\Delta_{i2}^0\neq0$. Hence the right-hand side of~\eqref{eq:lambda2_additive_alias} has an essential singularity at $u_\star$, and so $\bar\lambda_{i2}(u)$ must also be singular there. The alternative second path can be singular only at poles of its alternative scale transition or its alternative shape transition. Since the first identity already gave $\bar{\mathcal P}_{i1}=\mathcal P_{i1}^0$, this singularity cannot come from the alternative scale transition. It must therefore come from a nonconstant alternative shape transition, so
\begin{equation}
\bar\Delta_{i2}\neq0
\quad\text{and}\quad
\mathcal P_{i2}^0\subseteq\bar{\mathcal P}_{i2}.
\end{equation}

This containment forces the alternative shape pole set to have the same real location as the true shape pole set. Any additional pole in $\bar{\mathcal P}_{i2}\setminus\mathcal P_{i2}^0$ would therefore have real part $\mathfrak c_{i2}^0$. Since $\mathfrak c_{i2}^0\neq\mathfrak c_{i1}^0$, such an additional pole cannot belong to the true scale pole set $\mathcal P_{i1}^0$. It would thus create a singularity on the left-hand side of~\eqref{eq:lambda2_additive_alias} that is absent from its right-hand side. Hence no such additional poles can occur, and
\begin{equation}
\bar{\mathcal P}_{i2}=\mathcal P_{i2}^0.
\end{equation}
As with the scale pole sets, this equality and positivity of the steepness parameters imply
\begin{equation}
\bar{\mathfrak c}_{i2}=\mathfrak c_{i2}^0,
\quad
\bar\kappa_{i2}=\kappa_{i2}^0.
\end{equation}

Now take a point $u_\star\in\mathcal P_{i1}^0$. Since $\mathcal P_{i1}^0$ is disjoint from $\mathcal P_{i2}^0=\bar{\mathcal P}_{i2}$, the true and alternative shape-transition factors are holomorphic and nonzero in a neighborhood of $u_\star$. Define
\begin{equation}
w(u) = \exp\left\{ -\frac{1}{2}\Delta_{i1}^0\Lambda_{i1}^0(u)\right\}.
\end{equation}
The explicit formulas~\eqref{eq:cor2_explicit_paths}, the equality of the scale transition parameters, and~\eqref{eq:cor2_step3_lambda1_scaling} give
\begin{equation}
\lambda_{i2}^0(u)=A(u)w(u),
\quad
\bar\lambda_{i2}(u)=B(u)w(u),
\quad\text{and}\quad
\lambda_{i1}^0(u)=D(u)w(u)^2,
\end{equation}
where $A$ and $B$ are holomorphic and nonzero near $u_\star$, and $D(u)=-\exp\{-a_{i1}^0\}$ is a nonzero constant.

Substituting these local representations into~\eqref{eq:lambda2_additive_alias} gives, in a punctured neighborhood of $u_\star$,
\begin{equation}
B(u)w(u) = \check c_i^{-1}A(u)w(u) - 2\sqrt{\varepsilon}\,(1-\check c_i)\check c_i^{-2} D(u)w(u)^2 + K_{i2}.
\end{equation}
Equivalently,
\begin{equation}
-2\sqrt{\varepsilon}\,(1-\check c_i)\check c_i^{-2}D(u)w(u)^2 + \left\{\check c_i^{-1}A(u)-B(u)\right\}w(u) + K_{i2} = 0.
\end{equation}
Let
\begin{equation}
\alpha(u)=-2\sqrt{\varepsilon}\,(1-\check c_i)\check c_i^{-2}D(u)
\quad\text{and}\quad
\beta(u)=\check c_i^{-1}A(u)-B(u).
\end{equation}
Both functions are holomorphic near $u_\star$, and $\alpha(u_\star)\neq0$ because $\varepsilon>0$, $\check c_i\neq1$, and $D(u_\star)\neq0$. The preceding equation is
\begin{equation}
\alpha(u)w(u)^2+\beta(u)w(u)+K_{i2}=0.
\end{equation}
Since $\Delta_{i1}^0\neq0$, the function $w$ has an essential singularity at $u_\star$ and is unbounded in every punctured neighborhood. Choose a sequence $u_n\to u_\star$ such that $|w(u_n)|\to\infty$. The exponential function never vanishes, so division by $w(u_n)^2$ gives
\begin{equation}
\alpha(u_n)+\frac{\beta(u_n)}{w(u_n)}+\frac{K_{i2}}{w(u_n)^2}=0.
\end{equation}
Taking $n\to\infty$ yields $\alpha(u_\star)=0$, contradicting $\alpha(u_\star)\neq0$. Therefore~\eqref{eq:lambda2_additive_alias} cannot hold with $\check c_i\neq1$.

Consequently, for every component $i$ with $\theta_{i,\tau\rho}^0\notin\mathcal E_i$, no nontrivial transformed-scale representation of the form~\eqref{eq:natpar_nonclosure} exists within $\mathfrak L_{\mathrm{log}}$. Thus, the non-closure condition of Corollary~\ref{cor:nonclosure_ident} holds for component $i$.

Each $\mathcal E_i$ in~\eqref{eq:cor2_exceptional_component} is a finite union of codimension-one subsets of $\Theta_{\mathrm{log}}$ and is therefore Lebesgue-null. Its cylindrical lift to $\Theta_{\mathrm{log}}^d$ is likewise null by Fubini's theorem. Since $d$ is finite, the full set $\mathcal E$ in~\eqref{eq:cor2_exceptional_full} is also Lebesgue-null. Therefore, outside $\mathcal E$, the logistic class $\mathfrak L_{\mathrm{log}}$ satisfies the non-closure condition~\eqref{eq:natpar_nonclosure} for every component. By Step~1, Corollary~\ref{cor:nonclosure_ident} applies with $\mathfrak L=\mathfrak L_{\mathrm{log}}$, and hence any population maximizer satisfies
\begin{equation}
h_i^\ast(y_t,\boldsymbol y_{t-1}) = \iota_i g_{\pi(i)}(y_t,\boldsymbol y_{t-1}), \quad i=1,\ldots,d,
\end{equation}
on $\mathcal X_0$, where $\pi$ is a permutation of $\{1,\ldots,d\}$ and $\iota_i\in\{-1,+1\}$. This proves the corollary.\qed


\section{Monte Carlo details}\label{sec:mc_details}

This appendix describes the Monte Carlo experiments summarized in Section~\ref{sec:mc_evidence}. The objective of the experiments is to examine the finite-sample performance of the IIA-GCL estimator in recovering the structural shocks when the true data generating process (DGP) is fully nonlinear. 

\subsection{Data-generating processes}

All Monte Carlo designs are bivariate and use autoregressive order one:
\begin{equation}
y_t = f(y_{t-1},e_t), \quad t=1,\ldots,T,
\end{equation}
where $e_t=(e_{1t},e_{2t})$ contains the structural shocks and $f$ is a nonlinear function implemented as a feed-forward neural network (specified below). The auxiliary variable follows the autoregression
\begin{equation}
u_t = 0.65u_{t-1} + \xi_t, \quad \xi_t\sim N(0,1),
\end{equation}
after which the simulated path of $u_t$ is standardized to have sample mean zero and sample variance one. A burn-in of 200 observations is discarded before the estimation sample is formed.

The true structural map is the same in all three Monte Carlo designs. Let
\begin{equation}
x_t = (y_{1,t-1},y_{2,t-1},e_{1t},e_{2t})
\end{equation}
and let $\sigma_{\mathrm{S}}(x)=\{1+\exp(-x)\}^{-1}$ denote the sigmoid activation function applied componentwise. The true map is
\begin{align}
h_{1t} &= \sigma_{\mathrm{S}}(W_1x_t+b_1)-\sigma_{\mathrm{S}}(b_1),\\
h_{2t} &= \sigma_{\mathrm{S}}(W_2h_{1t}+b_2)-\sigma_{\mathrm{S}}(b_2),\\
y_t &= W_oh_{2t},
\end{align}
where
\begin{align}
W_1 &=
\begin{pmatrix}
1.76 & -1.67 & 1.30 & 0\\
1.06 & 2.25 & 1.44 & 0\\
0.68 & -0.69 & -1.24 & 1.20\\
-1.78 & 1.86 & 1.23 & 1.61
\end{pmatrix},
&
b_1 &= (-0.50,-0.43,-0.26,0.31),\\
W_2 &=
\begin{pmatrix}
2.42 & 2.61 & 0 & 0\\
2.73 & 1.75 & 0 & 0\\
1.28 & -0.01 & 2.17 & 1.87\\
-1.38 & -0.77 & 1.52 & 2.94
\end{pmatrix},
&
b_2 &= (-0.32,0.04,0.37,0.03).
\end{align}
The subtraction of $\sigma_{\mathrm{S}}(b_1)$ and $\sigma_{\mathrm{S}}(b_2)$ is a zero-at-zero normalization of the hidden-layer features. Equivalently, each hidden layer uses the transformed activation $\tilde\sigma_{\mathrm{S},b}(z)=\sigma_{\mathrm{S}}(z+b)-\sigma_{\mathrm{S}}(b)$, so that $f(0,0)=0$. This normalization removes the deterministic offset that the sigmoid hidden-layer biases would otherwise introduce into the recursive simulation, while the nonlinear dependence on both $y_{t-1}$ and $e_t$ is preserved through the shifted activation functions.

The weight matrices are also constructed so that the structural map is invertible with respect to the contemporaneous shocks. The triangular zero patterns in $W_1$ and $W_2$ guarantee a one-to-one, strictly monotone transmission from the two structural shocks to the two latent pathways; they are imposed to satisfy the invertibility condition and do not represent economic exclusion restrictions. In particular, the first two hidden features form a pathway that is strictly increasing in $e_{1t}$ and does not depend on $e_{2t}$, while the remaining features form a second pathway that is strictly increasing in $e_{2t}$ and may also depend on $e_{1t}$. The output matrix $W_o$ is defined as the product
\begin{equation}
\begin{aligned}
A&=
\begin{pmatrix}
1 & -1\\
1.60 & -0.20
\end{pmatrix},
\qquad
B=
\begin{pmatrix}
0.80 & 0.55 & 0 & 0\\
0 & 0 & 2.00 & 1.10
\end{pmatrix},\\
W_o&=AB=
\begin{pmatrix}
0.80 & 0.55 & -2.00 & -1.10\\
1.28 & 0.88 & -0.40 & -0.22
\end{pmatrix}.
\end{aligned}
\end{equation}
Because the two latent pathways have positive own-shock derivatives and $\det(A)>0$, the determinant of $\partial y_t/\partial e_t$ is positive for every finite input. The full matrix $A$ nevertheless mixes both latent pathways into both observed variables, so the observed-coordinate map is not triangular.

The benchmark shock design uses the exponential-family conditional distribution in Assumption~\ref{as:suffstats}. The conditional scale and shape functions in the correctly specified benchmark are
\begin{align}
\tau_i(u_t) &= \exp\left\{(1-\Lambda_{i1}(u_t))a_{i1}+\Lambda_{i1}(u_t)b_{i1}\right\},\\
\rho_i(u_t) &= \exp\left\{(1-\Lambda_{i2}(u_t))a_{i2}+\Lambda_{i2}(u_t)b_{i2}\right\},
\end{align}
where $\Lambda_{ij}(u_t)=\left[1+\exp\{-\kappa_{ij}(u_t-\mathfrak c_{ij})\}\right]^{-1}$. The endpoint values and transition parameters are reported in Table~\ref{tab:mc_dgp_parameters}. The two misspecified designs are included to examine whether shock recovery remains useful when the exact exponential-family shock density is violated. They keep the same structural map but replace the shock distribution by conditionally independent Student-$t$ or skewed-$t$ shocks \citep[see][for the latter]{Hansen:1994}.

In the Student-$t$ design, the shocks are generated as
\begin{equation}
e_{it}=\sqrt{v_i(u_t)}z_{it},\qquad
z_{it}\mid u_t\sim
\sqrt{\frac{\nu_i(u_t)-2}{\nu_i(u_t)}}\,t_{\nu_i(u_t)},
\end{equation}
where $v_i(u_t)$ is the conditional variance and $\nu_i(u_t)>2$ is the degrees-of-freedom parameter. Thus, $z_{it}$ has conditional mean zero and conditional variance one, and $\operatorname{Var}(e_{it}\mid u_t)=v_i(u_t)$. In the skewed-$t$ design,
\begin{equation}
e_{it}=\sqrt{v_i(u_t)}z_{it},\qquad
z_{it}\mid u_t\sim \operatorname{ST}_{\mathrm H}
\bigl(\nu_i(u_t),\zeta_i(u_t)\bigr),
\end{equation}
where $\operatorname{ST}_{\mathrm H}$ denotes Hansen's skewed-$t$ distribution standardized to conditional mean zero and conditional variance one, and $\zeta_i(u_t)$ is its skewness parameter. Consequently, also in this design $\operatorname{Var}(e_{it}\mid u_t)=v_i(u_t)$. Their dependence on $u_t$ is specified as follows. Letting $\Lambda(u;c,\kappa)=[1+\exp\{-\kappa(u-c)\}]^{-1}$, a positive quantity with low-$u$ and high-$u$ endpoints $q^{\mathrm L}$ and $q^{\mathrm H}$ is varied according to
\begin{equation}
q(u)=\exp\left\{[1-\Lambda(u;c,\kappa)]\log q^{\mathrm L}
+\Lambda(u;c,\kappa)\log q^{\mathrm H}\right\},
\end{equation}
as used for $v_i(u_t)$. The degrees of freedom and skewness instead use the arithmetic transition
\begin{equation}
q(u)=[1-\Lambda(u;c,\kappa)]q^{\mathrm L}
+\Lambda(u;c,\kappa)q^{\mathrm H}.
\end{equation}
The component-specific endpoints and transition parameters in these expressions are reported in Table~\ref{tab:mc_dgp_parameters}. 
The conditional variance $v_i(u_t)$ has a role analogous to the scale variation induced by $\tau_i(u_t)$, while $\nu_i(u_t)$ controls tail thickness and, in the skewed-$t$ design, $\zeta_i(u_t)$ additionally controls asymmetry. These are analogous sources of scale and shape variation, but there is no one-to-one parameter mapping between $(\tau_i,\rho_i)$ and $(v_i,\nu_i,\zeta_i)$.

\begin{table}[!ht]
\centering
\begin{tabular}{p{0.17\textwidth}p{0.24\textwidth}p{0.48\textwidth}}
\hline\\[-1.3ex]
DGP & Quantity & Parameter values \\
\hline\\[-1.3ex]
Exponential-family & Scale $\tau_i(u_t)$ &
Low endpoints $(0.22,0.30)$ and high endpoints $(6.50,5.50)$. Transition locations $(-0.70,0.35)$ and steepnesses $(1.7,3.1)$.\\[0.7em]
Exponential-family & Shape $\rho_i(u_t)$ &
Low endpoints $(0.35,0.45)$ and high endpoints $(16.00,14.00)$. Transition locations $(0.25,-0.45)$ and steepnesses $(3.4,1.9)$.\\[0.7em]
Student-$t$ & Variance &
Low endpoints $(0.45,0.55)$ and high endpoints $(2.80,2.40)$. Transition locations $(-0.65,0.30)$ and steepnesses $(1.8,2.8)$.\\[0.7em]
Student-$t$ & Degrees of freedom &
The paths are $\nu_1(u_t)=18-13\Lambda(u_t;0.20,2.6)$ and $\nu_2(u_t)=20-15\Lambda(u_t;-0.40,1.9)$, where $\Lambda(u;c,\kappa)=[1+\exp\{-\kappa(u-c)\}]^{-1}$. Thus, the degrees of freedom range from $5$ to $18$ and from $5$ to $20$, respectively.\\[0.7em]
Skewed-$t$ & Variance &
Low endpoints $(0.45,0.55)$ and high endpoints $(2.70,2.30)$. Transition locations $(-0.65,0.30)$ and steepnesses $(1.8,2.8)$.\\[0.7em]
Skewed-$t$ & Degrees of freedom &
The paths are $\nu_1(u_t)=20-15\Lambda(u_t;0.20,2.6)$ and $\nu_2(u_t)=21-16\Lambda(u_t;-0.40,1.9)$, giving ranges from $5$ to $20$ and from $5$ to $21$, respectively.\\[0.7em]
Skewed-$t$ & Skewness &
The paths are $\zeta_1(u_t)=-0.35+0.70\Lambda(u_t;0.20,2.6)$ and $\zeta_2(u_t)=0.35-0.70\Lambda(u_t;-0.40,1.9)$. Thus, each skewness parameter ranges from $-0.35$ to $0.35$, with the transition direction reversed across the two components.\\
\hline
\end{tabular}
\caption{Shock-distribution parameters in the Monte Carlo designs. All three designs use the same nonlinear structural map; only the conditional shock distribution differs across designs.}
\label{tab:mc_dgp_parameters}
\end{table}

\subsection{Estimation and evaluation}

For each DGP and sample size $T\in\{250,500,1000,2000\}$, the Monte Carlo experiment uses 100 replications. Within each replication, one path of length 2000 is generated and the shorter samples are formed as nested prefixes of that path. Before estimation, the observed variables and auxiliary variable are standardized within each sample to zero mean and unit variance. The IIA-GCL estimator is fitted with $p=1$ and eight independent optimization starts in each replication. The demixing network has one hidden layer with 12 neurons and sigmoid activation. The nuisance networks have one hidden layer with 8 neurons and leaky-ReLU activation. The training procedure follows Appendix~\ref{sec:iia_gcl_training_details}.

Shock recovery is measured after resolving the remaining sign and permutation indeterminacies. For each replication, the estimated shock matrix is aligned with the true shock matrix by choosing the permutation and componentwise signs that maximize the mean absolute componentwise correlation. Let $\hat\rho_{1r}$ and $\hat\rho_{2r}$ denote the resulting absolute correlations in replication $r$. The reported mean recovery is the Monte Carlo average of $(\hat\rho_{1r}+\hat\rho_{2r})/2$, while componentwise results report the Monte Carlo means and standard deviations of $\hat\rho_{1r}$ and $\hat\rho_{2r}$ separately. 

\begin{table}[!ht]
\centering
\begin{tabular}{llcccc}
\hline\\[-1.3ex]
DGP & $T$ & Shock 1 & Shock 2 & Mean & BCE \\
\hline\\[-1.3ex]
Exponential-family & 250 & 0.88 (0.04) & 0.86 (0.04) & 0.87 (0.04) & 0.60 \\
Exponential-family & 500 & 0.89 (0.03) & 0.87 (0.05) & 0.88 (0.04) & 0.59 \\
Exponential-family & 1000 & 0.90 (0.01) & 0.89 (0.01) & 0.89 (0.01) & 0.59 \\
Exponential-family & 2000 & 0.91 (0.02) & 0.89 (0.02) & 0.90 (0.02) & 0.59\\[1.0ex]
Student-$t$ & 250 & 0.88 (0.05) & 0.87 (0.05) & 0.87 (0.05) & 0.67 \\
Student-$t$ & 500 & 0.89 (0.03) & 0.88 (0.03) & 0.88 (0.03) & 0.67 \\
Student-$t$ & 1000 & 0.89 (0.02) & 0.88 (0.03) & 0.88 (0.02) & 0.67\\
Student-$t$ & 2000 & 0.90 (0.01) & 0.89 (0.02) & 0.89 (0.01) & 0.68\\[1.0ex]
Skewed-$t$ & 250 & 0.88 (0.06) & 0.87 (0.06) & 0.88 (0.06) & 0.67 \\
Skewed-$t$ & 500 & 0.88 (0.05) & 0.87 (0.06) & 0.88 (0.05) & 0.67\\
Skewed-$t$ & 1000 & 0.89 (0.03) & 0.88 (0.04) & 0.89 (0.03) & 0.67 \\
Skewed-$t$ & 2000 & 0.90 (0.02) & 0.89 (0.02) & 0.90 (0.02) & 0.67\\
\hline
\end{tabular}
\caption{Detailed Monte Carlo shock-recovery results under minimum-BCE selection. ``Shock 1'' and ``Shock 2'' report the componentwise aligned absolute correlations. ``Mean'' reports their within-replication average. Each correlation entry is the Monte Carlo mean with the Monte Carlo standard deviation in parentheses. ``BCE'' is the Monte Carlo mean of the deterministic full-sample terminal binary cross-entropy used to compare optimization rounds and shock basins.}
\label{tab:mc_detailed_results}
\end{table}

Table~\ref{tab:mc_detailed_results} shows that in the correctly specified design, the mean recovery increases from $0.87$ when $T=250$ to $0.90$ when $T=2000$, with the improvement taking place gradually as the sample size increases. The recovery is very similar for both shocks, although it is slightly higher for Shock~1. The MC standard deviations generally decrease with the sample size, although the decrease is not uniform. The terminal BCE remains stable at approximately $0.59$--$0.60$ across the considered sample sizes. Notably, although the shock distribution is correctly specified, the model is still limited by the capacity of the neural networks $\boldsymbol{h}$, $\phi$, and $\eta$ to approximate the corresponding functions of the true DGP. 

The recovery results under the two misspecified shock distributions are close to those obtained under the correct specification. The mean recovery increases from $0.87$ to $0.89$ under Student-$t$ shocks and from $0.88$ to $0.90$ under skewed-$t$ shocks as the sample size increases from $250$ to $2000$. The MC standard deviations also generally decrease, although the skewed-$t$ design exhibits greater dispersion in the shorter samples. The terminal BCE is approximately $0.67$--$0.68$ in the misspecified designs, indicating less clear separation between the joint and shuffled samples than under correct specification. Nevertheless, this difference in classifier fit does not translate into materially weaker shock recovery, suggesting robustness to the considered heavy-tailed and asymmetric shock distributions.

\subsection{Basin selection}\label{sec:basin_selection}

\begin{table}[!t]
\centering
{\small
\setlength{\tabcolsep}{4pt}
\begin{tabular}{llcccc}
\hline\\[-1.3ex]
DGP & $T$ & Minimum BCE & Mean BCE & BCE+diagnostic & Oracle best round\\
\hline\\[-1.3ex]
Exponential-family & 250 & 0.87 (0.04) & 0.87 (0.04) & 0.87 (0.03) & 0.89 (0.01)\\
Exponential-family & 500 & 0.88 (0.04) & 0.88 (0.04) & 0.89 (0.02) & 0.89 (0.01)\\
Exponential-family & 1000 & 0.89 (0.01) & 0.89 (0.02) & 0.89 (0.01) & 0.90 (0.01)\\
Exponential-family & 2000 & 0.90 (0.02) & 0.90 (0.02) & 0.90 (0.01) & 0.90 (0.01)\\[1.0ex]
Student-$t$ & 250 & 0.87 (0.05) & 0.87 (0.05) & 0.86 (0.06) & 0.89 (0.02)\\
Student-$t$ & 500 & 0.88 (0.03) & 0.88 (0.04) & 0.88 (0.04) & 0.90 (0.01)\\
Student-$t$ & 1000 & 0.88 (0.02) & 0.88 (0.03) & 0.88 (0.02) & 0.89 (0.01)\\
Student-$t$ & 2000 & 0.89 (0.01) & 0.89 (0.02) & 0.89 (0.02) & 0.90 (0.01)\\[1.0ex]
Skewed-$t$ & 250 & 0.88 (0.06) & 0.86 (0.08) & 0.86 (0.09) & 0.90 (0.01)\\
Skewed-$t$ & 500 & 0.88 (0.05) & 0.87 (0.06) & 0.87 (0.05) & 0.90 (0.01)\\
Skewed-$t$ & 1000 & 0.89 (0.03) & 0.88 (0.04) & 0.88 (0.05) & 0.90 (0.01)\\
Skewed-$t$ & 2000 & 0.90 (0.02) & 0.89 (0.02) & 0.90 (0.01) & 0.90 (0.01)\\
\hline
\end{tabular}
}
\caption{Monte Carlo shock-recovery correlations under alternative local-solution selection rules. Each entry reports the Monte Carlo mean of the average aligned componentwise absolute correlation, with the Monte Carlo standard deviation in parentheses. The mean-BCE rule averages the three lowest BCE values within each basin. The oracle best-round rule is infeasible in empirical work and is included only as a benchmark for the finite-sample selection problem.}
\label{tab:mc_basin_selection}
\end{table}

The IIA-GCL objective is nonconvex and contains nuisance networks that enter the binary classifier. Consequently, different optimization starts may converge to different local solutions. We use the term shock basin for a group of local solutions that recover essentially the same shock series after the sign and permutation indeterminacies have been aligned. Different starts may recover nearly the same shocks while obtaining slightly different terminal binary cross-entropies through different nuisance-network fits. Conversely, two starts with similar binary cross-entropies may recover materially different shocks. To make this issue explicit, we group optimization starts into shock basins using the sign- and permutation-aligned shock correlations described in Appendix~\ref{sec:iia_gcl_training_details}. Then, the task is to select the shock basin. Since the method of selecting the basin may affect the recovery, it is informative to consider various basin selection rules for comparison.

The benchmark Monte Carlo results use the minimum-BCE basin selector. The minimum-BCE rule selects the basin containing the global minimum and uses its lowest-BCE member directly, whereas the mean-BCE rule selects the basin with the smallest average terminal binary cross-entropy among its three lowest-BCE members (if there are fewer than three members in the basin, then among all its members). The BCE+diagnostic basin selector rule retains basins whose lowest terminal BCE is at most $0.005$ above the global minimum and selects the basin with the smallest weighted shock-diagnostic score described in Appendix~\ref{sec:iia_gcl_training_details}. The score is computed from the three lowest-BCE rounds in each basin (or from all the rounds in the basin if there are fewer than three rounds), while the lowest-BCE member represents the selected basin. For this comparison, a scored round is considered admissible when $S_0\leq0.5$ and $\kappa(\widehat\Sigma_e)\leq10$ (see Appendix~\ref{sec:iia_gcl_training_details}). If at least one BCE-competitive basin contains an admissible scored round, selection is restricted to such basins; otherwise, all BCE-competitive basins remain eligible. Finally, the oracle best-round rule selects the optimization start with the largest true shock-recovery correlation. The oracle rule is infeasible in empirical applications because the true shocks are unobserved, but it is informative as a best-of-eight benchmark for the finite-sample selection problem. It quantifies selection loss conditional on the eight solutions found and is not an upper bound on attainable recovery. 

Our Monte Carlo results based on each of the above-described basin selection rules are reported in Table~\ref{tab:mc_basin_selection}. 
They are intended to separate three finite-sample issues. First, the optimizer may fail to find a good shock representation. Second, the optimizer may find a good representation in at least one round, while the feasible basin selection rule chooses a different basin. Third, several feasible basins may be nearly equivalent in terms of binary cross-entropy but differ in their shock diagnostics. The oracle column shows whether the feasible selector misses a better representation among the eight solutions found, while the feasible alternative selectors indicate whether basin-level averaging or shock diagnostics can improve selection without using information about the true shocks.

The results in Table~\ref{tab:mc_basin_selection} are broadly similar across the feasible selection rules, particularly in the larger samples. In the correctly specified design, the BCE-plus-diagnostic rule very slightly improves the mean recovery and reduces its MC standard deviation at some sample sizes. Under the two misspecified shock distributions, the minimum-BCE rule generally performs at least as well as the alternative feasible rules, while the mean-BCE and BCE-plus-diagnostic rules provide no uniform improvement and perform somewhat worse in some of the shorter samples. The oracle best-round rule gives higher recovery in the smaller samples, but its advantage relative to minimum-BCE selection becomes small when $T=2000$. Thus, local-solution selection accounts for part of the finite-sample variation in recovery, whereas the feasible selection rules lead to similar conclusions in the larger samples under both correct specification and misspecification.

\FloatBarrier

\section{Training details}

\subsection{Details on IIA-GCL training}\label{sec:iia_gcl_training_details}

The IIA-GCL estimator is trained using the binary cross-entropy loss $\text{LOSS}_{\text{bin}}$ defined by Equation~\eqref{eq:lossbin} in Section~\ref{sec:IIA-GCL}. This loss corresponds to the empirical version of the binary classification problem that distinguishes the observed triples $(y_t,\boldsymbol y_{t-1},u_t)$ from the shuffled triples $(y_t,\boldsymbol y_{t-1},u_t^\ast)$, where $u_t^\ast$ is obtained by randomly permuting the observed auxiliary variable. The estimated shocks are given by the output of the demixing network $\tilde e_t=\boldsymbol h(y_t,\boldsymbol y_{t-1})$.

In finite samples, the nonconvex neural-network optimization problem may have several solutions with similar values of $\text{LOSS}_{\text{bin}}$ but noticeably different recovered shocks. Preliminary experiments indicated that training solely with $\text{LOSS}_{\text{bin}}$ could occasionally produce recovered shocks with substantial serial dependence, lagged cross-dependence, or ill-behaving fitted natural-parameter functions. We therefore use a phased training schedule in which temporary guidance penalties are used during the early stages of optimization in order to steer the optimization algorithm away from uninteresting regions of the parameter space. Then, in the final phase, these penalties are entirely removed so that the obtained local solution minimizes $\text{LOSS}_{\text{bin}}$ rather than its penalized version. 

Specifically, in phase $m$, the implemented loss is
\begin{align}
\text{LOSS}^{(m)} &= \text{LOSS}_{\text{bin}} + \text{PEN}^{(m)}, \ \ \text{where}\\
\text{PEN}^{(m)} &= \omega_m^{\mathrm{rec}}\text{PEN}^{\mathrm{rec}} + \omega_m^{\mathrm{mom}}\text{PEN}^{\mathrm{mom}} + \omega_m^{\mathrm{ac}}\text{PEN}^{\mathrm{ac}} + \omega_m^{\mathrm{xac}}\text{PEN}^{\mathrm{xac}} \nonumber\\
&\quad + \omega_m^{\mathrm{act}}\text{PEN}^{\mathrm{act}} + \omega_m^{\mathrm{xact}}\text{PEN}^{\mathrm{xact}} + \omega_m^{\mathrm{wd}}\text{PEN}^{\mathrm{wd}}.
\end{align}
The final phase sets all weights $\omega_m^{\mathrm{rec}}$, $\omega_m^{\mathrm{mom}}$, $\omega_m^{\mathrm{ac}}$, $\omega_m^{\mathrm{xac}}$, $\omega_m^{\mathrm{act}}$, $\omega_m^{\mathrm{xact}}$, and $\omega_m^{\mathrm{wd}}$ equal to zero, so that the final phase optimizes only $\text{LOSS}_{\text{bin}}$. 

The guidance penalties are summarized in Table~\ref{tab:iia_gcl_penalties}. The reconstruction penalty $\text{PEN}^{\mathrm{rec}}$ uses an auxiliary neural-network approximation of the structural mapping $f$ in~\eqref{eq:genvar}. This auxiliary network is used during IIA-GCL training to encourage the pair $(\tilde e_t,\boldsymbol y_{t-1})$ to retain information relevant for reconstructing $y_t$. It should be distinguished from the fitted structural function $f_{\hat\theta}$ used in the nonlinear SVAR after the shocks have been recovered. The moment, autocorrelation, and cross-autocorrelation penalties are included because the recovered components should have the properties required of structural shocks. 

\begin{table}[!p]
\centering
\begin{tabular}{p{0.15\textwidth}p{0.35\textwidth}p{0.40\textwidth}}
\hline\\[-1.3ex]
Penalty & Definition & Purpose during training \\
\hline\\[-1.3ex]
Reconstruction $\text{PEN}^{\mathrm{rec}}$ &
A scale-normalized Huber loss between $y_t$ and its reconstruction from $(\tilde e_t,\boldsymbol y_{t-1})$ using an auxiliary neural-network approximation of the structural mapping $f$. &
Encourages the recovered shocks together with the lagged observations to contain enough information to reconstruct the current observation. This helps avoid representations that are useful for the binary classifier but discard information needed for the dynamic mapping from $(\boldsymbol y_{t-1},e_t)$ to $y_t$.\\[0.8em]

Moment $\text{PEN}^{\mathrm{mom}}$ &
A penalty based on the sample mean vector and covariance matrix of the recovered shocks $\tilde e_t$. In particular, it penalizes deviations of the sample mean from zero and deviations of the sample covariance matrix from the identity matrix. &
Discourages recovered shocks with large sample means, highly unequal component scales, strong contemporaneous correlations, or nearly singular covariance matrices. These finite-sample pathologies can make the recovered components difficult to interpret as structural shocks and can destabilize training.\\[0.8em]

Autocorrelation $\text{PEN}^{\mathrm{ac}}$ &
A penalty based on the sample autocorrelations of each recovered shock component at lags $1,\ldots,p_{\mathrm{guid}}$.&
Discourages own-lag dependence in each recovered structural shock component $\tilde e_{it}$.\\[0.8em]

Cross-autocorrelation $\text{PEN}^{\mathrm{xac}}$ &
A penalty based on the lagged sample cross-correlations between different recovered shock components at lags $1,\ldots,p_{\mathrm{guid}}$.&
Encourages dynamic separation between the recovered shocks by penalizing lagged dependence between $\tilde e_{it}$ and $\tilde e_{j,t-\ell}$ for $i\neq j$.\\[0.8em]

Autocorrelation tail $\text{PEN}^{\mathrm{act}}$ &
A threshold penalty applied to same-component shock autocorrelations at lags $1,\ldots,p_{\mathrm{guid}}$ whose absolute values exceed the threshold $c_{\mathrm{tail}}$. &
Adds extra pressure against the largest remaining own-lag autocorrelations without strongly penalizing very small sample correlations.\\[0.8em]

Cross-autocorrelation tail $\text{PEN}^{\mathrm{xact}}$ &
A threshold penalty applied to lagged cross-correlations between different shock components at lags $1,\ldots,p_{\mathrm{guid}}$ whose absolute values exceed $c_{\mathrm{tail}}$. &
Discourages the optimizer from moving serial dependence into a small number of large lagged cross-correlations.\\[0.8em]

Weight decay $\text{PEN}^{\mathrm{wd}}$ &
An $\ell_2$ penalty on the trainable parameters during the guided phases. &
Acts as temporary shrinkage regularization to reduce overfitting and extreme parameter values during training.\\
\hline
\end{tabular}
\caption{Guidance penalties used during IIA-GCL training.}\label{tab:iia_gcl_penalties}
\end{table}

In the specification used in our computations, $p_{\mathrm{guid}}=12$, so that the autocorrelation-based penalties $\text{PEN}^{\mathrm{ac}}$, $\text{PEN}^{\mathrm{xac}}$, $\text{PEN}^{\mathrm{act}}$, and $\text{PEN}^{\mathrm{xact}}$ use lags $1,\ldots,12$. The tail penalties use threshold $c_{\mathrm{tail}}=0.1$. The choice is intended to detect remaining short- and medium-run serial dependence in monthly data while keeping the number of sample correlations moderate relative to the sample size.

The phase schedule is reported in Table~\ref{tab:iia_gcl_schedule}. The guidance weights are largest at the beginning of training, gradually reduced, and then set to zero in the final phase. The fifth phase retains a small amount of guidance with a lower neural-network learning rate, whereas the final phase removes the guidance terms and refines the solution under the original binary cross-entropy objective. The fitted natural-parameter functions $\lambda_{ij}(u_t)$ are trained using a separate learning-rate multiplier. Their learning rate is reduced substantially after the first phase because these parameters can otherwise drift toward extreme and empirically uninteresting modulation paths; the final-phase lambda learning rate remains below the phase-one value even though the custom guidance weights are zero. 

\begin{table}[!t]
\centering
\resizebox{\textwidth}{!}{%
\begin{tabular}{lcccccc}
\hline\\[-1.3ex]
& Phase 1 & Phase 2 & Phase 3 & Phase 4 & Phase 5 & Phase 6 \\
\hline\\[-1.3ex]
Epochs
& 150 & 150 & 100 & 75 & 100 & 75 \\
Reconstruction weight $\omega_m^{\mathrm{rec}}$ & 80 & 40 & 10 & 2 & 0.5 & 0 \\
Moment weight $\omega_m^{\mathrm{mom}}$ & 10 & 10 & 5 & 2 & 5 & 0 \\
Autocorrelation weight $\omega_m^{\mathrm{ac}}$ & 7.5 & 7.5 & 3 & 1 & 1.25 & 0 \\
Cross-autocorrelation weight $\omega_m^{\mathrm{xac}}$ & 7.5 & 7.5 & 3 & 1 & 1.25 & 0 \\
Autocorrelation tail weight $\omega_m^{\mathrm{act}}$ & 25 & 25 & 10 & 5 & 2.5 & 0 \\
Cross-autocorrelation tail weight $\omega_m^{\mathrm{xact}}$ & 25 & 25 & 10 & 5 & 2.5 & 0 \\
L2 weight decay $\omega_m^{\mathrm{wd}}$ & $5\cdot 10^{-6}$ & $5\cdot 10^{-6}$ & $5\cdot 10^{-6}$ & $5\cdot 10^{-6}$ & $5\cdot 10^{-6}$ & 0 \\
Learning-rate multiplier & 1 & 1 & 1 & 1 & 0.5 & 1 \\
Lambda learning-rate multiplier & 1 & 0.2 & 0.1 & 0.05 & 0.02 & 0.1 \\
\hline
\end{tabular}
}
\caption{Phased training schedule for the IIA-GCL specification used in the reported Monte Carlo and empirical computations.}\label{tab:iia_gcl_schedule}
\end{table}

The demixing network uses sigmoid activation functions in its hidden layer. The nuisance networks $\phi$ and $\eta$ use leaky-ReLU activation functions, and the temporary reconstruction decoder uses sigmoid activation functions. The neural networks are initialized with Xavier initialization. The demixing network uses hidden-layer gain $2$ and output-layer gain $0.05$, while the nuisance networks and the temporary reconstruction decoder use hidden-layer gain $0.25$ and output-layer gain $0.05$. In the computations reported below, the demixing network has one hidden layer with 12 neurons, the nuisance networks have one hidden layer with 8 neurons, and the temporary reconstruction decoder has one hidden layer with 16 neurons. 

The lambda-function parameters are initialized with small component-specific endpoint differences. In particular, the initial endpoint values are centered around $(-0.25,0.25)$ for the log-scale transition and around $(0.15,-0.15)$ for the log-shape transition, with small component-specific offsets. The transition locations are initialized around the empirical $0.4$ and $0.6$ quantiles of $u_t$, again with small component-specific offsets, and the transition steepness parameters are initialized at $\kappa_{ij}=2$. 

The IIA-GCL model is trained using the Adam optimizer with base learning rate $0.002$, minibatch size $64$, and gradient clipping at norm $5$. Reconstruction guidance uses Huber loss with threshold one. The learning-rate multipliers in Table~\ref{tab:iia_gcl_schedule} determine the learning rate in each phase, which is held constant within the phase, while the nuisance-network learning-rate multiplier is one throughout. The temporary $\ell_2$ weight decay coefficient is $5\cdot 10^{-6}$ in the first five phases and zero in the final phase. This can be interpreted as using standard weight decay to stabilize the guided optimization path while leaving the terminal phase as an unregularized binary-classification problem. Exponential moving averaging, early stopping, and best-checkpoint replacement are not used, so each round returns the endpoint of the complete phased schedule.

The binary cross-entropy loss $\text{LOSS}_{\text{bin}}$ is evaluated by comparing the observed triples $(y_t,\boldsymbol y_{t-1},u_t)$ with artificially shuffled triples $(y_t,\boldsymbol y_{t-1},u_t^\ast)$. The shuffled auxiliary variables are obtained by deranging the observed auxiliary-variable sample, so that the empirical marginal distribution of $u_t$ is preserved while the contemporaneous link between $(y_t,\boldsymbol y_{t-1})$ and $u_t$ is broken. 

In the final phase, in order to have a fixed final loss function, the auxiliary-variable shuffling used in the binary cross-entropy calculation is fixed before the optimization rounds are run. Equation~\eqref{eq:lossbin} writes the empirical loss with one shuffled auxiliary-variable draw $u_t^\ast$ per observation. To improve finite-sample performance of our method, we augment the data and use a multi-derangement version of this finite-sample criterion. Specifically, we draw $K=4$ derangements $s_1,\ldots,s_K$ of the aligned time indices, where a derangement means a permutation with no observation paired with itself, and minimize
\begin{equation}
\text{LOSS}_{\text{bin}}^{T,K}
=-\frac{1}{TK}\sum_{t=1}^T\sum_{k=1}^K
\left[
\log\Lambda^{\mathrm{std}}\!\left(r(y_t,\boldsymbol y_{t-1},u_t)\right)
+\log\left\{1-\Lambda^{\mathrm{std}}\!\left(r(y_t,\boldsymbol y_{t-1},u_{s_k(t)})\right)\right\}
\right].
\end{equation}
Thus each observation is paired with its observed auxiliary-variable value and with $K$ fixed shuffled auxiliary-variable values. The same $K=4$ derangements are reused for all of the final-phase epochs and all optimization rounds. This makes the terminal objective fixed across epochs and rounds, while reducing the finite-sample noise associated with a single random shuffle of the auxiliary-variable sample.\footnote{The population objective~\eqref{eq:popobj} is unchanged by this multi-derangement averaging. Let $\boldsymbol{y}_t^+\equiv(y_t,\boldsymbol y_{t-1})$, let $U^\ast_1,\ldots,U^\ast_K$ be independent draws from the marginal distribution of $u_t$, independent of $\boldsymbol{y}_t^+$. For any fixed classifier $r$, since the random vectors $(\boldsymbol{y}_t^+,U^\ast_k)$ are identically distributed across $k$, linearity of expectation gives
\begin{align}
\begin{aligned}
&\mathbb{E}_{\tilde{\boldsymbol{y}}_t}\left[\frac{1}{K}\sum_{k=1}^K\log \Lambda^{\text{std}}(r(\boldsymbol{y}_t^+,u_t))\right] + \mathbb{E}_{\tilde{\boldsymbol{y}}_t^\ast}\left[\frac{1}{K}\sum_{k=1}^K\log (1-\Lambda^{\text{std}}(r(\boldsymbol{y}_t^+,U^\ast_k)))\right]\\
&= \mathbb{E}_{\tilde{\boldsymbol{y}}_t}\big[\log \Lambda^{\text{std}}(r(\boldsymbol{y}_t^+,u_t))\big] + \mathbb{E}_{\tilde{\boldsymbol{y}}_t^\ast}\big[\log (1-\Lambda^{\text{std}}(r(\boldsymbol{y}_t^+,U^\ast_1)))\big] = \mathcal{L}(r).
\end{aligned}
\end{align}
Hence, $K=1$ and any fixed $K>1$ have the same population loss and thereby the same population maximizers. $K>1$ only changes the finite-sample approximation to this loss.} The preceding guided phases use fresh derangements across epochs.

\begin{table}[!t]
\centering
\begin{tabular}{p{0.14\textwidth}p{0.35\textwidth}p{0.35\textwidth}}
\hline\\[-1.3ex]
Selection rule & Definition & Interpretation \\
\hline\\[-1.3ex]
Minimum binary cross-entropy &
Select the basin whose representative round has the smallest terminal binary cross-entropy. The representative round is the lowest-loss round within the basin. &
Keeps the usual minimum-loss selection rule while making the recovered-shock basin explicit.\\[1.5em]
Minimum-BCE basin with diagnostic representative &
Select the minimum-BCE basin, retain its members whose binary cross-entropy is within a specified tolerance of the basin minimum, and use the retained member with the smallest shock-diagnostic score as the representative. &
Keeps basin selection tied to the classification objective while reducing sensitivity to nuisance-network differences among rounds that recover essentially the same shocks.\\[1.5em]
Mean binary cross-entropy &
Select the basin with the smallest mean terminal binary cross-entropy, using either all rounds in the basin or the best few rounds in the basin. &
Accounts for the possibility that the nuisance networks give slightly different binary cross-entropy values to rounds recovering the same shocks.\\[0.4em]
BCE plus diagnostic basin &
Retain basins whose minimum binary cross-entropy is within a specified tolerance of the global minimum. Apply the diagnostic-admissibility restrictions and minimize the weighted shock-diagnostic score among the remaining basins. &
Useful when several basins fit the classification problem nearly equally well but differ in whether the recovered components display the properties required of structural shocks. \\[6.5em]
Manual &
Select a user-specified basin. &
Used for robustness analysis and for reporting how the empirical conclusions change across near-best shock basins.\\
\hline
\end{tabular}
\caption{Alternative ex-post basin selection rules for multiround IIA-GCL estimation.}\label{tab:iia_gcl_basin_selection}
\end{table}

To reduce sensitivity to random initialization, the IIA-GCL model is estimated from multiple independent starting values. In the Monte Carlo computations, we use eight starting values, while in the empirical application we use 500 starting values. The nonconvexity of the problem implies that different rounds may converge to different local solutions. Moreover, due to the presence of the nuisance functions $\phi$ and $\eta$ in the classifier, two solutions may recover essentially the same shocks while having somewhat different binary cross-entropy values. Conversely, solutions with nearly identical binary cross-entropy values may recover materially different shocks. Therefore, the multiround output is grouped into shock basins such that the recovered shocks are similar across all solutions within each basin. Two rounds are assigned to the same basin when their recovered shocks are sufficiently similar after allowing for the sign and permutation indeterminacies remaining under Corollary~\ref{cor:logistic_ident}. Our benchmark thresholds require a mean absolute aligned component correlation of at least $0.95$ and a weakest-component aligned correlation of at least $0.90$.

After the basins have been formed, the final Stage-1 solution is determined by two distinct choices. First, a basin is selected by comparing the basins according to one of the rules summarized in Table~\ref{tab:iia_gcl_basin_selection}. Second, a representative round is selected from within that basin. Keeping these choices separate is useful because the criterion used to distinguish materially different recovered-shock solutions need not be the same as the criterion used to choose among rounds that recover essentially the same shocks.

The simplest basin selector uses only the classification objective. In particular, the minimum-BCE rule selects the basin containing the round with the smallest terminal BCE. The mean-BCE rule instead compares basin-level mean BCEs, calculated either from all members or from a specified number of the lowest-BCE members. These rules use no shock diagnostics. Diagnostic information enters only when the BCE-plus-diagnostic basin selector is used, or when a diagnostic representative is selected from an already chosen basin. We therefore define the diagnostic score before describing these two uses.

The score summarizes whether the recovered components have basic sample properties expected of structural shocks. It may be computed from a few with the lowest-BCE, all rounds, or only the lowest-BCE round of each basin. The default uses the three lowest-BCE rounds. Let $L=\min(24,T-1)$ and let $\widehat\rho_{jk}(\ell)=\operatorname{Corr}(\widehat e_{j,t},\widehat e_{k,t-\ell})$. For each scored round, define
\begin{align}
S_{\mathrm{acf}}
&=\left\{\frac{1}{L}\sum_{\ell=1}^{L}
\left(\frac{1}{d}\sum_{j=1}^{d}|\widehat\rho_{jj}(\ell)|\right)^2\right\}^{1/2},\\
S_{\mathrm{cross}}
&=\left\{\frac{1}{L}\sum_{\ell=1}^{L}
\left(\frac{1}{d(d-1)}\sum_{j\neq k}|\widehat\rho_{jk}(\ell)|\right)^2\right\}^{1/2},\\
S_{\max}
&=\max_{1\leq\ell\leq L}\left\{
\max_j|\widehat\rho_{jj}(\ell)|,
\max_{j\neq k}|\widehat\rho_{jk}(\ell)|\right\},\\
S_0&=\max_{j\neq k}|\widehat\rho_{jk}(0)|,\qquad
S_{\mu}=\frac{1}{d}\sum_{j=1}^{d}|\overline e_j|,\qquad
S_{\Sigma}=\log\kappa(\widehat\Sigma_e),
\end{align}
where $\kappa(\widehat\Sigma_e)$ is the condition number of the sample shock covariance matrix. The first two measures summarize persistent same-component and cross-component lag dependence, while $S_{\max}$ prevents an isolated large correlation from being hidden by averaging. The scalar score is
\begin{equation}
S=w_{\mathrm{acf}}S_{\mathrm{acf}}
+w_{\mathrm{cross}}S_{\mathrm{cross}}
+w_{\max}S_{\max}+w_0S_0+w_{\mu}S_{\mu}+w_{\Sigma}S_{\Sigma}.
\label{eq:bce_diagnostic_score}
\end{equation}
Our benchmark weights $(w_{\mathrm{acf}},w_{\mathrm{cross}},w_{\max},w_0,w_{\mu},w_{\Sigma})$ are $(1,1,0.25,1,0.25,0.05)$. The basin score is obtained by averaging each component across the rounds retained for scoring and then applying~\eqref{eq:bce_diagnostic_score}.

The BCE-plus-diagnostic selector restricts the diagnostic comparison to basins that fit the classification problem nearly as well as the best solution found. Specifically, basin $b$ is treated as BCE-competitive when
\begin{equation}
\min_{r\in b}\operatorname{BCE}_r
\leq \min_s\operatorname{BCE}_s+\delta_{\mathrm{BCE}},
\label{eq:bce_diagnostic_band}
\end{equation}
where $\delta_{\mathrm{BCE}}\geq0$ is the specified BCE tolerance and the minimum on the right-hand side is taken across all successful rounds. The diagnostic score is then used to rank only these BCE-competitive basins. Thus, BCE first limits attention to solutions that provide sufficiently similar classification fit, after which the recovered-shock diagnostics distinguish among them.

For the empirical application, we implement the BCE-plus-diagnostic basin selector with $\delta_{\mathrm{BCE}}=0.006$ and use the three lowest-BCE rounds in each basin for diagnostic scoring. A scored round is considered admissible when $S_{\mathrm{acf}}\leq0.08$, $S_{\mathrm{cross}}\leq0.08$, $S_{\max}\leq0.20$, $S_0\leq0.20$, $S_{\mu}\leq0.30$, and $\kappa(\widehat\Sigma_e)\leq5$. Within each BCE-competitive basin, if at least one scored round satisfies all these limits, only the admissible rounds contribute to the basin score; otherwise, all scored rounds are used. If at least one BCE-competitive basin contains an admissible round, only such basins remain eligible and the basin with the smallest weighted diagnostic score is selected. If none does, all BCE-competitive basins are ranked by their diagnostic scores, and the resulting choice is treated as a fallback rather than as satisfying the admissibility criteria. 

Once the basin has been selected, a representative round is chosen separately from among its members. Under the diagnostic representative rule used in the empirical application, the representative minimizes~\eqref{eq:bce_diagnostic_score} among the basin members whose BCE is within $\delta_{\mathrm{BCE}}$ of that basin's minimum. Alternative rules use the basin's lowest-BCE member, the member with the greatest average aligned-shock agreement with the other members, or a manually specified round. 

\FloatBarrier

\subsection{Details on SVAR training}\label{sec:svar_training_details}

After the IIA-GCL solution has been selected (Stage-1), the recovered shocks are held fixed during FFNN-SVAR estimation (Stage-2). Thus the FFNN-SVAR training rounds do not re-estimate the demixing network or the conditional shock distribution. Instead, each Stage-2 round fits only the prediction map $f_{\theta}$ in~\eqref{eq:nn_svar}, using the normalized recovered shocks $\tilde e_t$ and the lag vector $\boldsymbol y_{t-1}$ as inputs. This separation is important because the Stage-2 optimization problem is a nonlinear prediction problem conditional on a given shock path, whereas the Stage-1 problem is a contrastive shock-recovery problem.

The computations reported in this paper use a scale-normalized Huber prediction loss. Specifically, let $s_i$ denote the sample standard deviation of the $i$th dependent variable on the numerical scale used for training, and let
\begin{equation}
a_{it}(\theta)=\frac{y_{it}-f_{\theta,i}(\boldsymbol y_{t-1},\tilde e_t)}{s_i}
\end{equation}
be the componentwise scaled prediction error. For a threshold $\delta>0$, define
\begin{equation}
\ell_{\delta}(a)=
\begin{cases}
\frac{1}{2}a^2, & |a|\leq\delta,\\
\delta\left(|a|-\frac{1}{2}\delta\right), & |a|>\delta.
\end{cases}
\end{equation}
In phase $m$, the objective optimized over a minibatch $\mathcal B_m\subset\{p+1,\ldots,T\}$ is
\begin{equation}\label{eq:svar_training_loss}
\text{LOSS}_{\mathrm{SVAR}}^{(m)}(\theta)
=\frac{1}{|\mathcal B_m|d}\sum_{t\in\mathcal B_m}\sum_{i=1}^d
\ell_{\delta_m}\!\left(a_{it}(\theta)\right).
\end{equation}
The Huber loss is used to reduce the influence of unusually large prediction errors, which is useful in macroeconomic samples containing large outliers. The scale normalization prevents variables with larger numerical units from dominating the objective. 

The hidden layers of $f_{\theta}$ use leaky-ReLU activation functions. The FFNN-SVAR network is initialized with the following small Xavier initialization. Hidden-layer weights use Xavier initialization with gain $0.25$, the output layer uses Xavier initialization with gain $0.05$, and biases are initialized at zero. The main training stage uses the Adam optimizer with base learning rate $0.002$, minibatch size $64$, gradient clipping at norm $5$, and no weight decay. In addition to the phase-specific multipliers in Table~\ref{tab:svar_training_schedule}, the learning rate is multiplied by $0.5$ after every 300 optimizer steps. During minibatch training, we keep an exponential moving average (EMA) of the network parameters, updated after each optimizer step with decay $0.99$. When a new lowest training objective is reached, the corresponding averaged parameter vector is stored as the best checkpoint. After minibatch training, the network parameters are set to this stored averaged checkpoint, and the full-sample finalization stages are initialized from it rather than necessarily from the last minibatch-training iterate.

\begin{table}[!t]
\centering
\begin{tabular}{lcccc}
\hline\\[-1.3ex]
& Phase 1 & Phase 2 & Phase 3 & Phase 4 \\
\hline\\[-1.3ex]
Epochs & 160 & 80 & 80 & 80 \\
Learning-rate multiplier & 1 & 0.5 & 0.1 & 0.02 \\
Huber threshold $\delta_m$ & 0.5 & 0.5 & 0.5 & 0.5 \\
L2 weight decay & 0 & 0 & 0 & 0 \\
Minibatch size & 64 & 64 & 64 & 64 \\
\hline
\end{tabular}
\caption{Minibatch Adam training schedule for the FFNN-SVAR prediction map.}\label{tab:svar_training_schedule}
\end{table}

The minibatch Adam stage is followed by deterministic full-sample finalization. The first finalization stage continues from the selected minibatch-training checkpoint and applies Adam to the full aligned sample, using a smaller learning rate and a smaller Huber threshold. The second finalization stage applies a small number of full-sample limited-memory BFGS (LBFGS) steps. These finalization stages optimize the same prediction objective as~\eqref{eq:svar_training_loss}, with the minibatch $\mathcal B_m$ replaced by the full aligned sample. Their purpose is numerical refinement of the prediction map, especially reduction of the final gradient norm and removal of noise from the last stochastic minibatch updates.

\begin{table}[!t]
\centering
\begin{tabular}{lccc}
\hline\\[-1.3ex]
& Full-sample Adam 1 & Full-sample Adam 2 & LBFGS \\
\hline\\[-1.3ex]
Iterations & 80 epochs & 80 epochs & 6 outer steps \\
Learning rate & $10^{-5}$ & $2.5\cdot10^{-6}$ & 0.05 \\
Maximum inner iterations & - & - & 20 \\
Huber threshold $\delta$ & 0.25 & 0.25 & 0.25 \\
L2 weight decay & 0 & 0 & 0 \\
Checkpoint selection & Near-loss gradient & Near-loss gradient & Near-loss gradient \\
\hline
\end{tabular}
\caption{Full-sample finalization schedule for the FFNN-SVAR prediction map.}\label{tab:svar_finalization_schedule}
\end{table}

For each full-sample finalization stage, the selected checkpoint is the one with the smallest gradient norm among checkpoints whose Huber prediction loss is within
\begin{equation}
\max\{10^{-4},\,0.10\cdot\text{LOSS}_{\mathrm{Huber}}^{\mathrm{best}}\}
\end{equation}
of the best Huber prediction loss reached during that stage. This selection rule avoids choosing a substantially worse prediction map merely because it has a smaller gradient norm, while still emphasizing numerical stationarity among nearly equivalent fits.

The FFNN-SVAR optimization problem is nonconvex, and raw neural-network parameter vectors are not a useful basis for comparing solutions because different parameter values can represent essentially the same prediction function. We therefore estimate the Stage-2 model from 500 independent starting values and compare the resulting fitted-value paths $f_{\hat\theta}(\boldsymbol y_{t-1},\tilde e_t)$. The successful rounds are grouped into fitted-value basins using complete-link comparisons. Two rounds can be assigned to the same basin only if the correlation between their vectorized fitted-value paths is at least $0.98$ and their root mean squared difference, divided by the standard deviation of their pooled fitted values, is at most $0.10$. The correlation criterion measures similarity in the shape of the fitted paths, while the scale-normalized root mean squared difference rules out solutions that are highly correlated but materially different in level or scale.

We first retain the fitted-value basin containing the smallest Huber selection loss. Within that basin, rounds whose Huber selection loss is no more than $15\%$ above the basin minimum are considered as candidate representatives. The representative is selected by minimizing
\begin{equation}
D=A_{\mathrm{RMS}}+C_{\mathrm{RMS}}+|C_0|+\frac{1}{4}(A_{\max}+C_{\max}),
\end{equation}
where $A_{\mathrm{RMS}}$ is the root mean square of the own-component prediction-error autocorrelations over lags $1,\ldots,24$, $C_{\mathrm{RMS}}$ is the root mean square of the cross-component prediction-error correlations over the nonzero lags from $-24$ to $24$, $A_{\max}$ and $C_{\max}$ are the corresponding largest absolute correlations, and $C_0$ is the contemporaneous prediction-error correlation. The gradient norm and Huber loss are used as tie breakers. The Huber loss measures prediction-error magnitude robustly, but does not ensure that the prediction errors are serially or cross-sectionally uncorrelated. The dependence score is therefore used to distinguish among prediction maps with similar Huber losses that leave different amounts of systematic dynamic dependence unexplained.

\FloatBarrier

\section{Empirical application details}\label{sec:empapp_details}

\subsection{Shock diagnostics}\label{sec:empapp_shockdiag}

Direct diagnostic checks of the maintained assumptions are difficult in the present framework. Unlike in typical SVAR models, the assumptions on the shocks (Assumption~\ref{as:shocks}) are formulated conditional on the auxiliary variable $u_t$. Since $u_t$ is typically continuous and only a single observation is available at each time period, formal diagnostic checks of these conditional assumptions are difficult to construct. Nevertheless, Assumption~\ref{as:shocks} implies the absence of both serial and cross-sectional correlation in the shocks. Therefore, although not sufficient for validating the maintained assumptions, examination of the auto- and cross-correlation functions of the recovered shock series provides a useful diagnostic check in the spirit of conventional SVAR analysis.

\begin{figure}[t]
    \centerline{\includegraphics[width=\textwidth - 2cm]{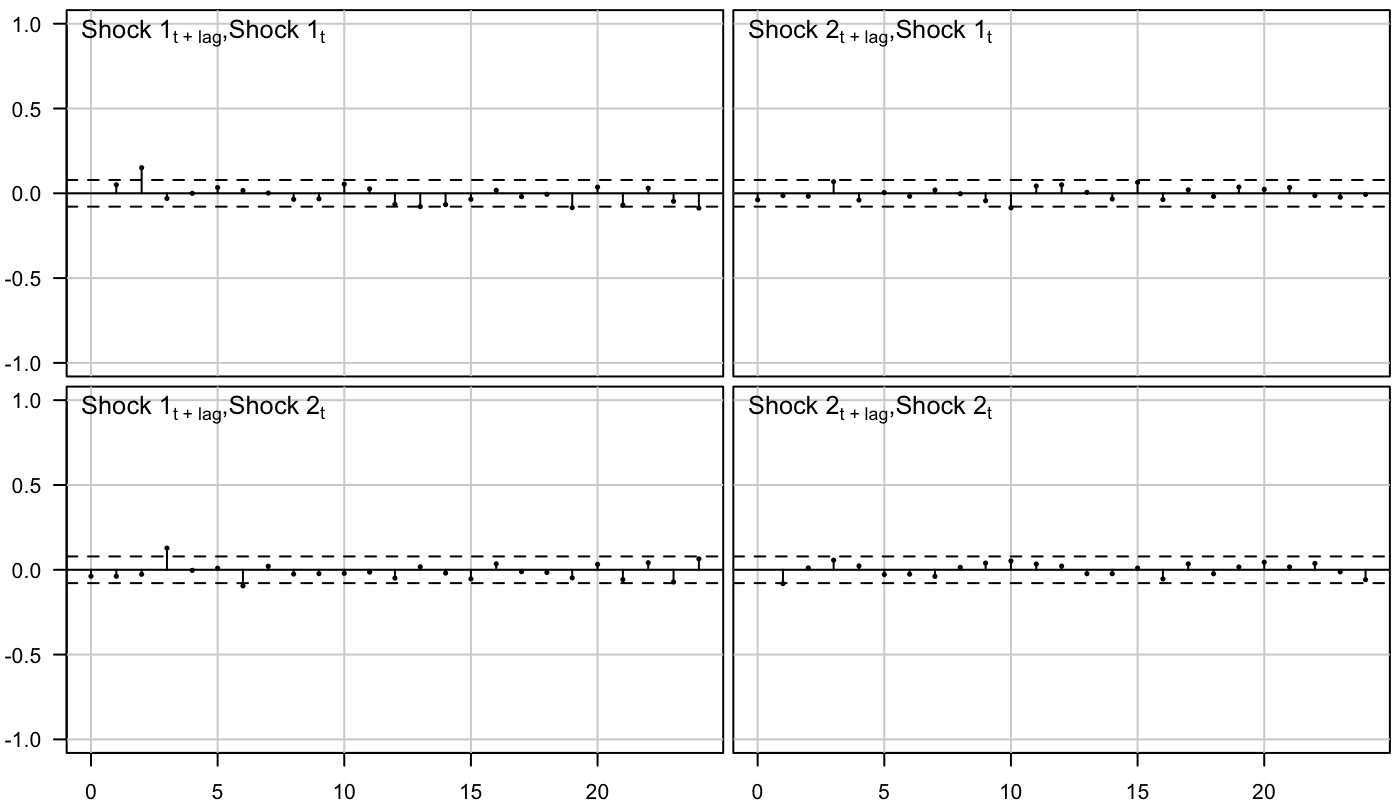}}
    \caption{Auto- and crosscorrelation functions of the recovered shocks of the IIA-GCL model fitted in Section~\ref{sec:empirical} for the lags $0,1,...,24$. The lag zero autocorrelation coefficients are omitted, as they are one by convention. The dashed lines are the $95\%$ bounds $\pm 1.96/\sqrt{T}$ for autocorrelations of IID observations.}
\label{fig:shock_acf}
\end{figure}

The auto- and cross-correlation functions of the shocks recovered in Section~\ref{sec:emp_training} are presented in Figure~\ref{fig:shock_acf} for the first $24$ lags. The figure shows that there is not much autocorrelation in the shocks, although there is a moderate correlation coefficient in second lag of Shock~1's autocorrelation function and in the fourth lag of the cross-correlation function between Shock~1 and Shock~2. Nevertheless, the auto- and crosscorrelation properties of the shocks seem overall reasonable. 

\subsection{SVAR fit diagnostics}\label{sec:svar_diagnostics}

To assess the adequacy of the fitted structural function $f_{\hat{\theta}}$, it is useful to examine the prediction errors $y_t - f_{\hat{\theta}}(\boldsymbol{y}_{t-1}, \tilde{e}_t)$ from the fitted SVAR model. These prediction errors are neither structural shocks nor reduced-form innovations, and they have no independent structural interpretation. Rather, they reflect a combination of finite-sample estimation error in the fitted function $f_{\hat{\theta}}$, approximation error of the chosen neural network architecture, and estimation error in the recovered shocks $\tilde{e}_t$ used as inputs. Their role is therefore purely diagnostic, providing an overall assessment of whether the fitted specification leaves substantial systematic structure unexplained. Importantly, these diagnostics do not allow one to disentangle whether potential inadequacies arise from the fitted mapping or from the recovered shocks.

\begin{figure}[p]
    \centerline{\includegraphics[width=\textwidth - 2cm]{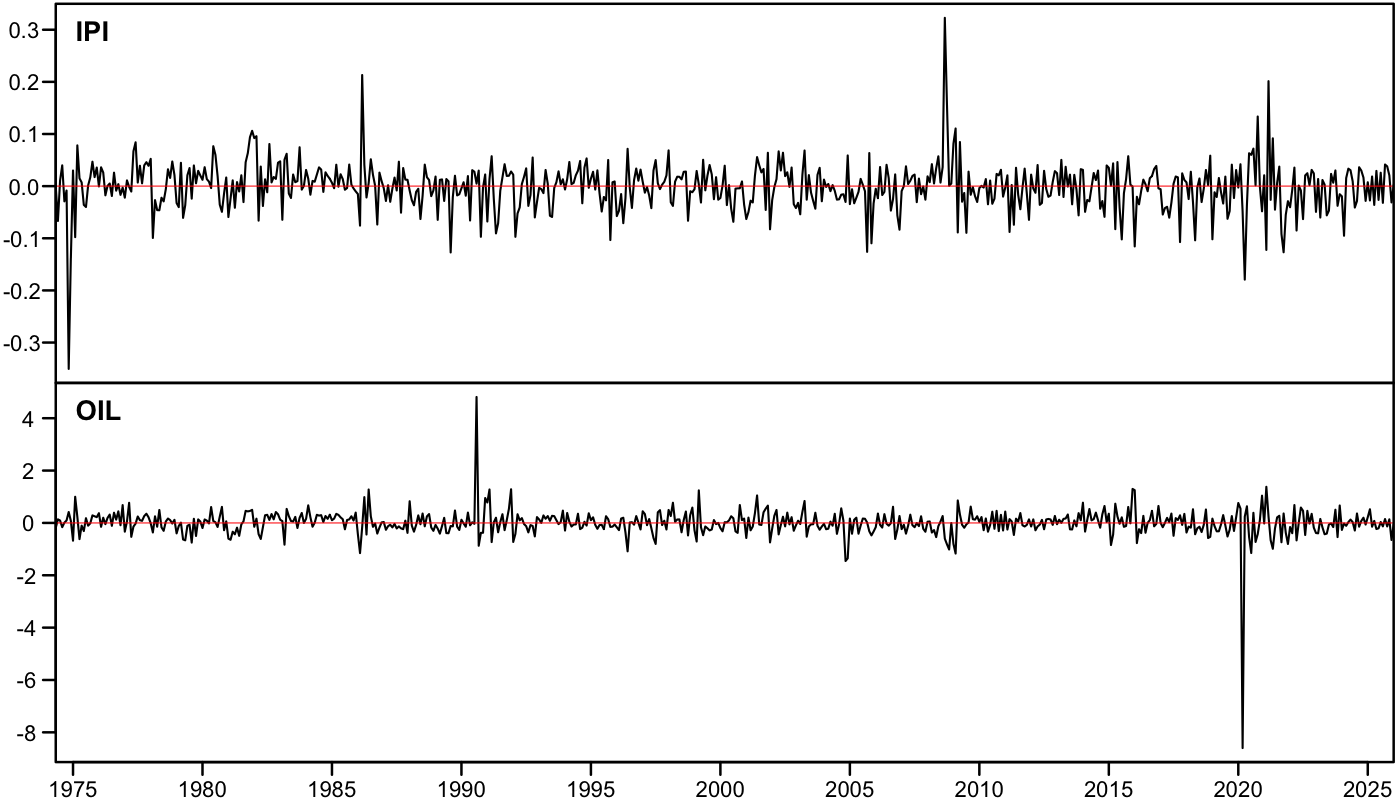}}
    \caption{Prediction errors of the FFNN-SVAR model fitted in Section~\ref{sec:empirical} shown in each panel for IPI and OIL, respectively, from top to bottom.}
\label{fig:pred_error_series}
\end{figure}

A first diagnostic examines the time series behavior of the prediction errors. Plotting the prediction error series over time provides an overall visual check of whether they exhibit obvious nonrandom structure. 
Figure~\ref{fig:pred_error_series} presents the prediction error series for our FFNN-SVAR model fitted in Section~\ref{sec:empirical}. The Gulf War oil shock in August 1990 and the COVID-19 shock in March 2020 are clearly visible for real oil price prediction errors (bottom panel). This presumably reflects the possibility that the recovered shock series (the second column in Figure~\ref{fig:shockplot}) does not capture these two shocks at magnitudes sufficient to predict the corresponding abrupt changes in oil prices. For IPI there are no particularly large prediction errors. Overall, there do not seem to be any obvious systematic patterns in the prediction errors.

\begin{figure}[p]
    \centerline{\includegraphics[width=\textwidth - 2cm]{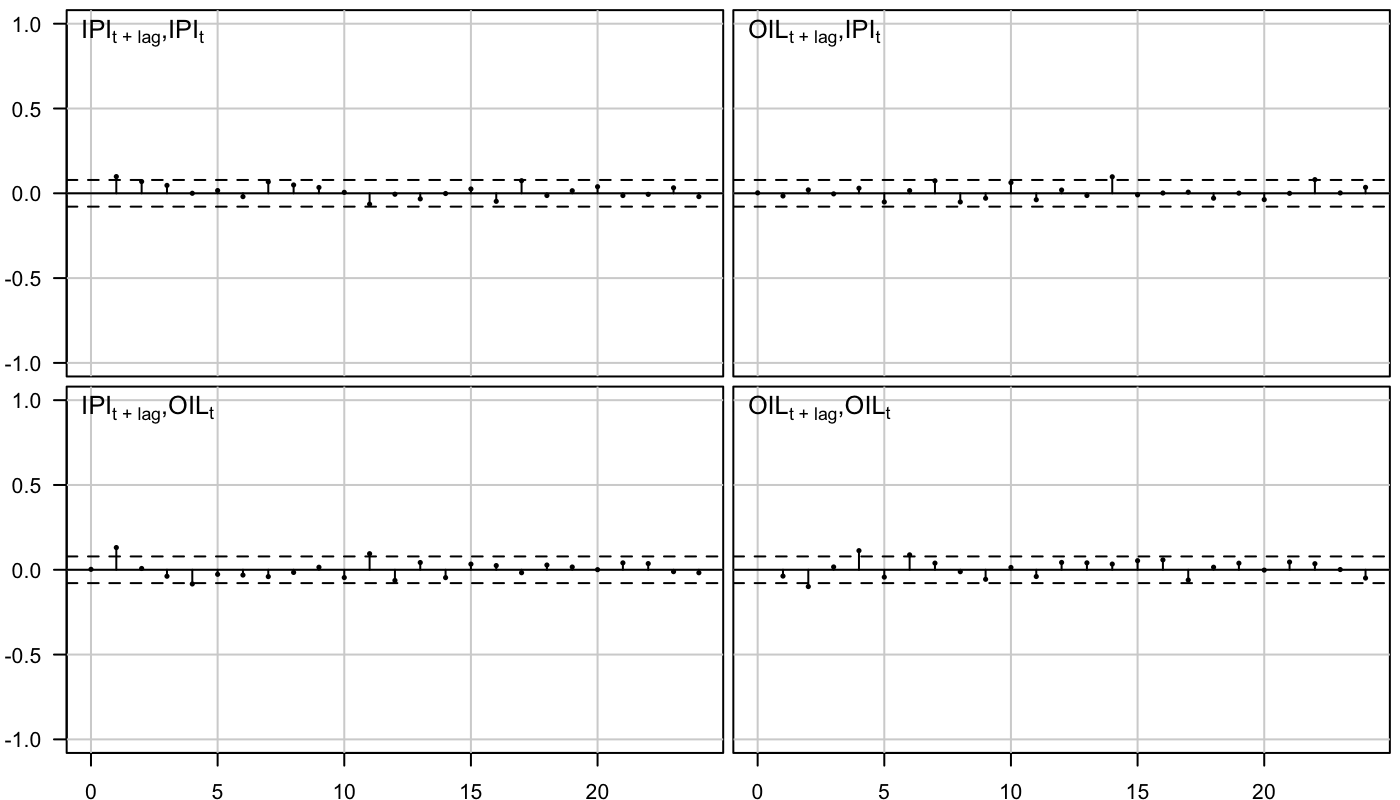}}
    \caption{Auto- and crosscorrelation functions of the prediction errors of the FFNN-SVAR model fitted in Section~\ref{sec:empirical} for the lags $0,1,...,24$. The lag zero autocorrelation coefficients are omitted, as they are one by convention. The dashed lines are the $95\%$ bounds $\pm 1.96/\sqrt{T}$ for autocorrelations of IID observations.}
\label{fig:pred_error_acf}
\end{figure}

A second diagnostic examines the temporal and cross-sectional dependence of the prediction errors using autocorrelation and cross-correlation functions. Strong and persistent autocorrelation in a given component suggests that the fitted specification does not fully capture the dynamic dependence in that equation. Similarly, substantial cross-correlation between prediction error components may indicate that the fitted system does not adequately account for cross-variable interactions. Figure~\ref{fig:pred_error_acf} presents the auto- and crosscorrelation functions for the prediction errors of our FFNN-SVAR model fitted in Section~\ref{sec:empirical}. The figure shows that there is little auto- or crosscorrelation remaining in the prediction errors. Thus, overall, the FFNN-SVAR fit seems reasonable.

\end{appendices}

\end{document}